\documentclass[%
 reprint,
superscriptaddress,
nobibnotes,
 amsmath,amssymb,
 aps,
 prx,
floatfix,
longbibliography,
raggedbottom
]{revtex4-2}

\usepackage{formatting}
\DeclareSIUnit{\inch}{in}
\usepackage{multirow}
\usepackage{listings}

\usepackage{hyperref}
\hypersetup{
    colorlinks   = true,
    linkcolor    = Maroon,
    citecolor    = Maroon
}

\newif\ifincludesupp
\includesupptrue

\renewcommand\op{\hat}

\defaultbibliography{citations}
\defaultbibliographystyle{apsrev4-2}

\makeatletter
\newcommand{\bibunitcontrol}[1]{%
    \@startbibunitorrelax
    \if@filesw
        \immediate\write\@bibunitaux{\string\citation{#1}}%
    \fi
}
\makeatother

\makeatletter
\def\maketitle{%
    \@author@finish
    \title@column\titleblock@produce
    \suppressfloats[t]}
\makeatother

\begin{document}

\preprint{APS/123-QED}

\title{Bosonic Error Correction with Fluxonium}


\author{Shantanu~R.~Jha}
\thanks{These authors contributed equally to this work. \\ Correspondence to: \href{mailto:shanjha@mit.edu}{shanjha@mit.edu}, \href{mailto:shoumikc@mit.edu}{shoumikc@mit.edu}}
\affiliation{Research Laboratory of Electronics, Massachusetts Institute of Technology, Cambridge, Massachusetts 02139, USA}
\affiliation{Department of Electrical Engineering and Computer Science, Massachusetts Institute of Technology, Cambridge, Massachusetts 02139, USA}

\author{Shoumik~D.~Chowdhury}
\thanks{These authors contributed equally to this work. \\ Correspondence to: \href{mailto:shanjha@mit.edu}{shanjha@mit.edu}, \href{mailto:shoumikc@mit.edu}{shoumikc@mit.edu}}
\affiliation{Research Laboratory of Electronics, Massachusetts Institute of Technology, Cambridge, Massachusetts 02139, USA}
\affiliation{Department of Electrical Engineering and Computer Science, Massachusetts Institute of Technology, Cambridge, Massachusetts 02139, USA}

\author{Gabriele~Rolleri}
\affiliation{Research Laboratory of Electronics, Massachusetts Institute of Technology, Cambridge, Massachusetts 02139, USA}
\affiliation{Department of Information Technology and Electrical Engineering, ETH Z\"{u}rich, ZH 8092, Switzerland}

\author{Anaida~Ali}
\affiliation{Départment de Physique and Institut Quantique,
Université de Sherbrooke, Sherbrooke J1K 2R1 Québec, CAN\looseness=-1}

\author{\\ Lev-Arcady~Sellem}
\affiliation{Départment de Physique and Institut Quantique,
Université de Sherbrooke, Sherbrooke J1K 2R1 Québec, CAN\looseness=-1}

\author{Réouven~Assouly}
\affiliation{Research Laboratory of Electronics, Massachusetts Institute of Technology, Cambridge, Massachusetts 02139, USA}

\author{David~Pahl}
\affiliation{Research Laboratory of Electronics, Massachusetts Institute of Technology, Cambridge, Massachusetts 02139, USA}
\affiliation{Department of Electrical Engineering and Computer Science, Massachusetts Institute of Technology, Cambridge, Massachusetts 02139, USA}

\author{Lukas~Pahl}
\affiliation{Research Laboratory of Electronics, Massachusetts Institute of Technology, Cambridge, Massachusetts 02139, USA}
\affiliation{Department of Electrical Engineering and Computer Science, Massachusetts Institute of Technology, Cambridge, Massachusetts 02139, USA}

\author{Junyoung~An}
\affiliation{Research Laboratory of Electronics, Massachusetts Institute of Technology, Cambridge, Massachusetts 02139, USA}
\affiliation{Department of Electrical Engineering and Computer Science, Massachusetts Institute of Technology, Cambridge, Massachusetts 02139, USA}

\author{Farid Hassani}
\affiliation{Research Laboratory of Electronics, Massachusetts Institute of Technology, Cambridge, Massachusetts 02139, USA}

\author{Hung-Yu~Tsao}
\affiliation{Research Laboratory of Electronics, Massachusetts Institute of Technology, Cambridge, Massachusetts 02139, USA}
\affiliation{Department of Electrical Engineering and Computer Science, Massachusetts Institute of Technology, Cambridge, Massachusetts 02139, USA}

\author{Chia-Chin~Tsai}
\affiliation{Research Laboratory of Electronics, Massachusetts Institute of Technology, Cambridge, Massachusetts 02139, USA}
\affiliation{Department of Materials Science and Engineering, Massachusetts Institute of Technology, Cambridge, Massachusetts 02139, USA}

\author{Aranya~Goswami}
\affiliation{Research Laboratory of Electronics, Massachusetts Institute of Technology, Cambridge, Massachusetts 02139, USA}

\author{Gabriel D. Cutter}
\affiliation{Research Laboratory of Electronics, Massachusetts Institute of Technology, Cambridge, Massachusetts 02139, USA}
\affiliation{Department of Electrical Engineering and Computer Science, Massachusetts Institute of Technology, Cambridge, Massachusetts 02139, USA}

\author{Jeremie~Boudreault}
\affiliation{Départment de Physique and Institut Quantique,
Université de Sherbrooke, Sherbrooke J1K 2R1 Québec, CAN\looseness=-1}

\author{Jeffrey~M.~Gertler}
\affiliation{Lincoln Laboratory, Massachusetts Institute of Technology, Lexington, Massachusetts 02421, USA}

\author{Michael~A.~Gingras}
\affiliation{Lincoln Laboratory, Massachusetts Institute of Technology, Lexington, Massachusetts 02421, USA}

\author{Bethany~M.~Niedzielski}
\affiliation{Lincoln Laboratory, Massachusetts Institute of Technology, Lexington, Massachusetts 02421, USA}

\author{Jeffrey~M.~Knecht}
\affiliation{Lincoln Laboratory, Massachusetts Institute of Technology, Lexington, Massachusetts 02421, USA}

\author{Mollie~E.~Schwartz}
\affiliation{Lincoln Laboratory, Massachusetts Institute of Technology, Lexington, Massachusetts 02421, USA}

\author{Kyle~Serniak}
\affiliation{Research Laboratory of Electronics, Massachusetts Institute of Technology, Cambridge, Massachusetts 02139, USA}
\affiliation{Lincoln Laboratory, Massachusetts Institute of Technology, Lexington, Massachusetts 02421, USA}

\author{Jeffrey~A.~Grover}
\affiliation{Research Laboratory of Electronics, Massachusetts Institute of Technology, Cambridge, Massachusetts 02139, USA}

\author{Baptiste~Royer}
\affiliation{Départment de Physique and Institut Quantique,
Université de Sherbrooke, Sherbrooke J1K 2R1 Québec, CAN\looseness=-1}

\author{Max~Hays}
\email{maxhays@mit.edu}
\affiliation{Research Laboratory of Electronics, Massachusetts Institute of Technology, Cambridge, Massachusetts 02139, USA}

\author{William~D.~Oliver}
\email{william.oliver@mit.edu}
\affiliation{Research Laboratory of Electronics, Massachusetts Institute of Technology, Cambridge, Massachusetts 02139, USA}
\affiliation{Department of Electrical Engineering and Computer Science, Massachusetts Institute of Technology, Cambridge, Massachusetts 02139, USA}
\affiliation{Department of Physics, Massachusetts Institute of Technology, Cambridge, Massachusetts 02139, USA}

\date{\today}

\begin{abstract}
Bosonic quantum error correction (QEC) offers a hardware-efficient route to fault-tolerant quantum computing. To date, however, superconducting circuit implementations of bosonic codes have utilized centimeter-scale 3D microwave cavities controlled by fixed-frequency transmon qubits, with logical lifetimes limited by transmon bit-flip errors. Here, we realize bosonic QEC in a fully planar architecture by pairing a heavy fluxonium, whose $\SI{451 \pm 70}{\micro\second}$ bit-flip lifetime exceeds that of any control qubit in previous demonstrations, with an on-chip Archimedean spiral resonator several orders of magnitude smaller in mode volume than prior 3D cavities. We prepare finite-energy Gottesman-Kitaev-Preskill (GKP) states and stabilize them using measurement-free error correction with rapid fluxonium reset, extending the logical lifetime by a factor of $1.59 \pm 0.05$. These results provide the first demonstration of resonator control using a weakly coupled fluxonium and, with it, the first realization of standalone bosonic QEC in a fully planar superconducting circuit architecture.
\end{abstract}

\maketitle


\begin{bibunit}[apsrev4-2]
\bibunitcontrol{apsrev42Control}

\renewcommand{\addcontentsline}[3]{}

\begin{figure}[ht]
    \centering
    \includegraphics[width=\linewidth]{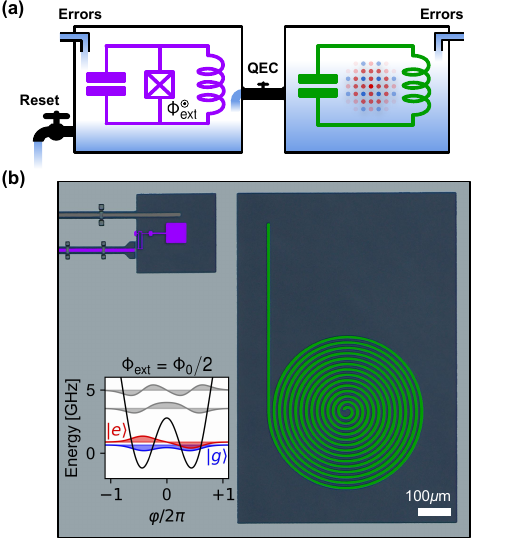}
    \caption{\textbf{Fluxonium-resonator device.} \textbf{(a)} Schematic of the fluxonium-resonator system, with the flow of liquid representing errors entering and leaving the system. The error-correction protocol maps errors from the resonator (green) to the fluxonium (purple) at a controllable rate, represented by the middle tunable valve. The fluxonium is then reset at a controllable rate, represented by the leftmost valve, removing the errors from the system. Error correction thus acts as engineered dissipation that restores the system to the code space. \textbf{(b)} False-colored optical micrograph of the device, with the fluxonium and its flux line in purple, ASR in green, readout resonator stub in dark grey, ground plane in light grey, and bare silicon in navy. Inset: fluxonium potential and wavefunctions plotted against the fluxonium phase variable $\varphi$ at an external flux bias of $\Phi_{\rm ext} = \Phi_0/2$, with the qubit $\{\ket{g},\ket{e}\}$ manifold highlighted in blue and red.}
    \label{fig: circuit diagram}
\end{figure}

Quantum error correction (QEC) protects quantum information through redundant encoding, but the number of physical components required to encode a logical qubit presents a major challenge to realizing useful quantum computation. Bosonic QEC offers a hardware-efficient route by exploiting the intrinsically large Hilbert space of a quantum harmonic oscillator to realize this redundancy within a single mode. Bosonic codes have been pursued across superconducting circuits~\cite{Ofek2016,Hu2019,CampagneIbarcq2020,Sivak2023,LachanceQuirion2024, Gertler2021}, trapped ions~\cite{Flhmann2019,deNeeve2022}, and integrated photonics~\cite{Larsen2025}, complementing progress with codes constructed from registers of physical qubits~\cite{Bluvstein2023,Hong2024,googleSCThreshold,Sivak2026}. Logical information can be protected within these bosonic systems by engineering a form of dissipation that returns the oscillator to the code space faster than errors drive it out, as illustrated in Fig.~\ref{fig: circuit diagram}(a).

Universal control of the logical information in a bosonic code can be achieved by coupling the harmonic oscillator mode, represented by its annihilation operator $\op{a}$, to an auxiliary nonlinear element, such as a control qubit with Pauli operator $\op{\sigma}_z$. In superconducting circuit implementations, this control qubit has been a fixed-frequency transmon~\cite{Koch2007}. Phase-flip errors generated by $\op{\sigma}_z$ commute with the primary dispersive coupling interaction $-\chi\op{a}^\dagger\op{a}\op{\sigma}_z/2$, whereas bit-flip errors generated by $\op{\sigma}_x$ and $\op{\sigma}_y$ do not. The encoded information is therefore relatively less sensitive to control qubit phase flips but inherits its bit-flip errors. Consequently, transmon bit-flip lifetimes of $T_1^q \approx \SI{200}{\micro\second}$ have limited the logical lifetimes in recent bosonic QEC demonstrations~\cite{CampagneIbarcq2020,Sivak2023,LachanceQuirion2024}. To address this limitation, it is desirable to use a longer-lived control element, such as a cat qubit~\cite{Lescanne2020,Grimm2020,Ding2025} or, as demonstrated here, a fluxonium qubit~\cite{Manucharyan2009,Nie2026,Ali2026}. However, unlike a fixed-frequency transmon, such qubits require flux control.

Superconducting-circuit demonstrations of bosonic codes have so far relied on high-$Q$ modes of 3D microwave cavities~\cite{Ofek2016,Hu2019,CampagneIbarcq2020,Gertler2021,Sivak2023}. While flux delivery is possible within 3D architectures~\cite{Gargiulo2021, Valadares2024, Atanasova2025}, it is comparatively simpler to incorporate broadband flux lines in planar circuits without also introducing channels for photon loss in the storage. These on-chip flux lines broaden the device design space to include real-time flux tunability and parametrically driven interactions. Together, these advantages motivate a monolithically integrated, fully planar architecture in which the 3D storage cavity is replaced by an on-chip resonator. Planar resonators also offer mode volumes several orders of magnitude smaller than those of 3D cavities and footprint scaling comparable to that of physical qubits in register-based architectures, providing a path toward densely integrated bosonic error-corrected processors. Considerable progress toward planar bosonic hardware has been made with two-legged cat encodings, including the realization of Kerr-cat qubits~\cite{Hajr2024}, autonomous stabilization of dissipative cats~\cite{Leghtas2015, Lescanne2020,Berdou2023,Rglade2024,Putterman2025} and concatenation with register-based repetition codes~\cite{Putterman2025_2}. To our knowledge, however, no standalone bosonic QEC code has been realized in a fully planar superconducting circuit.

In this work, we realize such a planar architecture for bosonic QEC by dispersively coupling a heavy fluxonium control qubit to an on-chip Archimedean spiral resonator (ASR), as depicted in Fig.~\ref{fig: circuit diagram}(b). The fundamental mode of the ASR serves as our storage oscillator, with a spiral geometry that reduces mode participation in lossy surface interfaces relative to coplanar waveguide resonators, enabling high quality factors in a planar device~\cite{Maleeva2015, Peruzzo2020,asrTominaga}. When biased at its half-flux sweet spot, our fluxonium has a time-averaged bit-flip lifetime of $\overline{T}_1^q = \SI{451 \pm 70}{\micro\second}$, exceeding that reported for any superconducting control qubit in prior experimental bosonic QEC demonstrations~\cite{Ofek2016,Hu2019,CampagneIbarcq2020,Gertler2021,Sivak2023,LachanceQuirion2024}. 

The fluxonium is weakly coupled to the storage mode with a dispersive coupling strength of $\chi/2\pi \approx \SI{23}{\kilo\hertz}$ to suppress the unwanted inherited storage nonlinearity to the hertz level, in contrast to prior work in the strongly coupled regime with kilohertz-level resonator nonlinearity~\cite{Nie2026}. Large storage resonator displacements then enhance this weak dispersive coupling to realize fast echoed conditional displacement (ECD) control~\cite{Eickbusch2022,Ali2026}. We use this control to prepare finite-energy Gottesman-Kitaev-Preskill (GKP) logical states in the ASR and stabilize them using measurement-free error correction with rapid pulsed fluxonium reset. The corrected logical lifetimes exceed their uncorrected counterparts along all three Pauli axes, yielding a Pauli-averaged enhancement factor of $1.59 \pm 0.05$. Together, these results realize a standalone bosonic error-correcting code in a fully planar superconducting circuit device.

\begin{figure}[t]
    \centering
    \includegraphics[width=\linewidth]{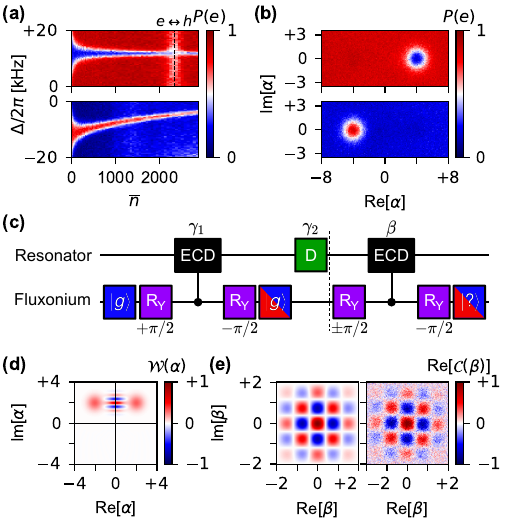}
    \caption{\textbf{Fluxonium-resonator characterization and control.} \textbf{(a)} Fluxonium excited-state population $P(e)$ after an out-and-back sequence versus the mean resonator photon number $\bar{n}$ axis (calibrated in Sec.~\ref{sec: Resonator Drive Strength Calibration}) and resonator detuning $\Delta/2\pi$ axis (calculated by dividing the back displacement phase by the preceding delay time) for the fluxonium initialized in $\ket{e}$ (top) and $\ket{g}$ (bottom). The dashed line marks a drive-induced $\ket{e}\!\leftrightarrow\! \ket{h}$ resonance feature (see main text and Sec.~\ref{sec: DUST Simulations of the Fluxonium-Resonator System}). \textbf{(b)} Husimi Q-functions of the resonator state after applying an ECD gate to vacuum with the fluxonium initialized in $\ket{e}$ (top) and $\ket{g}$ (bottom). This approximate measurement of the Q-function is achieved by scanning a resonator displacement and detecting the vacuum population using a long, vacuum-selective fluxonium $\pi$ pulse, with accuracy limited by the selectivity of the pulse. \textbf{(c)} Preparation and measurement circuit for a displaced cat state, with blue $\ket{g}$ blocks denoting fluxonium reset to the ground state, red/blue $\ket{g}$ blocks indicating postselection on $\ket{g}$, and red/blue $\ket{?}$ blocks representing a $\op{\sigma}_z$ measurement. The sequence after the dashed line is the characteristic-function measurement. \textbf{(d)} Simulated Wigner function $\mathcal{W}(\alpha)$ of the displaced cat state $\ket{\gamma_1/2 + \gamma_2} + \ket{-\gamma_1/2 + \gamma_2} = \ket{2 + 2i} + \ket{-2 + 2i}$. \textbf{(e)} Simulated (left) and measured (right) real part of the characteristic function ${\rm Re}[\mathcal{C}(\beta)]$ of the displaced cat state, taken near the origin for reasons detailed in the main text.}
    \label{fig: conditional displacement demo}
\end{figure}

\vspace{-2mm}
\section{Fast Resonator Control \\with a Weakly Coupled Fluxonium \label{sec: fluxonium control}}
\vspace{-2mm}

The fluxonium control qubit used in this work is a flux-tunable superconducting circuit comprising a capacitor and Josephson junction shunted by a large inductance realized using an array of Josephson junctions~\cite{Manucharyan2009}. When biased near half flux, $\Phi_\mathrm{ext} = \Phi_0/2$, the fluxonium exhibits a double-well potential [Fig.~\ref{fig: circuit diagram}(b), inset] whose confinement is set by the ratio of the Josephson energy $E_{J}^{q}$ to the charging energy $E_{C}^{q}$ with a correction from the inductive energy $E_{L}^{q}$. In the heavy fluxonium regime ($E_{J}^{q} \gg E_{C}^{q}$), the ground and first excited states are tightly confined in deep, disjoint wells, yielding large anharmonicity and bit-flip lifetimes that can approach several milliseconds when biased away from half flux~\cite{Earnest2018,Lin2018,Zhang2021,Ding2023}. With $E_{J}^{q}/E_{C}^{q} \approx 3.5$, the fluxonium in this work has a qubit transition frequency $\omega_{g,e}/2\pi \approx \SI{230}{\mega\hertz}$ at half flux, an order of magnitude smaller than its anharmonicity of $(\omega_{e,f} - \omega_{g,e})/2\pi \approx \SI{2.4}{\giga\hertz}$ as shown in the inset of Fig.~\ref{fig: circuit diagram}(b). Operated at the half-flux sweet spot to minimize sensitivity to flux noise and maximize coherence to $\overline{T}_{2E}^q = \SI{91\pm 10}{\micro\second}$ (limited by shot-noise dephasing to its readout resonator), the fluxonium exhibits a control-qubit bit-flip lifetime of $\overline{T}_1^q = \SI{451 \pm 70}{\micro\second}$.

The ASR has a frequency of $\SI{4.138}{\giga\hertz}$ and a single-photon lifetime of $T_1^a = \SI{103 \pm 2}{\micro\second}$, corresponding to an overall resonator quality factor of $Q = 2.65 \times 10^6$. At half flux, its weak coupling to the fluxonium results in a dispersive interaction of $\chi/2\pi = \SI{23}{\kilo\hertz}$. This yields Hz-level resonator self-Kerr nonlinearities $K_g/2\pi = \SI{2.4}{\hertz}$ and $K_e/2\pi = \SI{-0.007}{\hertz}$, which enter the system Hamiltonian via the terms ${\op{H}}_K/\hbar=\sum_s K_s(\op{a}^\dagger)^2\op{a}^2\ketbra{s}{s}$, with the index $s\in\{g,e\}$ denoting the fluxonium state.
Inspired by control techniques developed for the weakly coupled transmon-resonator system~\cite{CampagneIbarcq2020,Eickbusch2022,Sivak2023}, we achieve fast, high-fidelity control of the resonator via the fluxonium by enhancing the dispersive interaction with large displacements of the resonator. This is best illustrated by writing the effective Hamiltonian of the dispersive interaction in a reference frame displaced by $\alpha$ with respect to the lab frame and rotating at the resonator and qubit frequencies:
\begin{equation}
    {\op{H}}_{\rm disp} / \hbar = -\frac{\chi}{2} \op{\sigma}_z \left[\op{a}^\dagger \op{a}  + (\alpha \op{a}^\dagger + \alpha^* \op{a}) +  |\alpha|^2\right],\label{eq: displaced dispersive Hamiltonian}
\end{equation}
where we use the convention $\op{\sigma}_z=\ketbra{g}{g}-\ketbra{e}{e}$. The first term is the usual dispersive interaction, the second term is a resonator displacement with direction conditioned on the fluxonium state, and the final term is a Stark shift of the fluxonium due to photons in the resonator. The conditional displacement entangling interaction term scales as $\chi |\alpha|$, allowing large resonator displacements to enable fast resonator control~\cite{Eickbusch2022}.

In the ``out-and-back'' measurement~\cite{Eickbusch2022}, shown in Fig.~\ref{fig: conditional displacement demo}(a) and described in Sec.~\ref{sec: Measuring Resonator self-Kerr}, we first displace the resonator to a coherent state $\ket{\alpha}$ with average photon number $\bar{n}=|\alpha|^2$, allow it to acquire a fluxonium-state-dependent rotation for a fixed delay time, and then displace it back by $D(-\alpha e^{i\phi_j}e^{-\tau/(2T_1^a)})$ before measuring its return to vacuum with a long, frequency-selective fluxonium $\pi$ pulse. By scanning $\phi_j$ to maximize the return to vacuum, we determine the accumulated resonator rotation and extract the fluxonium-state-dependent rotation rate as a function of photon number, including its variation due to higher-order resonator nonlinearities. We efficiently and accurately model the effect of these higher-order nonlinearities on the out-and-back response up to mean photon numbers of $\bar n=40{,}000$ in Sec.~\ref{sec: Efficient Numerical Modeling of Out-and-Back Measurements}. Most importantly, the out-and-back measurement demonstrates that the storage resonator can support the large displacements required for the ECD protocol, with the resonator reaching more than 2000 photons before the onset of drive-induced unwanted state transitions (DUST) in the fluxonium, as depicted in Fig.~\ref{fig: conditional displacement demo}(a) and discussed in Sec.~\ref{sec: DUST Simulations of the Fluxonium-Resonator System}~\cite{Khezri2023,Dai2026}.

The ECD gate is thus our workhorse for resonator control, enabling state preparation, logical readout, and error correction. Its action on the system is captured by the unitary operation 
\begin{equation}
    {\rm ECD}(\beta) = {D}(+\beta/2) \ketbra{e}{g} + {D}(-\beta/2) \ketbra{g}{e},
\end{equation}
where ${D}(\alpha) = e^{\alpha \op{a}^\dagger - \alpha^* \op{a}}$ is the resonator displacement operator with creation and annihilation operators $\op{a}^\dagger$ and $\op{a}$. The $\pi$ pulse intrinsic to the ECD gate echoes away unconditional displacements and unwanted qubit-state-dependent resonator rotations during the gate, and it is accounted for when compiling state preparation and error correction circuits~\cite{Eickbusch2022}. We calibrate the gate to produce conditional displacements of arbitrary size $\beta$ by tuning the intermediate displacement amplitudes and wait times. Figure~\ref{fig: conditional displacement demo}(b) shows the action of the ECD gate on the vacuum state, and Fig.~\ref{fig: conditional displacement demo}(c-e) uses a sequence of ECD gates, qubit rotations, and measurement to prepare a displaced cat state and measure the characteristic function $\mathcal{C}(\beta) \coloneqq \ev{D(\beta)}$, whose Fourier transform is the Wigner quasiprobability distribution. The reduced contrast at large $|\beta|$ in the characteristic functions measured in this work, such as in Fig.~\ref{fig: conditional displacement demo}(e), is primarily a tomography artifact arising from reduced fluxonium purity during large-amplitude ECD measurements, as characterized by the cat-and-back experiment in Sec.~\ref{sec: Conditional Displacements with Fluxonium}. For this reason, the characteristic function in Fig.~\ref{fig: conditional displacement demo}(e) is measured near the origin and does not resolve the typical cat-state blobs that encode parity information. Nevertheless, the close agreement between the measured and simulated characteristic functions of the displaced cat state showcases the state-preparation and tomography capabilities enabled by the ECD gate.

\vspace{-2mm}
\section{Preparation of GKP Codewords \label{sec: gkp state}}
\vspace{-2mm}

The ideal square-lattice GKP codewords are superpositions of a grid of infinitely squeezed states, invariant under displacements in position and momentum by the lattice constant $\ell = \sqrt{2\pi}$~\cite{Gottesman2001}. As such, the ideal codewords are +1 eigenstates of the stabilizer generators ${S}_0^{\sX} = {D}(\ell)$ and ${S}_0^{\sZ} = {D}(i\ell)$, with logical Pauli operators defined as ${X}_L = {D}(\ell/2)$ and ${Z}_L = {D}(i\ell/2)$. To obtain a physical, finite-energy GKP code, we apply a normalizing envelope operator ${E}_{\sDelta} = \exp(-\Delta^2 \op{a}^\dagger \op{a})$ to the ideal codewords and logical operators. This produces a GKP grid with an overall Gaussian envelope of width $1/\Delta$, with the individual peaks having a width $\Delta$. For $\Delta \ll 1$, the finite-energy code approximates the ideal code, differing from it only by a coherent superposition of correctable errors; the ideal code is recovered asymptotically as $\Delta\to 0$~\cite{Royer2020}.

\begin{figure}[t]
    \centering
    \includegraphics[width=\linewidth]{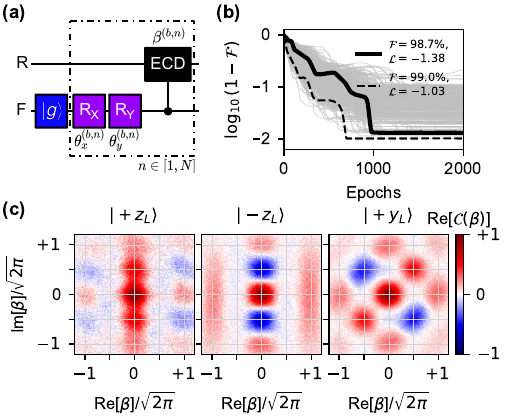}
    \caption{\textbf{GKP state preparation.} \textbf{(a)} Circuit for preparing an arbitrary fluxonium-resonator state using $N$ rounds of single-qubit rotations and ECD gates; $b$ is the batch index within the optimization. \textbf{(b)} Infidelity traces for a multi-start optimization of the preparation circuit for the $\ket{+y_L}$ GKP state with size $\Delta = 0.438$, $N = 7$ blocks, and regularization strength $\lambda = 0.5$. The minimum-loss parameter set (bold black line) does not coincide with the minimum-infidelity set (dashed black line), illustrating how regularization shapes the optimal circuit. \textbf{(c)} Characteristic functions of the prepared $\ket{+z_L}$, $\ket{-z_L}$, and $\ket{+y_L}$ GKP states at $\Delta = 0.438$.}
    \label{fig: gkp state prep}
\end{figure}

Any unitary on a qubit-resonator system can be approximated using a sequence of single-qubit rotations and ECD gates, which form a universal gate set~\cite{Eickbusch2022}. The circuit in Fig.~\ref{fig: gkp state prep}(a) uses this decomposition to prepare arbitrary resonator states, including the finite-energy GKP codewords used in this work. Building on numerical optimization techniques developed for qubit-resonator systems~\cite{Eickbusch2022}, we find efficient decompositions of GKP state preparation circuits. The optimization maximizes the fidelity between the prepared state and the target GKP codewords, with regularization penalizing large conditional displacements and a multi-start random seeding strategy mitigating local minima. To accommodate multiple optimization trajectories while rewarding the most promising, we use the batched loss function

\begin{align*}
    \mathcal{L}(\vec{\theta}_x, \vec{\theta}_y, \vec{\beta}) = \sum_{b=1}^{B} \Bigg\{ &\log_{10} \left[1 - \big|\mel{g,\psi_{\rm target}}{{U}^{(b)}}{g,0}\big|^2\right] \\
    &\quad+ \frac{\lambda}{N} \sum_{n=1}^{N} \Big| \beta^{(b,n)} \Big|^2 \Bigg\}. \numberthis
\end{align*}
Here ${U}^{(b)} \coloneqq {U}(\vec{\theta}_x^{(b)}, \vec{\theta}_y^{(b)}, \vec{\beta}^{(b)})$ is the unitary implementing $N$ rounds of single-qubit rotations set by $(\vec{\theta}_x^{(b)}, \vec{\theta}_y^{(b)})$ and ECD gates set by $\vec{\beta}^{(b)}$ for batch index $b$, $\lambda$ is the regularization strength on the conditional displacement amplitudes, $\ket{g,0}$ is the initial fluxonium-resonator state, and $\ket{g,\psi_{\rm target}}$ is the target state. The code used for the GKP state-preparation optimization is available through JAXQuantum~\cite{jha2024jaxquantum}, an open-source package developed by a subset of the authors. Built on JAX~\cite{jax2018github}, JAXQuantum provides auto-differentiable, hardware-accelerated tools for quantum hardware design, simulation, and control. The batched optimization ports to an NVIDIA H200 GPU, where it runs in minutes, yielding a 165$\times$ speedup over a 10th-generation Intel i7 CPU and producing the convergence trace in Fig.~\ref{fig: gkp state prep}(b).

Once an efficient compilation is found [see parameters listed in Table~\ref{tab: gkp state preparation compilations}], the corresponding sequence of ECD gates and single-qubit rotations is executed on the device. Figure~\ref{fig: gkp state prep}(c) shows the prepared finite-energy GKP codewords $\ket{+z_L}$, $\ket{-z_L}$, and $\ket{+y_L}$ with $\Delta = 0.438$. The remaining codewords ($\ket{+x_L}$, $\ket{-x_L}$, $\ket{-y_L}$) are obtained similarly or via a final resonator frame rotation. To our knowledge, this is the first experimental preparation of the logical states of the GKP code in a fully planar superconducting circuit device.

\vspace{-2mm}
\section{Measurement-Free\\Quantum Error Correction \label{sec: error correction}}
\vspace{-2mm}

Autonomously stabilizing the finite-energy GKP code requires cooling the storage resonator into the simultaneous $+1$ eigenspace of the two finite-energy stabilizers ${S}_\sDelta^\sX \coloneqq {E}_{\sDelta} {S}_0^{\sX} {E}_{\sDelta}^{-1}$ and ${S}_\sDelta^\sZ \coloneqq {E}_{\sDelta} {S}_0^{\sZ} {E}_{\sDelta}^{-1}$. Following Ref.~\cite{Royer2020}, this is achieved by engineering, for each stabilizer, an exchange interaction that transfers entropy to a cold bath (i.e.,\ a repeatedly reset control qubit) at a controllable rate,

\begin{figure}[ht]
    \centering
    \includegraphics[width=\linewidth]{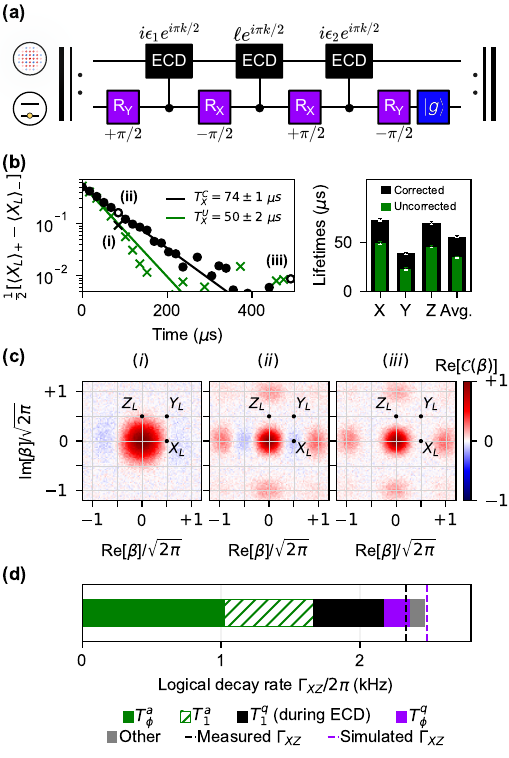}
    \caption{\textbf{Measurement-free GKP QEC.} \textbf{(a)} Circuit for a single stabilizer round of measurement-free GKP error correction, indexed by $k$, enabling cycling through $X$/$Z$ stabilizers along each cardinal direction in phase space. \textbf{(b)} Decay of the logical-state distinguishability $\tfrac{1}{2}\big[\ev{X_L}_{+} - \ev{X_L}_{-}\big]$ between the $\ket{\pm x_L}$ states, with and without error correction. As shown in the bar plot, the corrected logical lifetimes $T_X^C$, $T_Y^C$, and $T_Z^C$ exceed all corresponding uncorrected lifetimes $T_X^U$, $T_Y^U$, and $T_Z^U$, with $T_{\rm avg}^C/T_{\rm avg}^U=1.59\pm0.05$. The exact fitted values and ratios are listed in Supplemental Table~\ref{tab: gkp logical lifetimes}. \textbf{(c)} Characteristic function of the resonator state after \textit{(i)} 5 rounds of uncorrected decay, \textit{(ii)} 5 rounds of error correction, and \textit{(iii)} 29 rounds of error correction, as marked by a black cross and white-filled circles in (b). The logical Pauli labels indicate the points of the characteristic function that correspond to expectation values of these operators. \textbf{(d)} Logical error budget summary with simulated and measured logical error rates, defined as $\Gamma_{XZ}\coloneqq(\Gamma_X+\Gamma_Z)/2$, and simulated channel contributions based on a separate contemporaneous measurement of error channels, as detailed in Sec.~\ref{sec: logical error budget}.}
    \label{fig: error correction}
\end{figure}

\begin{equation}
    \op{H}_{j,\Delta}/\hbar = \sqrt{\frac{\Gamma_j}{\delta t}}\,\big(\op{d}_{j,\Delta}\,\op{\sigma}_{+} + \op{d}_{j,\Delta}^\dagger\,\op{\sigma}_{-}\big), \quad j \in \{x,p\},\label{eq: qec bath exchange}
\end{equation}
where $\delta t$ is the time between successive fluxonium resets, $\op{\sigma}_{-} = \ketbra{g}{e}$ and $\op{\sigma}_{+} = \ketbra{e}{g}$ are the lowering and raising operators of the fluxonium control qubit, and $\op{d}_{j,\Delta}$ is a modular squeezed-annihilation-like operator satisfying $\op{d}_{j,\Delta}\ket{\psi} = 0$ for any state $\ket{\psi}$ within the finite-energy GKP code space~\cite{Royer2020}. For the $x$ quadrature, we have $\op{d}_{x,\Delta} \propto \op{x}_{[\ell/2c_\Delta]}/\sqrt{t_\Delta} + i\op{p}\sqrt{t_\Delta}$, with $c_\Delta = \cosh(\Delta^2)$ and $t_\Delta = \tanh(\Delta^2)$. Here $\op{x}_{[m]} \coloneqq \op{x}\text{ mod }m$ denotes the modular position, which is taken in a symmetric interval centered around zero: $\op{x}_{[m]} \in (-m/2,m/2]$~\cite{Zak1967,Royer2020}. The use of the modular position imposes the required periodicity of the GKP code manifold under phase-space displacements, while the conjugate momentum term sets the envelope width $\sim 1/\Delta$ and ensures a manifold of normalizable states; $\op{d}_{p,\Delta}$ is obtained by the canonical exchange $\op{x}\leftrightarrow\op{p}$. Tracing out the fluxonium with frequent resets yields a dissipator $\Gamma_j\,\mathcal{D}[\op{d}_{j,\Delta}]$, autonomously cooling the resonator toward the finite-energy code space.

We implement this dissipation stroboscopically with the small-big-small (sBs) protocol~\cite{Royer2020,deNeeve2022,Sivak2023,LachanceQuirion2024}, in which a symmetric Trotter decomposition of the exchange interaction in Eq.~\eqref{eq: qec bath exchange} yields a qubit-controlled large displacement of length $\ell \cosh(\Delta^2) \approx \ell$ sandwiched between two small ECDs with amplitudes $\epsilon_1$ and $\epsilon_2$. In the symmetric construction, $\epsilon_1=\epsilon_2=(\ell/2)\sinh(\Delta^2)$. The big ECD gate stabilizes the lattice by mapping small displacement errors onto the control qubit phase, while the small ECD gates stabilize the envelope. The finite-energy squeezing parameter $\Delta$ of the stabilized GKP qubit sets the first amplitude; our control implementation uses the small-$\Delta$ approximation $\epsilon_1\approx\ell\Delta^2/2$ and independently tunes the ratio $r\coloneqq\epsilon_2/\epsilon_1$, finding the optimal ratio to be greater than one as in Refs.~\cite{Sivak2023,LachanceQuirion2024}. In practice, the manual parameter sweeps described in Sec.~\ref{sec: optimizing error correction} are used to select optimal $\Delta=0.42$ and $r=1.3$ values for the logical-lifetime measurements in Fig.~\ref{fig: error correction}. The two stabilizers differ only by a $\pi/2$ phase-space rotation of the controlled displacements, ${\rm ECD}(\beta)\to {\rm ECD}(i\beta)$, which maps ${x}$-quadrature kicks onto ${p}$-quadrature kicks. The repeated application of sBs rounds engineers a ``trickle-down'' dissipation involving directional hopping between error spaces back to the code space, implementing a low-control-overhead protocol that corrects the most probable small displacement errors in a single step while requiring multiple steps to correct larger errors~\cite{Sivak2023,ShraddhaPhDThesis}. To echo away residual unconditional displacement errors from imperfect ECD calibration, we cycle sBs rounds through the four cardinal directions of resonator phase space, $(+p,+x,-p,-x)$, by applying a phase factor $e^{i\pi k/2}$ to each ECD in the $k$th sBs round~\cite{Royer2020}, as shown in Fig.~\ref{fig: error correction}(a).

Measurement-free QEC replaces control qubit measurements and real-time feedback with engineered dissipation, reducing control latency and overhead~\cite{LachanceQuirion2024}. To realize these advantages, the fluxonium is reset autonomously using a pulsed cooling protocol~\cite{Magnard2018,Egger2018}, where the fluxonium–readout system is driven from $\ket{e,0}$ to $\ket{f,0}$ and then from $\ket{f,0}$ to $\ket{g,1}$, after which the readout excitation decays. This pulsed reset scheme reduces storage dephasing relative to its continuously driven counterpart~\cite{Magnard2018} and completes in \SI{236}{\nano\second}, \SI{5.6}{\percent} of the total sBs round time of \SI{4232}{\nano\second}.

Figure~\ref{fig: error correction}(b) shows the decay of logical information, with and without error correction, for a GKP qubit initialized in finite-energy logical states $\ket{\pm x_L}$. To measure this decay, we calculate the expectation of the infinite-energy logical Pauli operators~\cite{Sivak2023,LachanceQuirion2024}, defined as $\ev{X_L}_{\pm} = \mel{\pm x_L}{X_L}{\pm x_L}$. Although $\ev{X_L}_{\pm}$ decays exponentially to zero with error correction, the uncorrected logical expectation value $\ev{X_L}_{+}$ behaves non-monotonically and non-exponentially as the resonator decays out of the code space towards vacuum, since the vacuum state has a positive logical expectation value $\mel{0}{X_L}{0} > 0$. The corrected lifetime $T_X^C$ and uncorrected lifetime $T_X^U$ are thus extracted from zero-offset exponential fits to $\tfrac{1}{2}\big[\ev{X_L}_{+} - \ev{X_L}_{-}\big]$, which measures logical information as the distinguishability of $\ket{+x_L}$ from $\ket{-x_L}$ and decays exponentially to zero with time for both the corrected and uncorrected cases. The characteristic functions in Fig.~\ref{fig: error correction}(c) show that without stabilization the resonator decays toward vacuum, while with stabilization it remains in the code space even as logical information is gradually lost. Averaged across Pauli axes, the corrected GKP qubit lifetime exceeds the uncorrected one by a factor of $1.59 \pm 0.05$, as depicted in the bar plot in Fig.~\ref{fig: error correction}(b).

\vspace{-2mm}
\section{Discussion and Outlook\label{sec: discussion}}
\vspace{-2mm}

Prior demonstrations of bosonic QEC in superconducting 3D cavities have shown that, once intrinsic resonator loss and dephasing are sufficiently suppressed, errors and nonlinearities inherited from the control qubit become dominant. In this regime, the logical lifetime of GKP qubits will benefit directly from the fluxonium's capacity for improved bit-flip lifetimes and $\chi/K_g$ and $\chi/K_e$ ratios. Notably, threading the ASR frequency between the fluxonium plasmon transitions produces a zero crossing of the dispersive interaction near half flux. This tunability~\cite{Atanasova2025} could be used to decouple the fluxonium from the storage resonator during operations such as fluxonium reset, reducing backaction on the encoded information~\cite{Chowdhury2024Thesis}.

While we demonstrate an enhancement of the GKP logical lifetime, the logical qubit does not yet exceed break-even, as both the Fock qubit, measured in Sec.~\ref{sec: Fock Qubit Lifetimes}, and the fluxonium have longer Pauli-averaged lifetimes. The error budget in Fig.~\ref{fig: error correction}(d) [detailed further in Sec.~\ref{sec: logical error budget}] agrees closely with the measured logical error rate and identifies the dominant channel as the resonator's intrinsic dephasing, $T_{\phi}^{a} = \SI{1.08\pm0.16}{\milli\second}$, likely arising from dispersive dephasing through coupled two-level systems such as surface defects~\cite{Burnett2014,deGraaf2018, Niepce2021}. Surface treatments~\cite{deGraaf2018}, resonator geometry optimization~\cite{asrTominaga}, and materials engineering such as tantalum metallization~\cite{Bland2025} offer promising routes for improving the intrinsic loss and dephasing of planar resonators. Leading superconducting qubits such as the transmon are limited by similar loss mechanisms, and as their fabrication improves, planar resonator internal quality factors $Q_i$ are expected to follow, approaching the bulk silicon substrate limit of $Q_i \sim 10^8$~\cite{Checchin2022,Zhang2024}.

Finally, the compact on-chip GKP qubit demonstrated here could be used to implement the inner layer of a concatenated QEC scheme, with a higher-level code such as the surface code protecting against residual logical errors. Recent estimates suggest that such a scheme could reach algorithmically relevant logical error rates with substantially fewer hardware resources (physical qubits and their associated controls) than a surface code constructed from bare two-level qubits~\cite{Noh2022, Borah2025}. Within this concatenated approach, the outer decoder could also make use of soft information from the inner GKP error correction, such as the fluxonium state measured before reset in each sBs round~\cite{Fukui2017, Fukui2018, Berent2024, Hopfmueller2024}; this is analogous to how erasure information can improve surface-code decoding~\cite{Kubica2023}. Altogether, these considerations position on-chip bosonic qubits as an extensible platform for hardware-efficient quantum error correction.

\vspace{-2mm}
\section*{Acknowledgments}
\vspace{-2mm}

We are grateful to Marco Paradina, Miuko Tanaka, Agustin Di Paolo, Patrick Harrington, Melvin Mathews, Ben Freiman, Alec Emser, Neill Warrington, Alec Eickbusch, Volodymyr Sivak, Luigi Frunzio, Zaki Leghtas, and Philippe Campagne-Ibarcq for invaluable discussions at different stages of this project. This research was funded in part by the Army Research Office under Award No. W911NF-23-1-0045 (Extensible and Modular Advanced Qubits), in part by the AWS Center for Quantum Computing, and in part under Air Force Contract No. FA8702-15-D-0001. S.R.J. and S.D.C. acknowledge support from the National Science Foundation Graduate Research Fellowship under Grant No. 1745302 and from the Doc Bedard Fellowship Program through the MIT Center for Quantum Engineering.
J.A. acknowledges support from Korea Foundation for Advanced Studies.
M.H. acknowledges funding from the IC Postdoctoral Fellowship.
We acknowledge Google for supporting JAXQuantum through the TPU Research Cloud (TRC) program. The views and conclusions contained in this document are those of the authors and should not be interpreted as representing the official policies, either expressed or implied, of the U.S. Air Force or the U.S. Government. 

\vspace{-2mm}
\section*{Author Contributions}
\vspace{-2mm}
S.R.J., S.D.C., and M.H. conceptualized the project idea. S.R.J. performed the experiments and analyzed data with feedback from S.D.C. and M.H. S.R.J., S.D.C., M.H., B.R., G.R., A.A., L.-A.S., and J.B. developed the theory and numerical simulations. S.D.C. and S.R.J. designed the fluxonium-resonator devices, which were then fabricated by M.A.G., B.M.N., and J.M.K. at MIT Lincoln Laboratory under the supervision of M.E.S. and K.S. We thank the packaging team at MIT LL for their technical assistance. S.D.C. designed, fabricated, and measured the resonator-only devices with assistance from F.H., H.-Y.T., C.-C.T., A.G., and G.D.C. who all developed the fabrication and packaging processes in MIT.nano. R.A., D.P., L.P., J.A., and J.M.G. provided assistance with various aspects of device design and measurement. S.R.J. and S.D.C. wrote the manuscript, with feedback from all authors. M.H., B.R., J.A.G., and W.D.O. supervised the project. All authors contributed to the discussion of the results and the manuscript.

\newpage


\putbib[citations]
\end{bibunit}


\ifincludesupp
    \onecolumngrid

    \clearpage

    \twocolumngrid


\setcounter{equation}{0}
\setcounter{figure}{0}
\setcounter{table}{0}
\setcounter{section}{0}

\renewcommand{\theequation}{S\arabic{equation}}
\renewcommand{\thefigure}{S\arabic{figure}}
\renewcommand{\thetable}{S\arabic{table}}
\renewcommand{\thesection}{S\arabic{section}}

\title{Supplementary Material:\\ ``Bosonic Error Correction with Fluxonium''}

\clearpage
\maketitle

\onecolumngrid
\setcounter{tocdepth}{2}
\tableofcontents
\bigskip

\begin{bibunit}[apsrev4-2]
\bibunitcontrol{apsrev42Control}

\newpage

\begin{figure}[h]
    \centering
    \includegraphics[width=0.77\linewidth]{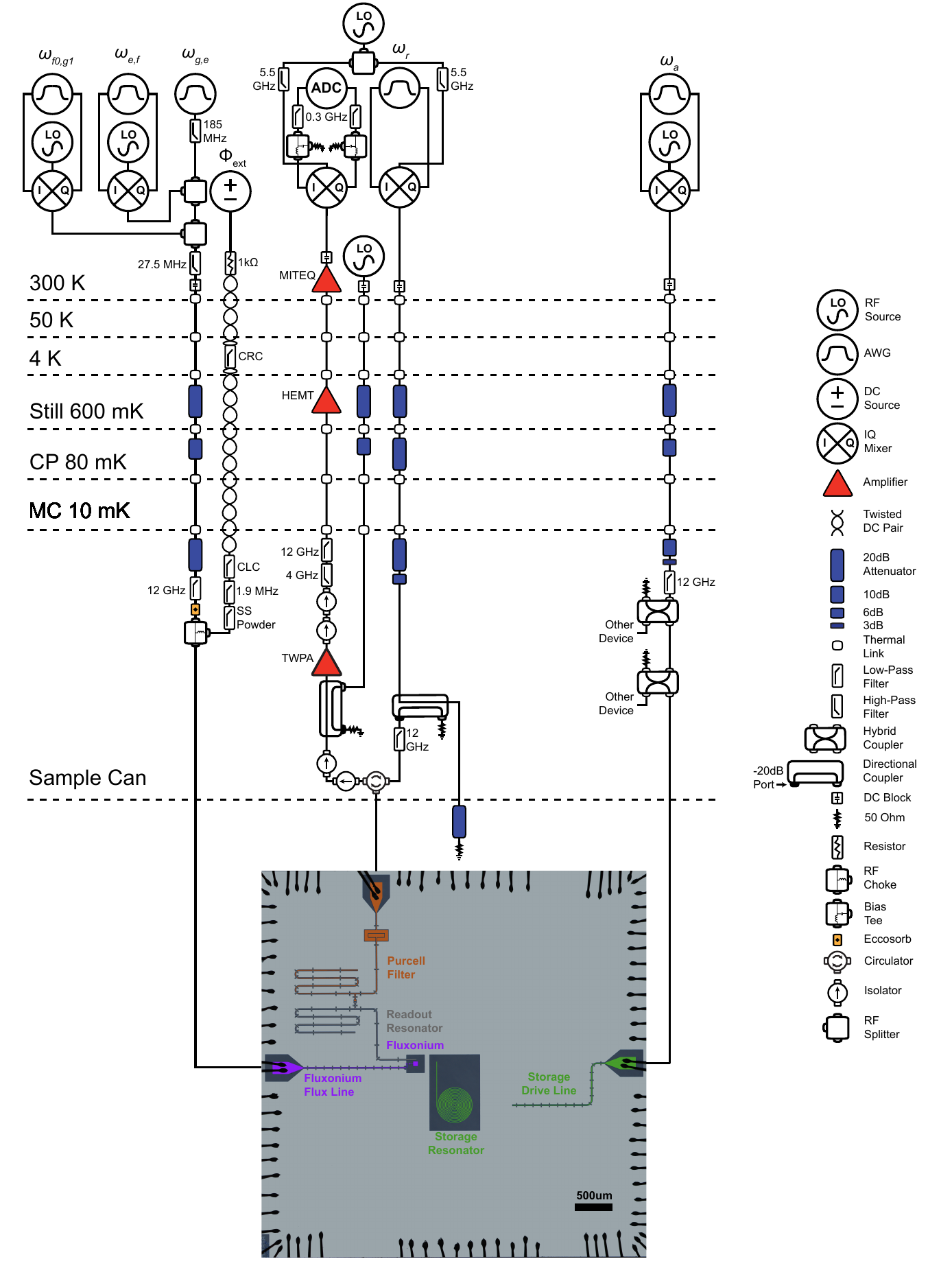}
    \caption{\textbf{Wiring diagram.} Here, we depict the room-temperature and cryogenic wiring diagram of our experiment, with a false-colored optical micrograph of the device shown at the bottom. The final \SI{20}{\decibel} of attenuation in the readout input chain is thermalized to the sample can via a directional coupler, reducing the effective readout resonator temperature. A single drive line delivers storage resonator displacement pulses to several devices through hybrid couplers. The sample can is surrounded by a shield made of mu-metal, and the mixing-chamber (MC) can is also made of mu-metal.}
    \label{si-fig: wiring diagram}
\end{figure}

\newpage 

\section{Experimental Setup and Device Parameters \label{sec: experimental setup and device parameters}}
\begin{table}[t]
    \centering
    \setlength{\tabcolsep}{12pt} 
    \begin{tabular}{lll}
        \hline\hline
        \textbf{Component}             & \textbf{Manufacturer}     & \textbf{Model}   \\
        \hline
        Dilution Refrigerator & Leiden Cryogenics  & CS-CF81-1500 \\
        RF Source ($\times 3$\verythinspace)   & Rohde \& Schwarz & SGS100A \\
        RF Source ($\times 2$\verythinspace)   & Agilent Technologies         & E8267D PSG  \\
        AWG \& ADC     & Quantum Machines (QM)         & OPX+  \\
        DC Source & QDevil (acquired by QM) & QDAC-II \\
        \hline\hline
    \end{tabular}
    \caption{\textbf{Summary of equipment.}
    The manufacturers and model numbers of experimental hardware used in this work.}
    \label{tab:control-equipment}
\end{table}

\begin{table}
    \centering
    \newcommand{\topspacing}{\rule{0pt}{2.4ex}}
    \newcommand{\bottomspacing}{\rule[-1.0ex]{0pt}{0.6ex}}
    \begin{tabular}{|c@{\hspace{8pt}}|l@{\hspace{8pt}}|l@{\hspace{8pt}}|}
    \hline
    \multirow{9}{*}{\textbf{$\,\,\,$Fluxonium$\,\,\,$}}
    & \topspacing Charging Energy & $E_{C}^{q} = h \times \SI{0.789}{\giga\hertz}$ \tabularnewline
    & Josephson Energy & $E_{J}^{q} = h \times \SI{2.782}{\giga\hertz}$ \tabularnewline
    & Inductive Energy & $E_{L}^{q} = h \times \SI{0.372}{\giga\hertz}$ \tabularnewline
    & Half-Flux $\ket{g}\leftrightarrow \ket{e}$ Frequency & $\omega_{g,e}=2\pi\times \SI{0.233}{\giga\hertz}$ \tabularnewline
    & Half-Flux $\ket{e}\leftrightarrow \ket{f}$ Frequency & $\omega_{e,f} =2\pi\times\SI{2.667}{\giga\hertz}$ \tabularnewline
    & Half-Flux $\ket{f,0}\leftrightarrow \ket{g,1}$ Frequency & $\omega_{f0,g1} =2\pi\times \SI{3.295}{\giga\hertz}$ \tabularnewline
    & Half-Flux Relaxation& $\overline{T}_{1}^{\,q} = \SI{451\pm 70}{\micro\second}$ \tabularnewline
    & Half-Flux Coherence (Ramsey) & $\overline{T}_{2R}^{\,q}=\SI{85\pm 11}{\micro\second}$ \tabularnewline
    & \bottomspacing Half-Flux Coherence (Echo) & $\overline{T}_{2E}^{\,q}=\SI{91\pm 10}{\micro\second}$ \tabularnewline
    \hline
    \multirow{8}{*}{\textbf{$\,\,\,$Storage Resonator$\,\,\,$}}
    & \topspacing Frequency & $\omega_{a}=2\pi\times\SI{4.138}{\giga\hertz}$ \tabularnewline 
    & Dispersive Coupling to Fluxonium & $\chi=2\pi\times\SI{23}{\kilo\hertz}$  \tabularnewline
    & Effective Kerr (Fluxonium in $\ket{g}$) & $\widetilde{K}_g=2\pi\times\SI{1.7}{\hertz}$ \tabularnewline
    & Effective Kerr (Fluxonium in $\ket{e}$) & $\widetilde{K}_e=-2\pi\times \SI{0.05}{\hertz}$ \tabularnewline
    & Low-Photon Kerr (Fluxonium in $\ket{g}$) & $K_g=2\pi\times\SI{2.411}{\hertz}$ \tabularnewline
    & Low-Photon Kerr (Fluxonium in $\ket{e}$) & $K_e=-2\pi\times\SI{7.19}{\milli\hertz}$ \tabularnewline
    & Relaxation & $T_{1}^{\,a}=\SI{103\pm 2}{\micro\second}$\tabularnewline
    & \bottomspacing Intrinsic Pure Dephasing & $T_\phi^a=\SI{1.08\pm0.16}{\milli\second}$ \tabularnewline
    \hline
    \multirow{4}{*}{\textbf{$\,\,\,$Readout resonator$\,\,\,$}}
    & \topspacing Frequency & $\omega_{r}=2\pi\times\SI{6.170}{\giga\hertz}$ \tabularnewline 
    & Dispersive Coupling to Fluxonium & $\chi_{qr}=2\pi\times\SI{7.0}{\mega\hertz}$ \tabularnewline
    & Feedline Coupling Strength & $\kappa_{c}=2\pi\times\SI{2.42}{\mega\hertz}$\tabularnewline
    & \bottomspacing Internal Loss & $\kappa_{i}=2\pi\times\SI{0.41}{\mega\hertz}$\tabularnewline
    \hline
    \end{tabular}
    \caption{\textbf{Measured and derived system parameters.} The effective coefficients $\widetilde{K}_s$ are extracted from linear out-and-back fits, whereas $K_s$ denotes the low-photon Kerr derived from numerical diagonalization. This distinction is discussed in Sec.~\ref{sec: Measuring Resonator self-Kerr}.}
    \label{tab: summary of parameters}
\end{table}

\textbf{Control.} In contrast to the hybrid planar-3D architectures used in past demonstrations of bosonic error correction, our system is a fully planar device with on-chip control and readout lines. As depicted in Fig.~\ref{si-fig: wiring diagram}, the fluxonium qubit is controlled using an on-chip flux line, through which a DC current bias, a baseband signal for single-qubit rotations, and two upconverted signals for fluxonium reset are combined and delivered. All baseband signals are generated by a Quantum Machines OPX+ arbitrary waveform generator (AWG). The DC bias is provided by a QDAC-II, while two separate Rohde \& Schwarz SGS100A local oscillators are used to up-convert each of the $\omega_{f0,g1}$ and $\omega_{e,f}$ fluxonium reset signals. An Agilent E8267D PSG Vector Signal Generator is used as a local oscillator to both up-convert the readout baseband pulse and later down-convert the returning signal to the same intermediate frequency. The down-converted signal is then digitized by the OPX+ analog-to-digital converter (ADC), digitally demodulated, and integrated to obtain the readout IQ values. The readout signal is first amplified at the mixing-chamber stage by a traveling-wave parametric amplifier (TWPA) provided by MIT Lincoln Laboratory, pumped by an Agilent E8267D PSG vector signal generator, followed by a high-electron-mobility transistor (HEMT) amplifier at the \SI{4}{\kelvin} stage and a MITEQ RF amplifier at room temperature. The storage baseband pulses are up-converted with a Rohde \& Schwarz SGS100A local oscillator and sent down the on-chip storage drive line to displace the storage resonator state. We summarize the equipment used in Table~\ref{tab:control-equipment}.

\textbf{Fabrication.} The fluxonium-resonator device used in this work (Fig.~\ref{si-fig: wiring diagram}) was fabricated at MIT Lincoln Laboratory on a high-resistivity 8-inch-wide, \SI{725}{\micro\metre}-thick silicon wafer with a patterned \SI{250}{\nano\metre}-thick MBE-grown aluminum film on top, using a fabrication process similar to the process of record in Ref.~\cite{Gingras2026}. Larger structures (e.g., resonators, drive lines, qubit capacitor pads, and the ASR) were patterned via optical lithography, while smaller structures (e.g., the Dolan-style Josephson junction and junction arrays) were defined using electron-beam lithography, followed by double-angle evaporation of aluminum to form the junctions. Different sections of the ground plane metal were connected via aluminum airbridge crossovers in a separate fabrication step. The wafer was finally coated in a protective resist, diced into \SI{5}{\milli\metre}$\times$\SI{5}{\milli\metre} dies, and cleaned with NMP, acetone, and isopropanol. The individual chip was then packaged into an aluminum sample holder with a Rogers TMM10i PCB and mounted to the mixing chamber stage of a Leiden Cryogenics CF-CS81-1500 dilution refrigerator.

\textbf{Summary of Device Parameters.} The measured and derived parameters of the fluxonium-resonator system are summarized in Table~\ref{tab: summary of parameters}. Experimental parameters are based on the measurements in Sec.~\ref{sec: Characterization and Calibration Experiments}, while the low-photon Kerr coefficients are derived by the numerical diagonalization in Sec.~\ref{sec: numerical diagonalization of the fluxonium resonator system}. For the fluxonium lifetimes $\overline{T}_j^q$, we quote the mean and sample standard deviation of measurements taken over several hours in a single day, as shown in Fig.~\ref{si-fig: fluxonium spec and coherence}(c).

\section{Engineering High-Coherence Planar Resonators \label{sec: engineering better planar resonators}}

\subsection{Designing Fluxonium-Resonator Devices with an Archimedean Spiral Resonator Geometry}

A central challenge in implementing GKP error correction on chip is realizing a sufficiently long-lived storage mode. Previous demonstrations of bosonic error correction with superconducting circuits have utilized 3D microwave cavity resonators, with the electromagnetic fields residing predominantly in vacuum. As a result, these 3D resonators have low energy participation in lossy dielectrics. By contrast, lithographically defined planar resonators concentrate a larger fraction of their electric field energy near lossy dielectrics, such as the material interfaces between the substrate, metal, and vacuum, as well as at the conductor edges~\cite{Wenner2011, Calusine2018, Woods2019, Melville2020, Crowley2023, Ganjam2024}. As a result, planar resonators tend to be more sensitive to dielectric loss than their 3D counterparts, resulting in shorter single-photon lifetimes. This is particularly consequential for stabilized GKP memories: in the parameter regime relevant to our implementation, the logical lifetime under sBs error correction is predicted to scale super-linearly with the physical storage mode lifetime~\cite{Royer2020}. Therefore, even modest increases in storage lifetime can yield a disproportionately large improvement in logical memory performance. 

A recent study identified circular Archimedean spiral resonators (ASRs) as a promising planar geometry for reducing surface participation ratios with an improved distribution of electric fields compared to the more widely used coplanar waveguide (CPW) geometry~\cite{asrTominaga}. The authors of Ref.~\cite{asrTominaga} measured ASR and quarter-wave CPW resonators cofabricated from a titanium nitride film on a silicon substrate and found that the ASRs outperformed the CPWs by a factor of 2--4, with the highest-performing ASR achieving a single-photon internal quality factor of $Q_i = (9.6 \pm 1.5)\times 10^6$. These results motivate ASR geometries as an effective strategy for extending the coherence of planar resonators.

\begin{figure}[b]
    \centering
    \includegraphics[width=\linewidth]{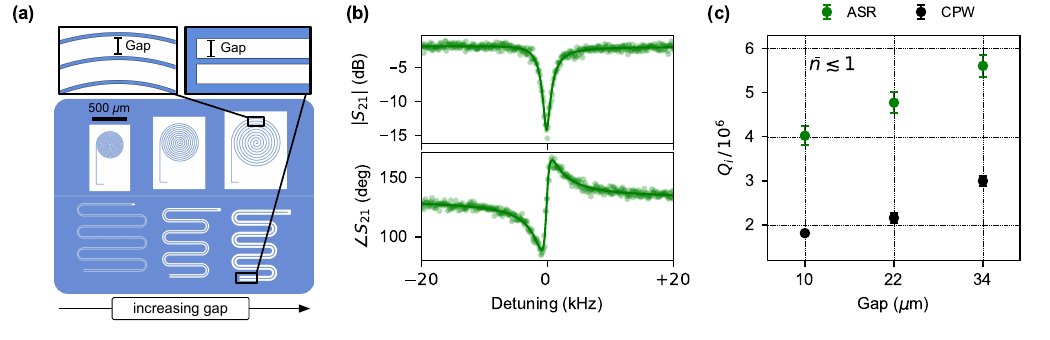}
    \caption{\textbf{ASR vs. CPW resonator quality factors.} \textbf{(a)} GDSII design image of a separate resonator test device comprising three coplanar waveguide (CPW) resonators and three Archimedean spiral resonators (ASRs) of varying gap sizes: 10 $\mu$m, 22 $\mu$m, and 34 $\mu$m. As shown in the inset, we define the ASR gap as the separation between successive rungs of the spiral and the CPW gap as the distance from the center conductor trace to the ground plane. \textbf{(b)} Transmission signal $S_{21}$ measured at the lowest output power for the ASR with 34 $\mu$m gap [see text for details]. The solid lines denote a fit to Eq.~\eqref{eq:supp-S21-resonator-trace}. \textbf{(c)} Extracted low-power internal quality factors $Q_i$ as a function of gap size for ASRs (green) and CPWs (black). We observe a trend of $Q_i$ increasing with gap size for both ASRs and CPWs, with the ASRs on this test chip outperforming the CPWs by a factor of approximately two at each gap size. }
    \label{si-fig: asr q factors}
\end{figure}

In our work here, we designed the storage mode to utilize an ASR geometry, as shown in Fig.~\ref{fig: circuit diagram}(b) of the main text. We measured a single-photon lifetime of $T_1^a = 103$ $\mu$s, corresponding to a resonator quality factor of $Q = 2.65 \times 10^6$ at our storage frequency of $\omega_a/2\pi = 4.138$ GHz. Our storage ASR used 14 spiral turns and had a wire trace width of 8 $\mu$m, a gap between spiral rungs of 8 $\mu$m, and an inner radius of 6 $\mu$m. Through microwave reflection measurements on all ports of a nominally identical device from the same fabrication run, we estimate the coupling quality factors of the ASR to the storage drive line, the qubit drive line, and the readout line to be $Q_c^{\rm s} = 1.11 \times 10^7$, $Q_c^{\rm q} = 3.74 \times 10^7$, and $Q_c^{\rm r} = 1.51 \times 10^9$, respectively. We can therefore use the relation $1/Q = 1/Q_i + \sum_j 1/Q_c^j$ to estimate the internal quality factor $Q_i$ of the storage resonator of the device we predominantly report on here to be $Q_i \approx 3.85 \times 10^6$.

We note that in a previous design iteration of our fluxonium-resonator device, we implemented the storage resonator using a CPW geometry with a 10 $\mu$m center trace width and a large coplanar gap of 60 $\mu$m and measured a resonator quality factor of $Q = 0.72 \times 10^6$ at a similar resonator frequency of 4.18 GHz. This CPW-storage device was fabricated at MIT Lincoln Laboratory in a prior fabrication run with nominally identical process parameters on a wafer that also contained an ASR-storage device with the same design as the main device shown in Fig.~\ref{fig: circuit diagram}(b); we measured this prior ASR-storage device to have a resonator quality factor of $Q = 2.29\times 10^6$. Since all devices were packaged identically (with identical qubit and readout designs), we therefore attribute the approximately threefold increase in $Q$ between our specific design iterations primarily to the change from the CPW to the ASR geometry. These results corroborate those of Ref.~\cite{asrTominaga} and indicate that the gains in quality factor from using ASRs over CPW resonators also translate to our aluminum-on-silicon devices.

\subsection{Resonator-Only Study: Identifying Routes to Improve Planar Resonator Coherence}

To identify routes to further improve planar storage mode coherence, we performed a separate resonator-only study using our niobium-on-silicon fabrication process developed in the MIT.nano cleanroom at MIT. As shown in Fig.~\ref{si-fig: asr q factors}(a), the $5\times5~\mathrm{mm}^2$ test chip comprises three ASRs and three quarter-wave CPW resonators coupled to a common feedline, with one resonator of each type at 10 $\mu$m, 22 $\mu$m, and 34 $\mu$m gap sizes. The CPWs and ASRs have trace widths of 10 $\mu$m and 6 $\mu$m, respectively. The CPWs serve here as on-chip control resonators, while sweeping the gap enables us to reduce electric field participation at lossy interfaces. The resonators were patterned via optical photolithography on a \SI{275}{\nano\metre} sputtered niobium film atop a 500 $\mu$m-thick silicon substrate and defined using a chlorine-based reactive-ion etch. Following etching, the 2-inch wafer was coated with protective photoresist and diced into individual chips. The protective resist was subsequently stripped in NMP, followed by rinses in acetone and isopropanol. Immediately before packaging, the chip was acid cleaned in a buffered oxide etch (BOE) solution for 20 minutes to reduce niobium surface oxides, and was subsequently cooled down within 24 hours of this treatment.

At a cryostat base temperature of 14 mK, we measured the transmitted signal $S_{21}$ using a vector network analyzer (VNA) and fitted the data to a hanger resonator fit function \cite{Probst2015}:
\begin{equation}
    S_{21}(\omega) = A e^{i\alpha} e^{-i\omega \tau}\Bigg[\frac{2i(\omega - \omega_r) + \kappa - \kappa_c(1 + i\tan(\phi_0))}{2i(\omega - \omega_r) + \kappa}\Bigg],
    \label{eq:supp-S21-resonator-trace}
\end{equation}
where $\omega_r$ is the resonator frequency, $\kappa$ is the total linewidth, and $\kappa_c = {\rm Re}(\tilde{\kappa}_c)$ is the real coupling loss with $\tilde{\kappa}_c = |\tilde{\kappa}_c|e^{i\phi_0}$ being the complex-valued coupling constant that accounts for Fano asymmetries in the measured lineshape. We also fit the electrical delay $\tau$, phase offset $\alpha$, and overall signal amplitude $A$ using the prefactors above. The internal loss rate $\kappa_i$ can be derived via $\kappa_i = \kappa - \kappa_c$, and the various quality factors are defined as $Q_\bullet = \omega_r / \kappa_\bullet$. Using the coupling quality factor $Q_c$ and the internal quality factor $Q_i$, we can also then estimate the circulating photon number
\begin{equation}
    \bar{n} = \frac{4}{\omega_r Q_c}\left(\frac{1}{Q_c} + \frac{1}{Q_i}\right)^{-2} \frac{P_{\rm in}}{\hbar\omega_r},
\end{equation}
which is proportional to the input power $P_{\rm in}$ incident on the device. We calibrated the line attenuation using a through measurement in a separate cooldown and estimated -82 dB of input attenuation at the relevant frequencies. As a result, at the lowest applied VNA power corresponding to -172 dBm at the mixing chamber, the fitted resonator parameters give $\bar{n} \ll 1$. A representative low-power $S_{21}$ response of the 34 $\mu$m gap ASR is shown in Fig.~\ref{si-fig: asr q factors}(b) with fitted parameters $\kappa_c/2\pi = 2.53$ kHz and $\kappa_i/2\pi = 0.807$ kHz. In Fig.~\ref{si-fig: asr q factors}(c), we plot the extracted internal quality factors for all six resonators. Across the three tested gap values, $Q_i$ increases with gap for both resonator families, consistent with reduced electric-field participation at lossy surfaces and interfaces as the gap is widened. At each gap, the ASR exhibits an internal quality factor approximately twice that of the corresponding CPW resonator, with the ASRs reaching low-power values of $Q_i\simeq(4\text{--}6)\times10^6$. The results of this study demonstrate a promising path toward improving the coherence of our planar storage resonators by increasing the ASR gap size in future design iterations. Additionally, we anticipate that further gains in the resonator quality factor can be achieved through the use of niobium or tantalum for the base layer metallization (as compared to the aluminum metallization used in the main device reported on in this work), since these refractory metals are compatible with more aggressive surface cleaning treatments~\cite{Place2021, Crowley2023, Bland2025}.

\newpage
\section{Fluxonium-Resonator Hamiltonian}

\subsection{Fluxonium Circuit Hamiltonian\label{sec: Fluxonium Hamiltonian}}

The fluxonium is a superconducting qubit composed of a Josephson junction shunted by a large inductance and a capacitance, as depicted in Fig.~\ref{fig: circuit diagram}(a). In practice, an array of Josephson junctions is used to realize a large and nearly linear superinductance which, together with a single Josephson junction, forms a superconducting loop that can be flux biased to control the fluxonium potential and properties. The fluxonium is modeled via the Hamiltonian

\begin{equation}\label{eq: fluxonium hamiltonian}
    \op{H}_{\rm F} = 4E_{C}^{q} \op{n}^2 + \frac{1}{2} E_{L}^{q} \op{\varphi}^2 - E_{J}^{q} \cos\left(\op{\varphi} - \frac{2\pi\Phi_{\rm ext}}{\Phi_0}\right),
\end{equation}
where $E_{C}^{q}$ is the charging energy, $E_{L}^{q}$ is the inductive energy, $E_{J}^{q}$ is the Josephson energy, $\op{\varphi}$ is the superconducting phase across the junction, $\op{n}$ is the charge operator conjugate to the phase, and $\Phi_{\rm ext}$ is the external flux bias. The fluxonium Hamiltonian can be diagonalized numerically to obtain the eigenenergies and eigenstates of the system.

\subsection{Fluxonium-Resonator Circuit Hamiltonian}

The Hamiltonian of the fluxonium-resonator (FR) circuit follows from Eq.~\eqref{eq: fluxonium hamiltonian}, with the addition of the resonator and coupling terms:

\newcommand{\PhantomVspace}{%
  \vphantom{i\hbar g\frac{\hat n}{n_{\rm zpf}}
  \left(\hat a^\dagger-\hat a\right)}}
\begin{equation}
\op{H}_{\rm FR}
=
\underbrace{
  \PhantomVspace
  4E_C^q\hat n^2
  +\frac{1}{2}E_L^q\hat{\varphi}^2
  -E_J^q\cos\left(
    \hat{\varphi}-\frac{2\pi\Phi_{\rm ext}}{\Phi_0}
  \right)
}_{\hat{H}_{\rm F}}
\,+\,
\underbrace{
  \PhantomVspace
  \hbar\omega_a\hat a^\dagger\hat a
}_{\hat{H}_{\rm R}}
\,+\,
\underbrace{
  \PhantomVspace
  i\hbar g\frac{\hat n}{n_{\rm zpf}}
  \left(\hat a^\dagger-\hat a\right)
}_{\hat{ H}_{\rm coupling}},
\label{eq:fluxonium-resonator-hamiltonian}
\end{equation}
where $\op{a}$ is the annihilation operator for the storage resonator, $\omega_a$ is its frequency, $n_{\rm zpf}=(E_L^q/32E_C^q)^{1/4}$ is the zero-point charge fluctuation of the fluxonium, and $g$ is the storage-fluxonium coupling strength. The fluxonium phase $\op{\varphi}$ is a noncompact variable within a confining potential and obeys nonperiodic boundary conditions. A harmonic-oscillator Fock basis is therefore an appropriate and efficient basis for representing the fluxonium~\cite{Devoret2021}, while the storage is represented directly in its Fock basis.

\subsection{Hierarchical Numerical Diagonalization\label{sec: numerical diagonalization of the fluxonium resonator system}}

The numerical calculations use hierarchical diagonalization. Each mode is first represented in a sufficiently large basis, diagonalized and truncated to only retain its lowest eigenstates. The charge-charge coupling in Eq.~\eqref{eq:fluxonium-resonator-hamiltonian} is then represented in these truncated eigenbases, and the full coupled Hamiltonian is diagonalized. Each dressed eigenstate is assigned the label $\ket{n,s}$ of the uncoupled product state with which it has the largest overlap, with corresponding energy $E_{n,s}$; here, $n$ is the storage occupation and $s$ is a fluxonium eigenstate. For the calculations reported here, the storage uses and retains eight oscillator Fock states. The fluxonium Hamiltonian is represented using 100 levels of its $E_C^q$-$E_L^q$ Fock basis and truncated to its 25 lowest eigenstates. The final coupled Hamiltonian therefore has dimension $8\times25=200$, rather than the $8\times100=800$ dimension of the corresponding untruncated product basis. This retains the nonlinear low-energy structure of the fluxonium while reducing the composite diagonalization cost. JAXQuantum~\cite{jha2024jaxquantum} makes this staged construction straightforward and allows the basis and truncation to be selected independently for each circuit element.

The photon-number-dependent dispersive shift and nonlinearities are evaluated directly from finite differences of these dressed energies:
\begin{align}
    \chi(n)&=\frac{1}{\hbar}\left[\left(E_{n+1,e}-E_{n,e}\right)-\left(E_{n+1,g}-E_{n,g}\right)\right],\label{eq: full numerical eigenenergy finite differences chi}\\
    K_s(n)&=\frac{E_{n+2,s}-2E_{n+1,s}+E_{n,s}}{2\hbar}, \label{eq: full numerical eigenenergy finite differences kerr}\\
    K_s^{(m)}(n)&=\frac{1}{\hbar m!}\sum_{j=0}^{m}(-1)^{m-j}\binom{m}{j}E_{n+j,s}.
    \label{eq: full numerical eigenenergy finite differences higher kerr}
\end{align}
In this work, $\chi\coloneqq\chi(0)$ and $K_s\coloneqq K_s^{(2)}(0)$, where $s \in \{g,e\}$ represents the ground and excited states of the fluxonium. The factor of $1/2$ in the definition of $K_s$ follows directly from the convention $\op{H}_{K,s}/\hbar=K_s\op a^\dagger{}^2\op a^2\ketbra{s}{s}$. The final line generalizes the same construction to all higher-order photon-nonlinearity coefficients of $\op a^\dagger{}^m\op a^m$. These quantities make no approximation beyond Hilbert-space truncations, which are tested for convergence with the basis size.

\subsection{Fluxonium-Resonator Hamiltonian in the Weak Dispersive Regime}

In the weak dispersive regime, where the fluxonium-resonator detuning is much larger than the coupling strength $g$, the system Hamiltonian can be approximated as~\cite{Eickbusch2022}
\begin{equation}\label{eq: dispersive system hamiltonian}
   \op{\tilde{H}}_{\rm FR}/\hbar \,\approx\, \Delta_a \op{a}^\dagger \op{a} - \frac{\chi}{2} \op{a}^\dagger \op{a} \op{\sigma}_z + \frac{K_g}{2} \op{a}^\dagger{}^2 \op{a}^2 (1 + \op{\sigma}_z) + \frac{K_e}{2} \op{a}^\dagger{}^2 \op{a}^2 (1 - \op{\sigma}_z),
\end{equation}
where $\op{\tilde{H}}_{\rm FR}$ is expressed in the rotating frame of the fluxonium and resonator. Here, $\op{\sigma}_z$ is the Pauli-Z operator in the fluxonium qubit manifold, and $\Delta_a$ is the storage-drive detuning. The parameters $\chi$, $K_g$, and $K_e$ denote the dispersive coupling and the fluxonium-state-dependent storage self-Kerr nonlinearities. These values are obtained from the hierarchical full-system diagonalization and energy differences in Eqs.~(\ref{eq: full numerical eigenenergy finite differences chi}--\ref{eq: full numerical eigenenergy finite differences higher kerr}). Experiments instead extract the effective Kerr coefficients $\widetilde{K}_g$ and $\widetilde{K}_e$, as detailed in Sec.~\ref{sec: Measuring Resonator self-Kerr}.

\section{Characterization and Calibration Experiments \label{sec: Characterization and Calibration Experiments}}

\begin{figure}[h]
    \centering
    \includegraphics[width=\linewidth]{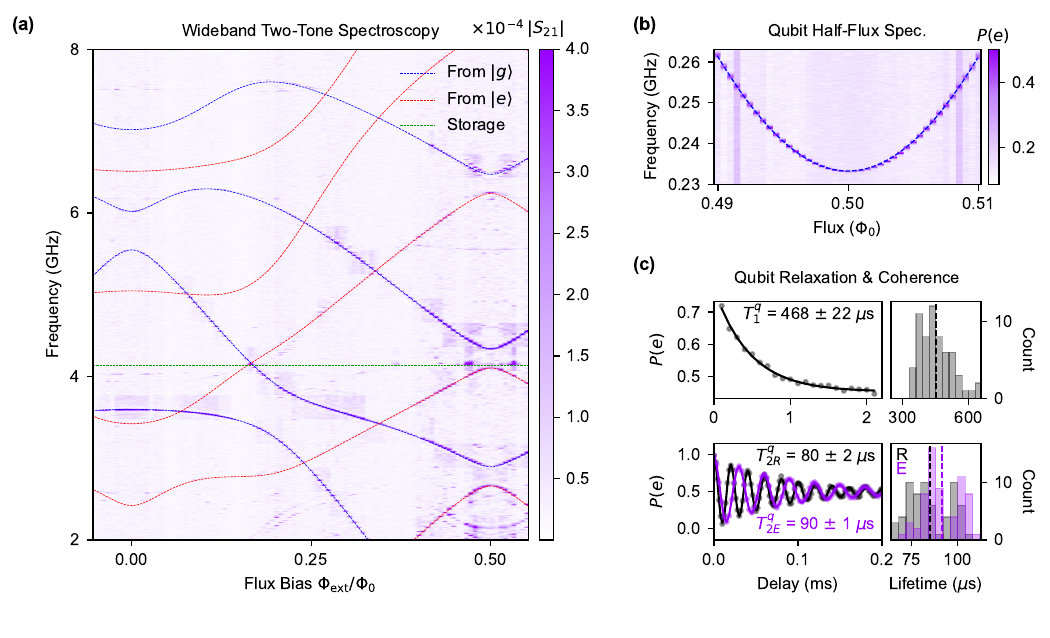}
    \caption{\textbf{Fluxonium spectroscopy and lifetimes.} \textbf{(a)} Wideband two-tone (fluxonium drive and readout drive) spectroscopy of the fluxonium transitions versus the external flux bias $\Phi_{\rm ext}$, with the numerically diagonalized spectrum overlaid. \textbf{(b)} Low-frequency spectroscopy of the fluxonium $\ket{g}\leftrightarrow \ket{e}$ transition near half flux ($\Phi_{\rm ext} = \Phi_0/2$), performed after preparing the fluxonium ground state, with the numerically diagonalized spectrum overlaid. \textbf{(c)} Half-flux fluxonium relaxation and coherence lifetime example traces and histograms of measurements taken over several hours in a single day.}
    \label{si-fig: fluxonium spec and coherence}
\end{figure}

\subsection{Fluxonium Characterization and Calibration \label{sec: Fluxonium Characterization and Calibration}}

\subsubsection{Fluxonium Spectroscopy and Lifetimes}

The fluxonium Hamiltonian is specified by the circuit parameters $E_{C}^{q}$, $E_{L}^{q}$, and $E_{J}^{q}$, as introduced in Sec.~\ref{sec: Fluxonium Hamiltonian}. The flux-bias axis $\Phi_{\rm ext}$ is calibrated using the two symmetry points of the measured spectrum. The circuit parameters are then varied to match the numerically diagonalized spectrum to the high- and low-frequency spectroscopy in Fig.~\ref{si-fig: fluxonium spec and coherence}(a,b), yielding the fitted parameters listed in Table~\ref{tab: summary of parameters}.

The fluxonium is biased at the half-flux sweet spot ($\Phi_{\rm ext} = \Phi_0/2$) for the remainder of the experiment, where insensitivity to first-order flux noise maximizes coherence, as shown in Fig.~\ref{si-fig: fluxonium tuneup}(a). Histograms and example traces of the fluxonium relaxation and coherence lifetimes at half flux are shown in Fig.~\ref{si-fig: fluxonium spec and coherence}, with averaged lifetimes quoted in Table~\ref{tab: summary of parameters} and the main text.

In our device, the fluxonium coherence at half flux is likely limited by photon-shot-noise dephasing from the readout resonator~\cite{Gambetta2006}. Assuming this, we estimate the resonator temperature from the averaged pure-dephasing time, $\overline{T}_{\phi}^{\,q}=[1/\overline{T}_{2E}^{\,q}-1/(2\overline{T}_{1}^{\,q})]^{-1}$, using
\begin{equation}
    \bar{n}_{\rm th}=\frac{1/\overline{T}_{\phi}^{\,q}}{\kappa_c\chi_{qr}^{2}/(\kappa_c^{2}+\chi_{qr}^{2})}, \qquad T_r=\frac{\hbar\omega_r}{k_{\rm B}\ln(1+1/\bar{n}_{\rm th})}.
    \label{eq: photon shot noise dephasing}
\end{equation}
Given the measured parameters in Table~\ref{tab: summary of parameters}, we find an average readout resonator temperature of $T_r=\SI{41\pm 1}{\milli\kelvin}$.

\begin{figure}[h]
    \centering
    \includegraphics[width=\linewidth]{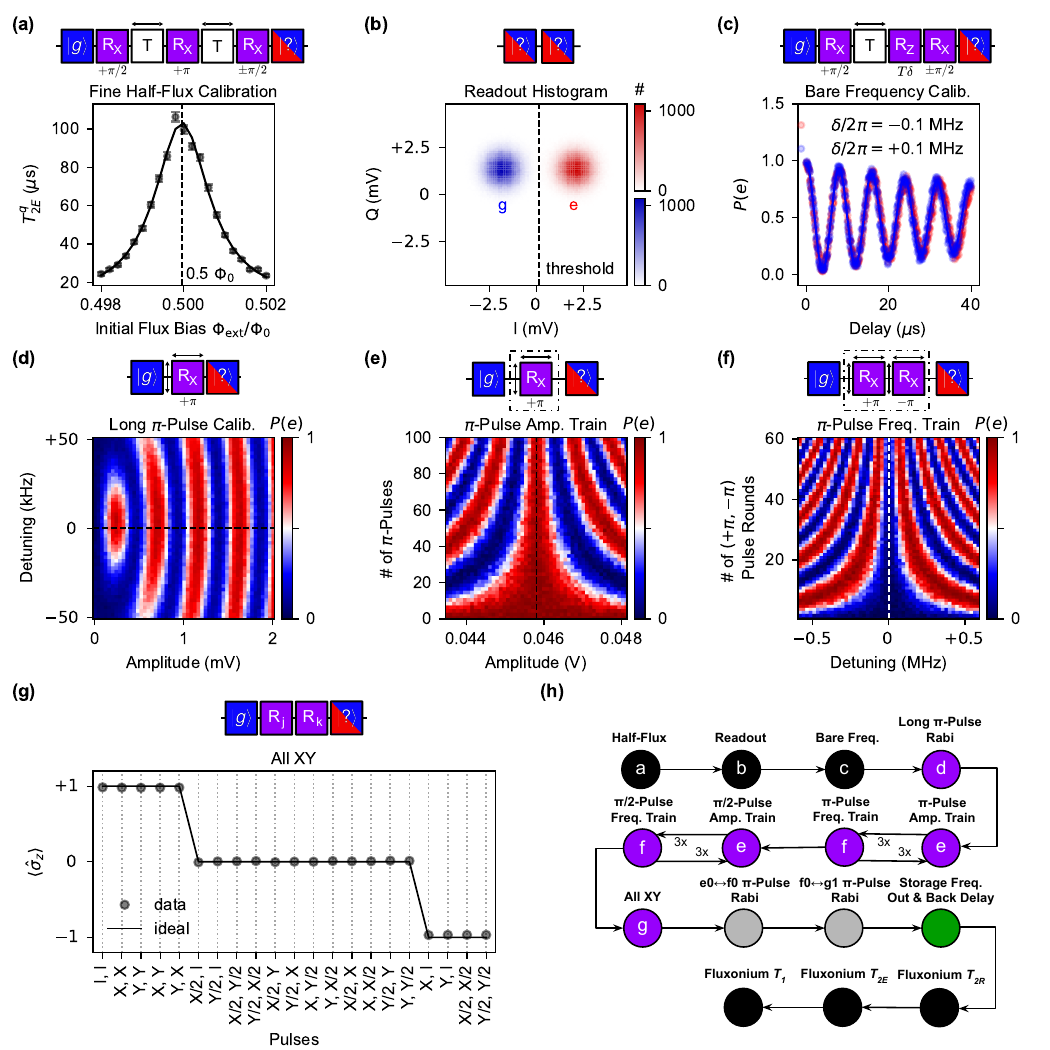}
    \caption{\textbf{Fluxonium daily tune-up.} \textbf{(a)} Fluxonium echo coherence time $T_{2E}^q$ versus initial flux bias, used to locate half flux. \textbf{(b)} Integrated readout histogram for the $\ket{g}$ and $\ket{e}$ states, with the discrimination threshold indicated by the dashed line. \textbf{(c)} Bare $\ket{g}\leftrightarrow\ket{e}$ transition frequency calibration using two detuned time-proportional phase increment (TPPI) Ramsey sequences. \textbf{(d)} Long $\pi$-pulse amplitude and frequency calibration. \textbf{(e, f)} Repeated $\pi$-pulse amplitude (e) and frequency (f) calibrations. Each iteration of the $\pi$- and $\pi/2$-pulse amplitude trains contains $2n+1$ and $4n+2$ pulses, respectively, where $n$ is a nonnegative integer. \textbf{(g)} AllXY validation of the single-qubit gate calibration. \textbf{(h)} Flow of full daily calibration sequence.}
    \label{si-fig: fluxonium tuneup}
\end{figure}

\subsubsection{Fluxonium Daily Tune-Up\label{sec: Fluxonium Daily Tune-Up}}

The daily tune-up, inspired by Ref.~\cite{Ding2023}, begins by locating the half-flux sweet spot from the maximum of the fluxonium echo coherence time $T_{2E}^q$ versus flux bias, as shown in Fig.~\ref{si-fig: fluxonium tuneup}(a). A histogram of the integrated readout signal then determines the optimal readout angle and threshold separating the $\ket{g}$ and $\ket{e}$ states along the real ($I$) axis [Fig.~\ref{si-fig: fluxonium tuneup}(b)].

The bare $\ket{g}\leftrightarrow \ket{e}$ transition frequency is calibrated using two time-proportional phase increment (TPPI) Ramsey sequences detuned by $\delta/2\pi=\pm \SI{0.1}{\mega\hertz}$, as shown in Fig.~\ref{si-fig: fluxonium tuneup}(c). Each oscillation frequency combines the TPPI detuning and the offset from the bare $\ket{g}\leftrightarrow \ket{e}$ transition frequency. For example, a bare-frequency miscalibration of $\delta_0=\SI{+0.05}{\mega\hertz}$ produces oscillations at $|0.1+\delta_0|\,\si{\mega\hertz}=\SI{0.15}{\mega\hertz}$ and $|-0.1+\delta_0|\,\si{\mega\hertz}=\SI{0.05}{\mega\hertz}$. These two frequencies determine the magnitude and sign of $\delta_0$, allowing correction of the bare-frequency calibration.

Three types of pulses are calibrated for the fluxonium $\ket{g}\leftrightarrow \ket{e}$ transition: \textit{(i)} a long $\SI{20}{\micro\second}$ $\pi$ pulse with a $\SI{1}{\micro\second}$ cosine rise and fall, and \textit{(ii, iii)} short $\SI{144}{\nano\second}$ $\pi$ and $\pi/2$ pulses, each with a $\SI{44}{\nano\second}$ cosine rise and fall. The frequency selectivity of the long $\pi$ pulse detects photons in the dispersively coupled storage resonator: storage photons result in a Stark shift of the fluxonium $\ket{g}\leftrightarrow \ket{e}$ transition frequency, causing the long $\pi$ pulse to fail. The short $\pi$ and $\pi/2$ pulses provide all other fluxonium control operations. Several rounds of amplitude and frequency $\pi$-pulse trains separately calibrate the amplitude and frequency of these short pulses, as described in Fig.~\ref{si-fig: fluxonium tuneup}(e,f,h). The frequency frame of the AWG (Quantum Machines OPX+) remains fixed at the bare $\ket{g}\leftrightarrow\ket{e}$ transition frequency, while envelope detunings target the optimal Stark-shifted frequency. After calibration, the AllXY method identifies errors in the $X$, $Y$, $X/2$, and $Y/2$ pulses generated from the calibrated $\pi$ and $\pi/2$ pulses~\cite{MattReedPhDThesis}.

The daily tune-up concludes with calibration of the $\ket{f, 0}\leftrightarrow\ket{g, 1}$ reset pulses described in Sec.~\ref{sec: Pulsed f0g1 Reset of Fluxonium}, calibration of the storage frequency described in Sec.~\ref{sec: Measuring Fluxonium-Resonator Dispersive Coupling}, and measurement of the fluxonium lifetimes [Fig.~\ref{si-fig: fluxonium tuneup}(h)].

\begin{figure}[h]
    \centering
    \includegraphics[width=\linewidth]{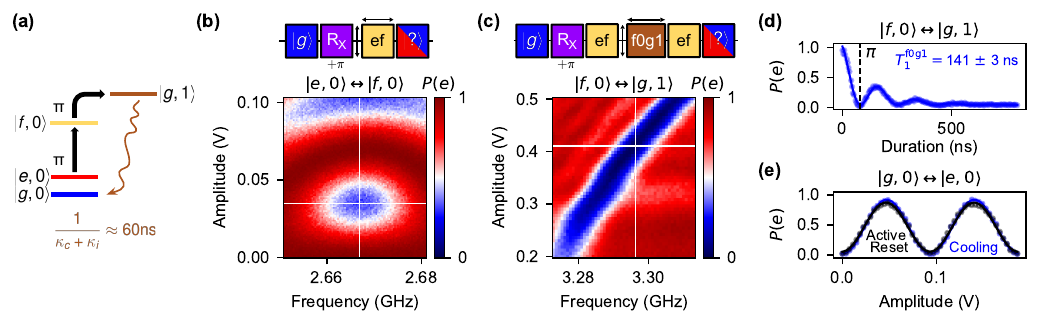}
    \caption{\textbf{Fluxonium pulsed reset.} \textbf{(a)} Reset sequence transferring a fluxonium excitation to the readout resonator through the $\ket{f,0}\leftrightarrow\ket{g,1}$ transition. \textbf{(b, c)} Power-Rabi calibration of the $\ket{e,0}\leftrightarrow\ket{f,0}$ (b) and $\ket{f,0}\leftrightarrow\ket{g,1}$ (c) reset pulses. \textbf{(d)} Time-domain $\ket{f,0}\leftrightarrow\ket{g,1}$ Rabi oscillations, with the calibrated $\pi$-pulse duration indicated by the dashed line. \textbf{(e)} Fluxonium $\ket{g,0}\leftrightarrow\ket{e,0}$ Rabi oscillations following measurement-based active reset (black) or pulsed cooling (blue).}
    \label{si-fig: fluxonium reset}
\end{figure}

\subsubsection{Pulsed $\ket{f,0}\leftrightarrow\ket{g,1}$ Reset of Fluxonium\label{sec: Pulsed f0g1 Reset of Fluxonium}}

The fluxonium is reset to the ground state using a $\SI{48}{\nano\second}$ $\ket{e,0}\leftrightarrow\ket{f,0}$ $\pi$ pulse followed by an $\SI{88}{\nano\second}$ $\ket{f,0}\leftrightarrow\ket{g,1}$ $\pi$ pulse, transferring the fluxonium excitation to the readout resonator [Fig.~\ref{si-fig: fluxonium reset}(a)]~\cite{Magnard2018,Egger2018}. An additional $\SI{100}{\nano\second}$ delay allows the readout resonator to decay into the environment, leaving both modes in the ground state $\ket{g,0}$. Power-Rabi chevrons are used to calibrate the two reset pulses [Fig.~\ref{si-fig: fluxonium reset}(b,c)].

The reset mechanism is characterized using time-domain Rabi oscillations of the $\ket{f,0}\leftrightarrow\ket{g,1}$ transition [Fig.~\ref{si-fig: fluxonium reset}(d)]. The fluxonium is prepared in $\ket{f,0}$ before driving the $\ket{f,0}\leftrightarrow\ket{g,1}$ transition. A subsequent $\ket{e,0}\leftrightarrow\ket{f,0}$ $\pi$ pulse maps the remaining population in state $\ket{f,0}$ to the state $\ket{e,0}$, providing high readout contrast between $\ket{f,0}$ and $\ket{g,1}$. The Rabi oscillations decay with a time constant $T_1^{{\rm f0g1}}\approx\SI{140}{\nano\second}$, which is approximately twice the readout-resonator decay time of $1/(\kappa_c+\kappa_i)\approx\SI{60}{\nano\second}$. This is expected because the excitation spends half of each oscillation in the lossy readout resonator and half in the fluxonium.

Fluxonium $\ket{g,0}\leftrightarrow\ket{e,0}$ power-Rabi oscillations are used to compare the performance of ground-state initialization using active reset versus the pulsed $\ket{f,0}\leftrightarrow\ket{g,1}$ cooling scheme [Fig.~\ref{si-fig: fluxonium reset}(e)]. The two methods achieve similar performance, but the measurement-free pulsed cooling sequence completes in $\SI{236}{\nano\second}$. Measurement-based active reset, on the other hand, requires a $\SI{1.5}{\micro\second}$ readout pulse, hundreds of nanoseconds of feedback latency, and potentially a final $\SI{144}{\nano\second}$ fluxonium $\pi$ pulse. In principle, the pulsed $\ket{f,0}\leftrightarrow\ket{g,1}$ cooling scheme is limited by the thermal population of the readout resonator, whereas active reset is limited by readout fidelity.

To estimate the reset fidelity, we use the raw IQ Rabi data acquired after interleaved cooling-based and measurement-based resets. A two-component Gaussian-mixture fit separates the ground- and non-ground-state readout clouds, while the ratio of the Rabi contrast after each of the two reset methods provides an estimate of reset fidelity that does not require choosing an IQ-classification threshold. For the dataset used in the logical-error-budget analysis of Fig.~\ref{si-fig: repeated ecd error budget}, this procedure gives a ground-state population of $\SI{96.59}{\percent}$ after the $\SI{484}{\nano\second}$ reset sequence, corresponding to a reset error of $\SI{3.41}{\percent}$.

\subsection{Fluxonium-Resonator Characterization\label{sec: Fluxonium-Resonator Characterization}}

The low-energy fluxonium-resonator interaction Hamiltonian is given in Eq.~\eqref{eq: dispersive system hamiltonian}. Because the storage resonator is designed to be weakly coupled to its drive line, the fluxonium must be used to characterize the resonator properties. The calibration in Sec.~\ref{sec: Measuring Fluxonium-Resonator Dispersive Coupling} sets the storage-drive detuning $\Delta_a$ to zero and measures $\chi$ as defined in Eq.~\eqref{eq: full numerical eigenenergy finite differences chi}. The measurements of the effective Kerr coefficients $\widetilde{K}_g$ and $\widetilde{K}_e$ are described in Sec.~\ref{sec: Measuring Resonator self-Kerr}.

\begin{figure}[h]
    \centering
    \includegraphics[width=\linewidth]{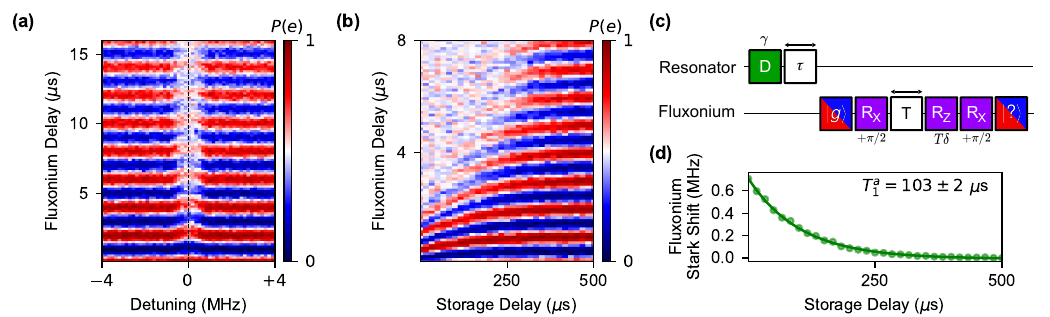}
    \caption{\textbf{Storage resonator spectroscopy and relaxation.} \textbf{(a)} Fluxonium Ramsey response versus storage-drive detuning, used to locate the storage resonance. \textbf{(b)} Fluxonium Ramsey response after coherent-state preparation versus storage delay $\tau$ and fluxonium delay $T$, corresponding to the circuit diagram in \textbf{(c)}. \textbf{(d)} Extracted fluxonium Stark shift versus storage delay, with an exponential fit yielding $T_1^a = \SI{103 \pm 2}{\micro\second}$.}
    \label{si-fig: storage spec relaxation}
\end{figure}

\subsubsection{Resonator Spectroscopy and Relaxation\label{sec: Resonator Spectroscopy and Relaxation}}

The dispersive coupling in Eq.~\eqref{eq: dispersive system hamiltonian} produces a photon-number-dependent ac Stark shift of the fluxonium transition frequency. For a coherent state with amplitude $\alpha$ in the storage resonator, the mean shift is $\chi|\alpha|^2$, as shown in Eq.~\eqref{eq: displaced dispersive Hamiltonian}. Photon-number fluctuations caused by loss therefore produce fluctuations in the fluxonium transition frequency and dephase the fluxonium. When $|\chi|T_1^a \gg 1$, the fluxonium phase accumulated between photon-loss events is highly sensitive to their random timing. Averaging over these random loss trajectories gives a photon-shot-noise dephasing rate of $\Gamma_\phi \approx \bar{n}/T_1^a=|\alpha|^2/T_1^a$~\cite{Gambetta2006}. Since a fluxonium Ramsey measurement is sensitive to both the mean Stark shift and shot-noise dephasing, it can therefore be used for storage resonator spectroscopy [Fig.~\ref{si-fig: storage spec relaxation}(a)].

After locating the storage resonance, a coherent state is prepared and allowed to decay toward vacuum for a variable storage delay. The Ramsey response in Fig.~\ref{si-fig: storage spec relaxation}(b) measures the residual Stark shift $\chi |\alpha|^2 = \chi \bar{n}$, which decays with the average resonator photon number $\bar{n}$. An exponential fit in Fig.~\ref{si-fig: storage spec relaxation}(c) gives $T_1^a = \SI{103 \pm 2}{\micro\second}$, corresponding to a quality factor of $Q = \omega_a T_1^a = 2.65 \times 10^6$.

\begin{figure}[h]
    \centering
    \includegraphics[width=\linewidth]{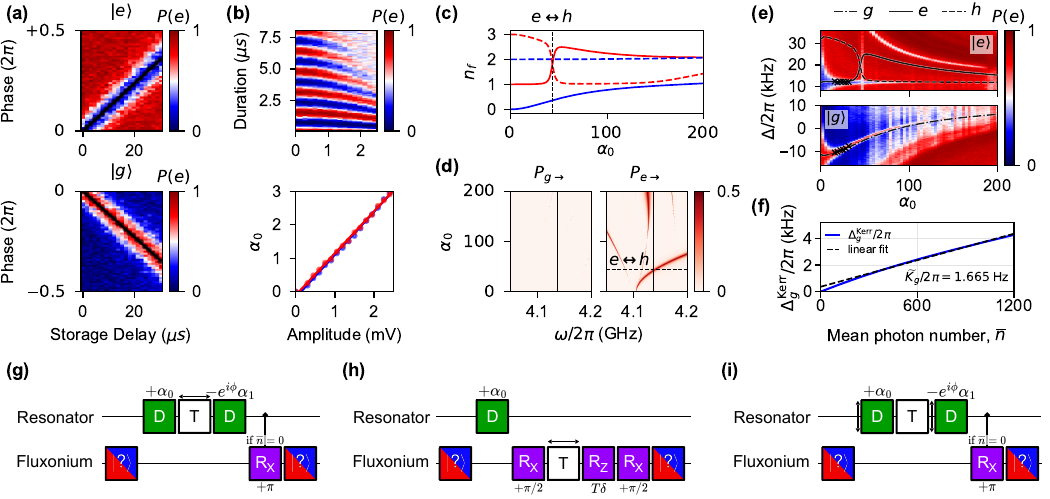}
    \caption{\textbf{Out-and-back characterization.} \textbf{(a)} Out-and-back measurements with fixed $\alpha_0$ versus storage delay time for the fluxonium initialized in $\ket{e}$ (top) and $\ket{g}$ (bottom). \textbf{(b)} Fluxonium Ramsey calibration of the coherent-state magnitude $\alpha_0$ versus storage-drive amplitude (top) and the extracted linear calibration (bottom). \textbf{(c)} Calculated Floquet-averaged fluxonium excitation number $n_f$ versus $\alpha_0$ for branches connected to the dressed fluxonium states $\ket{g}$ (solid blue), $\ket{e}$ (solid red), $\ket{f}$ (dashed blue), and $\ket{h}$ (dashed red). \textbf{(d)} Calculated hybridization indicators $P_{g\rightarrow}$ (left) and $P_{e\rightarrow}$ (right) versus drive frequency and $\alpha_0$. The solid vertical black line marks the storage resonator frequency, while the dashed horizontal black line marks the $\alpha_0$ value corresponding to the DUST feature. \textbf{(e)} Out-and-back measurements with fixed storage delay time and large displacement sweeps for the fluxonium initialized in $\ket{e}$ (top) and $\ket{g}$ (bottom). The black crosses mark peaks corresponding to the optimal return phase at six amplitudes for each $s\in\{g,e\}$. Dividing these phases by the $\SI{6.000}{\micro\second}$ out-and-back delay time gives $\Delta_s^{\rm plot}(\alpha_0)$. These 12 points lie below the $\ket{e}$--$\ket{h}$ exchange and are used to determine the common rotating-frame offset $\Delta_{\rm frame}$ described in Sec.~\ref{sec: Efficient Numerical Modeling of Out-and-Back Measurements}. The thin black curves are the pulse-integrated Floquet predictions at $\Phi_{\rm ext}/\Phi_0=0.499$, a flux bias at which $\chi(\bar n)$ changes little from its half-flux value while the $\ket{e}$--$\ket{h}$ DUST feature becomes clearly visible. The solid $\ket{e}$- and dashed $\ket{h}$-connected branches are overlaid on the upper map, and the dash-dotted $\ket{g}$-connected branch is overlaid on the lower map. Residual storage photons left by incorrect \textit{back} displacements lead to DUST during the subsequent fluxonium readout, producing a partially red background in both maps. \textbf{(f)} Kerr contribution $\Delta_g^{\rm Kerr}$ to the ground-state coherent-state rotation rate reconstructed through fourth order. A linear fit over $200\leq\bar n\leq1200$ gives $\widetilde{K}_g/2\pi=\SI{1.665}{\hertz}$. \textbf{(g)} Pulse sequence for the delay-swept out-and-back measurement in (a). \textbf{(h)} Fluxonium Ramsey sequence used for the displacement calibration in (b). \textbf{(i)} Pulse sequence for the amplitude-swept out-and-back measurement in (e). In (g,i), $\alpha_1=\alpha_0e^{-\tau/(2T_1^a)}$ is the loss-compensated return-displacement magnitude after storage-evolution time $\tau$.}
    \label{si-fig: out and back}
\end{figure}

\subsubsection{Measuring Fluxonium-Resonator Dispersive Coupling\label{sec: Measuring Fluxonium-Resonator Dispersive Coupling}}

The long $\SI{20}{\micro\second}$ $\pi$ pulse calibrated in Sec.~\ref{sec: Fluxonium Daily Tune-Up} is sensitive to shifts of the fluxonium transition frequency, including Stark shifts from storage resonator photons. The pulse therefore serves as a sensitive probe of whether the storage resonator is in the vacuum state.

In an ``out-and-back'' measurement, the fluxonium is first prepared in state $j \in \{g,e\}$. A displacement $D(\alpha)$ then prepares a coherent state in the storage resonator. During the subsequent delay $\tau$, the state rotates at the fluxonium-state-dependent rate $\Delta_j$ set by Eq.~\eqref{eq: dispersive system hamiltonian} and decays toward vacuum at the photon-loss rate $1/T_1^a$. A final displacement $D(-\alpha e^{i\phi_j}e^{-\tau/(2T_1^a)})$ accounts for this decay and returns the resonator to vacuum only when the back displacement angle $\phi_j$ satisfies $\phi_j = \Delta_j\tau$.

The fluxonium-state-dependent resonator rotation rate is approximately

\begin{align*}
    \Delta_g &\approx \Delta_a - \frac{\chi}{2} + 2\widetilde{K}_g |\alpha_0|^2,\\
    \Delta_e &\approx \Delta_a + \frac{\chi}{2} + 2\widetilde{K}_e |\alpha_0|^2.\numberthis\label{eq: fluxonium-state-dependent resonator rotation}
\end{align*}
Here $\alpha_0 = |\alpha|$ is the magnitude of the coherent state prepared in the storage resonator. The coefficients $\widetilde{K}_g$ and $\widetilde{K}_e$ are effective slopes over the fitted photon-number range and are measured to be on the order of $2\pi \times \SI{1}{\hertz}$ in our system. Meanwhile, the dispersive shift $\chi$ is on the order of $2\pi \times \SI{10}{\kilo\hertz}$, and thus the dispersive contribution dominates the fluxonium-state dependence of $\Delta_j$ at small $\alpha_0$.

The first out-and-back measurement uses a small displacement amplitude $\alpha_0$, as confirmed in Sec.~\ref{sec: Resonator Drive Strength Calibration}, and sweeps the delay $\tau$, as shown for both initial fluxonium states in Fig.~\ref{si-fig: out and back}(a). At each delay, a vertical linecut is fit to a Lorentzian to extract the optimal return angle $\phi_j$ that brings the resonator back to vacuum, identified by success of the long $\pi$ pulse. Fitting these angles to the zero-intercept model $\phi_j = \Delta_j\tau$ yields $\Delta_j$. In the small-$\alpha_0$ regime, the fluxonium-state-dependent rotation rates $\Delta_g$ and $\Delta_e$ are used to determine $\chi$ and the storage detuning $\Delta_a$ through

\begin{equation}
    \chi \approx \Delta_e - \Delta_g, \qquad \Delta_a \approx \frac{\Delta_g + \Delta_e}{2}.\label{eq: fluxonium-state-dependent resonator rotation, low alpha_0}
\end{equation}
After this calibration, we adjust the resonator drive frame such that $\Delta_a$ is zero in subsequent measurements. The extracted value of $\chi$ is listed in Table~\ref{tab: summary of parameters}.

\subsubsection{Resonator Drive Strength Calibration\label{sec: Resonator Drive Strength Calibration}}

A fluxonium Ramsey measurement determines the Stark shift $\chi |\alpha|^2$ produced by a coherent state in the storage resonator. Combined with the measured $\chi$ from Sec.~\ref{sec: Measuring Fluxonium-Resonator Dispersive Coupling}, this shift gives the coherent-state magnitude $\alpha_0 = |\alpha|$ as a function of the resonator displacement-drive voltage, as shown in Fig.~\ref{si-fig: out and back}(b). This calibration also confirms that the dispersive-coupling measurement in Sec.~\ref{sec: Measuring Fluxonium-Resonator Dispersive Coupling} was performed at $\alpha_0 \approx 3$, well within the small-$\alpha_0$ regime.

The lower panel of Fig.~\ref{si-fig: out and back}(b) shows the resonator drive-strength calibrations with the fluxonium initialized in $\ket{g}$ (blue) and $\ket{e}$ (red). The overlap of the two curves indicates that the $\SI{28}{\nano\second}$ pulse, with a $\SI{4}{\nano\second}$ cosine rise and fall, is sufficiently short to displace the storage resonator irrespective of the fluxonium state.

\subsubsection{Measuring Resonator Self-Kerr\label{sec: Measuring Resonator self-Kerr}}

With the storage resonator drive strength calibrated, we perform a second out-and-back measurement which fixes $\tau$ and sweeps $\alpha_0$ out to values at which the fluxonium-state-dependent nonlinearity affects the resonator rotation $\Delta_j$. Fitting Eq.~\eqref{eq: fluxonium-state-dependent resonator rotation} over this finite photon-number range gives the effective coefficients $\widetilde{K}_g/2\pi=\SI{1.7}{\hertz}$ and $\widetilde{K}_e/2\pi=\SI{-0.05}{\hertz}$ listed in Table~\ref{tab: summary of parameters}. These values differ from the low-photon numerically obtained Kerr nonlinearities $K_g/2\pi=\SI{2.411}{\hertz}$ and $K_e/2\pi=\SI{-7.19}{\milli\hertz}$ because the linear fit for the approximate relation in Eq.~\eqref{eq: fluxonium-state-dependent resonator rotation} folds higher-order nonlinearities into its effective slope (calculated in the range $200<\bar n<1200$).

We can explain this discrepancy as follows. At half flux, numerical diagonalization gives $K_g^{(2)}/2\pi=\SI{2.41108}{\hertz}$, $K_g^{(3)}/2\pi=\SI{-0.52169}{\milli\hertz}$, and $K_g^{(4)}/2\pi=\SI{+0.10797}{\micro\hertz}$, where the higher nonlinearities $K_s^{(m)}$ are defined in Eq.~\eqref{eq: full numerical eigenenergy finite differences higher kerr}. With these coefficients, we can numerically calculate the Kerr contribution $\Delta_g^{\rm Kerr}(\bar n)$ to the coherent-state rotation rate plotted in Fig.~\ref{si-fig: out and back}(f) as follows:
\begin{equation}
    \Delta_g^{\rm Kerr}(\bar n)
    \coloneqq\Delta_g(\bar n)-\Delta_g(0)
    \approx2K_g^{(2)}\bar n+3K_g^{(3)}\bar n^2+4K_g^{(4)}\bar n^3.
\end{equation}
Fitting $\Delta_g^{\rm Kerr}$ to a linear model $\Delta_g^{\rm Kerr}(\bar n)=A+2\widetilde{K}_g\bar n$ over $200\leq\bar n\leq1200$ gives an effective slope of $\widetilde{K}_g/2\pi=\SI{1.665}{\hertz}$, which matches well the experimentally obtained value from out-and-back [Fig.~\ref{fig: conditional displacement demo}(a)]. The corresponding excited-state coefficients remain strongly suppressed: $K_e^{(2)}/2\pi=\SI{-7.186}{\milli\hertz}$, $K_e^{(3)}/2\pi=\SI{+0.0918}{\micro\hertz}$, and $K_e^{(4)}/2\pi=\SI{-0.0475}{\micro\hertz}$. We therefore see that the alternating higher-order terms flatten the photon-number dependence of the rotation rate, explaining both the distinction between the effective coefficients $\widetilde{K}_s$ and the low-photon coefficients $K_s$ (for $s \in \{g, e\}$) and the nearly constant excited-state rotation rate [see Figs.~\ref{fig: conditional displacement demo}(a) and \ref{si-fig: out and back}(e) demonstrating this suppression of the Kerr nonlinearity when the fluxonium is in state $\ket{e}$]. Finally, we note that in Sec.~\ref{sec: Efficient Numerical Modeling of Out-and-Back Measurements}, we extend this calculation out to $\bar n=40{,}000$ using Floquet modeling of the fluxonium-resonator dynamics.

\begin{figure}[b]
    \centering
    \includegraphics[width=\linewidth]{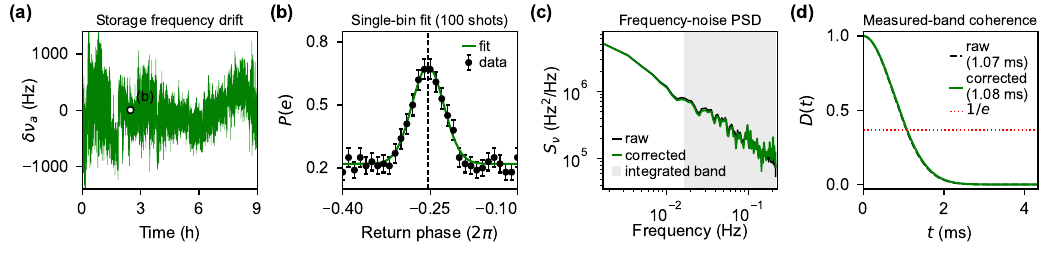}
    \caption{\textbf{Intrinsic storage-resonator dephasing.} \textbf{(a)} Ground-state storage-frequency fluctuations extracted from repeated fixed-delay linecuts of the out-and-back sequence shown in Fig.~\ref{si-fig: out and back}(g), with the point used in (b) indicated. \textbf{(b)} Representative 100-shot linecut and maximum-likelihood fit; the dashed line marks the fitted return-phase center. \textbf{(c)} Raw frequency-noise power spectral density (PSD) and the fit-noise-corrected split-shot cross-PSD. The shaded region marks the integrated frequency band. \textbf{(d)} Measured-band coherence envelopes calculated from the raw and fit-noise-corrected PSDs. The dotted line marks $1/e$.}
    \label{si-fig: storage dephasing}
\end{figure}

\subsubsection{Intrinsic Resonator Dephasing\label{sec: Intrinsic Resonator Dephasing}}

The intrinsic storage-resonator frequency noise is measured by repeatedly acquiring a fixed-delay linecut of the out-and-back duration measurement described in Sec.~\ref{sec: Measuring Fluxonium-Resonator Dispersive Coupling}. The fluxonium is initialized in $\ket{g}$, the storage is displaced to a fixed $\alpha_0$, and the coherent state evolves for $\tau=\SI{20}{\micro\second}$ before the phase of the return displacement is swept. The data from the full acquisition are first combined to determine the width, contrast, and baseline of the out-and-back linecut. These fit parameters are then held fixed while each linecut, formed from 100 repetitions at each return phase, is fitted only for its center $\phi_g$. Its fluctuation gives the instantaneous ground-state storage-frequency fluctuation, $\delta\nu_a(t)=\delta\phi_g(t)/(2\pi\tau)$. The resulting nine-hour frequency trace and a representative out-and-back linecut are shown in Fig.~\ref{si-fig: storage dephasing}(a,b).

This measurement tracks the center of the ground-state return feature rather than its contrast. A fluxonium bit flip transfers weight away from the ground-state trajectory and therefore reduces the return contrast [magnitude of $P(e)$ in Fig.~\ref{si-fig: storage dephasing}(b)], but does not shift the ground-state center used to infer $\nu_a$. The extracted frequency noise consequently probes pure dephasing of the storage resonator. In contrast, the transverse coherence of the Fock qubit is directly sensitive to fluxonium bit flips: a flip changes the storage frequency by the dispersive shift $\chi$, causing a superposition of $\ket{0}$ and $\ket{1}$ to accumulate a stochastic phase.

Fig.~\ref{si-fig: storage dephasing}(c) shows the one-sided power spectral density (PSD) $S_\nu(f)$ of the fitted center sequence. The raw PSD contains both storage-frequency noise and statistical noise from the center fits. To remove the latter without subtracting an assumed white-noise floor, the shots in each bin are separated into interleaved even and odd subsets and their centers are fit independently. The physical frequency fluctuation is common to both estimates, whereas fitting noise is uncorrelated, so their cross-PSD removes the fit-noise contribution. The cross-PSD is then divided by the finite-bin boxcar response $\operatorname{sinc}^2(f\Delta t_{\rm bin})$, where $\Delta t_{\rm bin}$ is the duration of each 100-shot bin.

The PSD is integrated from $f_{\rm low}=1/T_{\rm recalib}=\SI[parse-numbers=false]{(1/60)}{\hertz}$ to $f_{\rm high}=1/(2\Delta t_{\rm bin})=\SI{0.2165}{\hertz}$, where $T_{\rm recalib}=\SI{60}{\second}$ is on the order of the storage-frequency recalibration period. The lower edge excludes quasistatic drift removed by this recalibration, while the upper edge is the Nyquist frequency for $\Delta t_{\rm bin}=\SI{2.309}{\second}$ and therefore includes all available high-frequency data. For a storage free-evolution time $t$, the accumulated phase is $\phi(t)=2\pi\int_0^t\delta\nu_a(t')\,dt'$. Assuming zero-mean Gaussian frequency noise, the measured-band coherence envelope and dephasing time satisfy~\cite{Krantz2019}
\begin{align}
    \ev{\phi^2(t)} &= (2\pi)^2t^2 \int_{f_{\rm low}}^{f_{\rm high}} S_\nu(f)\operatorname{sinc}^2(ft)\,df,\nonumber\\
    D(t) &\equiv \big|\ev{e^{i\phi(t)}}\big|=e^{-\ev{\phi^2(t)}/2},\nonumber\\
    \ev{\phi^2(T_\phi^a)} &= 2,\label{eq: intrinsic storage dephasing}
\end{align}
where $\operatorname{sinc}(x)=\sin(\pi x)/(\pi x)$, $D(t)$ is the normalized storage coherence envelope, and the measured-band dephasing time $T_\phi^a$ is defined by $D(T_\phi^a)=1/e$, which gives the final line of Eq.~\eqref{eq: intrinsic storage dephasing}. Frequencies above $f_{\rm high}$ are not resolved and may contribute additional dephasing. The measured-band $T_\phi^a$ is therefore an upper bound on the storage coherence time under periodic storage-frequency recalibration, which removes the dephasing contribution of slow frequency drift. The fit-noise-corrected envelope in Fig.~\ref{si-fig: storage dephasing}(d) gives $T_\phi^a=\SI{1.08\pm0.16}{\milli\second}$, where the uncertainty is the standard error across calculations of $T_\phi^a$ that each omit a different portion of the acquired frequency record.

\subsection{Conditional Displacements with Fluxonium\label{sec: Conditional Displacements with Fluxonium}}

\begin{figure}[h]
    \centering
    \includegraphics[width=\linewidth]{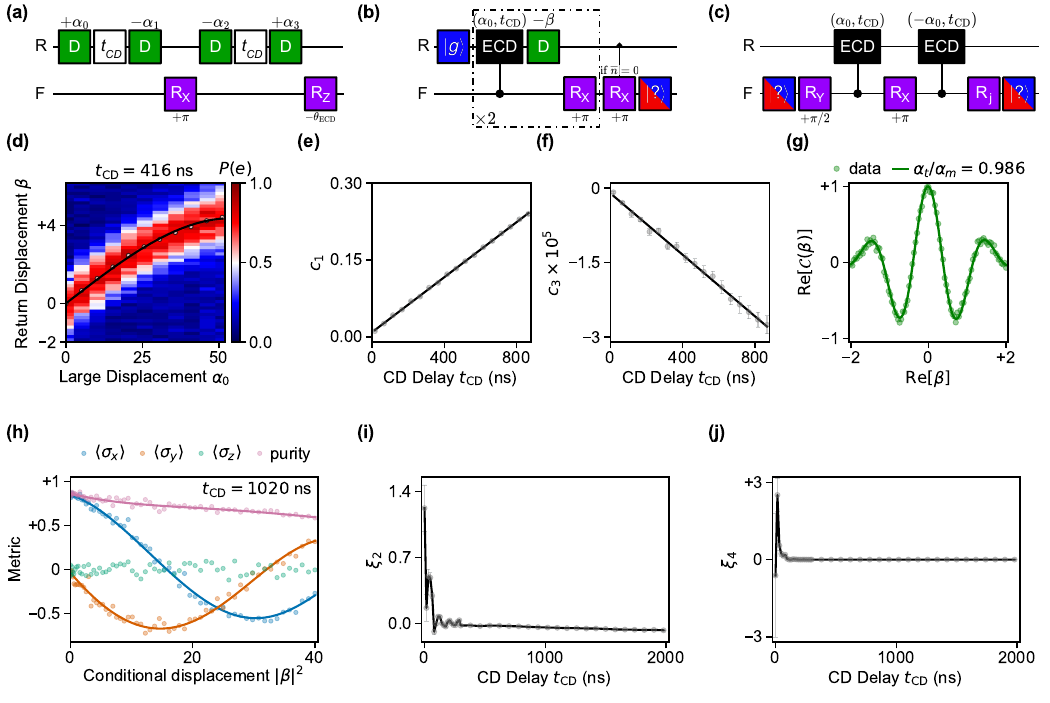}
    \caption{\textbf{Echoed conditional-displacement calibration.} \textbf{(a)} ECD pulse sequence, comprising four storage displacements, two equal dispersive-evolution intervals of duration $t_{\rm CD}$, an echoing fluxonium $\pi$ pulse, and a virtual qubit-phase correction. \textbf{(b)} Length-calibration sequence using a variable return displacement and vacuum-selective fluxonium $\pi$ pulse. \textbf{(c)} Cat-and-back sequence used to measure the ECD-induced qubit phase, with the first fluxonium measurement used to herald $\ket{g}$ or $\ket{e}$. To obtain $\ev{\op{\sigma}_z}$, $\ev{\op{\sigma}_y}$, and $\ev{\op{\sigma}_x}$, the final readout pulse is preceded by either no tomography pulse, $X/2$, or $-Y/2$, respectively. \textbf{(d)} Fluxonium excited-state probability $P(e)$ versus the large displacement $\alpha_0$ and return displacement $\beta$ at $t_{\rm CD}=\SI{416}{\nano\second}$; the black curve is a joint linear-plus-cubic fit to the centers of the Gaussian return peaks. \textbf{(e, f)} Extracted linear coefficient $c_1$ (e) and cubic coefficient $c_3$ (f) versus $t_{\rm CD}$, with linear fits. \textbf{(g)} Real-axis characteristic-function cut of a coherent state and fitted fringe, yielding the ratio of target to measured displacement scales $\alpha_t/\alpha_m$. \textbf{(h)} Cat-and-back fluxonium tomography and purity versus $|\beta|^2$ for $t_{\rm CD}=\SI{1020}{\nano\second}$; the curves are the simultaneous phase fit described in the text, while $\ev{\op{\sigma}_z}$ is shown without a fit. \textbf{(i, j)} Extracted quadratic (i) and quartic (j) ECD phase coefficients versus $t_{\rm CD}$, with smooth interpolants.}
    \label{si-fig: ecd tuneup}
\end{figure}

We implement the echoed conditional displacement (ECD) introduced in Refs.~\cite{CampagneIbarcq2020, Eickbusch2022} and used for real-time bosonic error correction in Refs.~\cite{CampagneIbarcq2020, Sivak2023, LachanceQuirion2024}. In the convention of the main text, its ideal action is
\begin{equation}
    {\rm ECD}(\beta) = D(+\beta/2)\ketbra{e}{g} + D(-\beta/2)\ketbra{g}{e}.
    \label{eq:si ecd definition}
\end{equation}
The large resonator displacement converts the weak dispersive interaction into the conditional force in Eq.~\eqref{eq: displaced dispersive Hamiltonian}, with a rate that scales as $\chi|\alpha|$. The fluxonium $\pi$ pulse exchanges the two conditional trajectories halfway through the gate, echoing the unconditional displacement and the qubit-state-dependent resonator rotation while retaining the conditional displacement. Together with arbitrary fluxonium rotations, ECD gates provide universal control of the joint fluxonium-resonator system~\cite{Eickbusch2022}.

The pulse construction used here is shown in Fig.~\ref{si-fig: ecd tuneup}(a). Each free-evolution interval has duration $t_{\rm CD}$, and $\kappa=1/T_1^a$ is the storage energy-decay rate. For a target $\beta=|\beta|e^{i\theta_\beta}$, we choose
\begin{equation}
    \alpha_0=|\alpha_0|e^{i(\theta_\beta-\pi/2)}.
\end{equation}
The four displacements in Fig.~\ref{si-fig: ecd tuneup}(a) are $D(+\alpha_0)$, $D(-\alpha_1)$, $D(-\alpha_2)$, and $D(+\alpha_3)$, with
\begin{align}
    \alpha_1=\alpha_2
    &=\frac{\alpha_0}{2}\left[
        \sqrt{1-\sin^2\left(\frac{\chi t_{\rm CD}}{2}\right)e^{-\kappa t_{\rm CD}}}
        +\cos\left(\frac{\chi t_{\rm CD}}{2}\right)e^{-\kappa t_{\rm CD}/2}
    \right], \nonumber\\
    \alpha_3
    &=-\alpha_0 e^{-\kappa t_{\rm CD}}
        +2\alpha_1e^{-\kappa t_{\rm CD}/2}
        \cos\left(\frac{\chi t_{\rm CD}}{2}\right),
    \label{eq:si ecd pulse amplitudes}
\end{align}
following the derivations in Ref.~\cite{Ali2026}. The expressions in Eq.~\eqref{eq:si ecd pulse amplitudes} include compensation factors for resonator single-photon loss during the ECD sequence while neglecting Kerr and higher-order nonlinearities.

For fixed $|\alpha_0|$, the time required for the target conditional displacement is obtained numerically~\cite{Ali2026} from
\begin{equation}
    |\beta|=4|\alpha_1|\sin\left(\frac{\chi t_{\rm CD}}{2}\right)e^{-\kappa t_{\rm CD}/2},
    \label{eq:si ecd beta analytic}
\end{equation}
with $\alpha_1$ evaluated using Eq.~\eqref{eq:si ecd pulse amplitudes}. These expressions provide the initial gate parameters; the measurements below calibrate their realized length and qubit phase.

\textbf{Conditional-displacement length.}
We first calibrate the relation between $\alpha_0$, $t_{\rm CD}$, and $\beta$ using the sequence in Fig.~\ref{si-fig: ecd tuneup}(b). After the ECD, we scan an unconditional return displacement and use a vacuum-selective fluxonium $\pi$ pulse to measure the return probability. Gaussian fits along the return-displacement axis determine $\beta$ for each $\alpha_0$, as shown in Fig.~\ref{si-fig: ecd tuneup}(d). At each $t_{\rm CD}$, the measured lengths are described by
\begin{equation}
    |\beta|=c_1(t_{\rm CD})|\alpha_0|+c_3(t_{\rm CD})|\alpha_0|^3,
    \qquad
    c_j(t_{\rm CD})=m_jt_{\rm CD}+b_j,\quad j\in\{1,3\}.
    \label{eq:si ecd length calibration}
\end{equation}
The extracted coefficients $c_1, c_3$ and their delay dependence are shown in Figs.~\ref{si-fig: ecd tuneup}(e) and~\ref{si-fig: ecd tuneup}(f), respectively. During gate compilation, Eq.~\eqref{eq:si ecd length calibration} maps a chosen $\alpha_0$ to $\beta$ while preserving the requested phase; the inverse map is obtained from the non-negative real root of the equation $c_3|\alpha_0|^3+c_1|\alpha_0|-|\beta|=0$ that is continuously connected to $|\alpha_0|=0$.

The powers in Eq.~\eqref{eq:si ecd length calibration} follow from the fluxonium-resonator Hamiltonian written in the displaced frame. The dispersive term written under the displaced-frame transformation $\op{a}\rightarrow\op{a}+\alpha$ produces a linear conditional displacement term
\begin{equation}
    \frac{\op{H}_{\chi}^{(1)}}{\hbar}
    =-\frac{\chi}{2}\op{\sigma}_z
    \left(\alpha\op{a}^\dagger+\alpha^*\op{a}\right),
\end{equation}
and hence a contribution proportional to $\alpha$. The fluxonium-state-dependent storage Kerr contains the differential term
\begin{equation}
    \frac{\op{H}_{K,z}}{\hbar}
    =\frac{K_g-K_e}{2}\op{\sigma}_z\op{a}^\dagger{}^2\op{a}^2.
\end{equation}
Under $\op{a}\rightarrow\op{a}+\alpha$, the above term also produces a conditional displacement:
\begin{equation}
    \frac{\op{H}_{K,z}^{(1)}}{\hbar}
    =(K_g-K_e)|\alpha|^2\op{\sigma}_z
    \left(\alpha\op{a}^\dagger+\alpha^*\op{a}\right),
\end{equation}
which scales as $\alpha|\alpha|^2$ and motivates the leading cubic correction.

\textbf{Displacement-scale refinement.}
We next refine the unconditional-displacement scale by preparing a coherent state of target amplitude $i\alpha_t$ and measuring a real-axis cut of its characteristic function, as in Fig.~\ref{si-fig: ecd tuneup}(g). For a displaced thermal state
\begin{equation}
    \rho=D(i\alpha_t)\rho_{\rm th}D^\dagger(i\alpha_t),
\end{equation}
the cut at real $\beta=x$ is
\begin{equation}
    {\rm Re}[\mathcal{C}(x)]
    =e^{-(\bar{n}_{\rm th}+1/2)x^2}\cos(2\alpha_t x).
    \label{eq:si coherent cf cut}
\end{equation}
We fit the data to $A\exp[-x^2/(2\sigma^2)]\cos(\omega x+\phi)+b$ and define the measured scale by $\alpha_m=\omega/2$. Because the same displacement calibration sets both the prepared coherent-state amplitude and the characteristic-function sampling coordinate, their common scale error enters the fringe frequency twice. If $r_\alpha$ is the conversion from drive voltage to displacement amplitude, we therefore update it as $r_\alpha^{\rm new}=r_\alpha^{\rm old}\sqrt{\alpha_m/\alpha_t}$. Thermal occupation changes the Gaussian envelope in Eq.~\eqref{eq:si coherent cf cut}, but not its fringe frequency, making this calibration insensitive to whether the resonator begins in vacuum or a thermal state. In contrast, a vacuum Q-function calibration depends on both the resonator thermal population and the selectivity of the vacuum-selective fluxonium $\pi$ pulse.

\textbf{ECD qubit phase.}
The photon-number-dependent fluxonium frequency produces a residual qubit phase during an ECD. We measure it with the cat-and-back sequence in Fig.~\ref{si-fig: ecd tuneup}(c), in which opposite ECDs recombine the resonator trajectories while their qubit phases add~\cite{Eickbusch2022,Sivak2023}. The first measurement heralds the fluxonium in $\ket{g}$ or $\ket{e}$; for each tomography axis $i$, we analyze the half-difference $\left[\ev{\op{\sigma}_i}_g-\ev{\op{\sigma}_i}_e\right]/2$ between the two heralded datasets. We parameterize the phase of one ECD as
\begin{equation}
    \theta_{\rm ECD}(\beta,t_{\rm CD})
    =\xi_2(t_{\rm CD})|\beta|^2+\xi_4(t_{\rm CD})|\beta|^4.
    \label{eq:si ecd phase}
\end{equation}
The dispersive ac Stark term is proportional to $\chi|\alpha(t)|^2\op{\sigma}_z$; integrating it along a trajectory whose scale is proportional to $|\beta|$ gives the quadratic term. The differential Kerr shift is proportional to $(K_g-K_e)|\alpha(t)|^4\op{\sigma}_z$ and gives the leading quartic correction.

Ideally, the two ECDs in the cat-and-back sequence yield
\begin{equation}
    \ev{\op{\sigma}_x}=\cos\left[2\theta_{\rm ECD}(\beta,t_{\rm CD})\right],\qquad
    \ev{\op{\sigma}_y}=\sin\left[2\theta_{\rm ECD}(\beta,t_{\rm CD})\right],
\end{equation}
independent of the initial resonator state. To account for contrast loss, we fit the transverse Bloch vector to
\begin{equation}
    \begin{pmatrix}
        \ev{\op{\sigma}_x}\\
        \ev{\op{\sigma}_y}
    \end{pmatrix}
    =
    \left(1-p[\beta]\right)
    \begin{pmatrix}
        \cos\left[2\theta_{\rm ECD}(\beta,t_{\rm CD})\right]\\
        \sin\left[2\theta_{\rm ECD}(\beta,t_{\rm CD})\right]
    \end{pmatrix}.
    \label{eq:si cat and back fit}
\end{equation}
Following Ref.~\cite{Sivak2023}, $\sqrt{1-p[\beta]}$ is the uniform contraction of the qubit Bloch vector per ECD gate. We model the displacement dependence as $p[\beta]=\eta_0+\eta_2|\beta|^2+\eta_4|\beta|^4+\eta_6|\beta|^6$. The constant term $\eta_0$ captures $\beta$-independent contrast loss, including state-preparation and measurement errors, qubit decoherence, and displacement-independent ECD errors. The photon-shot-noise dephasing described in Sec.~\ref{sec: Resonator Spectroscopy and Relaxation} gives $\Gamma_\phi\approx|\alpha|^2/T_1^a$; because $\alpha\propto\beta$ at fixed $t_{\rm CD}$, this produces the leading $|\beta|^2$ dependence. More generally, the $n$th-order qubit-state-dependent frequency shift scales as $|\alpha|^{2n}\propto|\beta|^{2n}$, with $n=1$ corresponding to the dispersive shift, $n=2$ to the qubit-state-dependent Kerr, and $n>2$ to higher-order nonlinearities. These shifts can produce the corresponding even-power contributions to the dephasing. Assuming the contractions from the two ECD gates are uncorrelated and independent of phase-space direction, their product gives the factor $1-p[\beta]$ in Eq.~\eqref{eq:si cat and back fit}. For each $t_{\rm CD}$ dataset, we independently fit the coefficients $\eta_0$, $\eta_2$, $\eta_4$, and $\eta_6$, together with $\xi_2$, $\xi_4$, and an additive offset $b$ shared by the $\ev{\op{\sigma}_x}$ and $\ev{\op{\sigma}_y}$ quadratures.
An example fit is shown in Fig.~\ref{si-fig: ecd tuneup}(h); $\ev{\op{\sigma}_z}$ is measured as a consistency check but is not included in the fit. Repeating the measurement versus $t_{\rm CD}$ gives the lookup tables for $\xi_2$ and $\xi_4$ shown in Figs.~\ref{si-fig: ecd tuneup}(i) and~\ref{si-fig: ecd tuneup}(j). After every ECD, we apply the virtual correction $R_Z[-\theta_{\rm ECD}(\beta,t_{\rm CD})]$ shown in Fig.~\ref{si-fig: ecd tuneup}(a).

\subsection{Fock Qubit Lifetimes\label{sec: Fock Qubit Lifetimes}}

\begin{figure}[h]
    \centering
    \includegraphics[width=\linewidth]{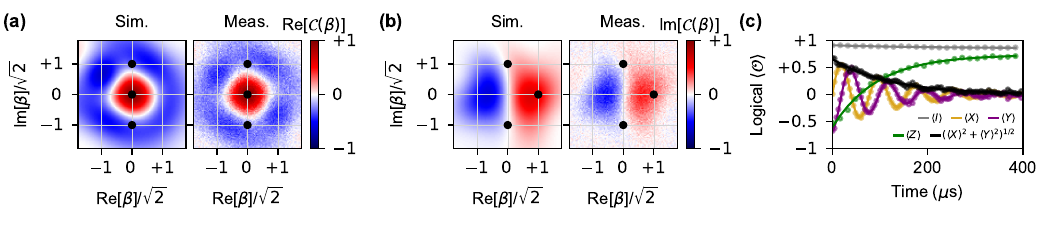}
    \caption{\textbf{Fock-qubit state preparation and lifetimes.} \textbf{(a, b)} Simulated (left) and measured (right) characteristic functions of the optimized $\ket{-z_L}=\ket{1}$ (a) and $\ket{-y_L}=(\ket{0}-i\ket{1})/\sqrt{2}$ (b) states, showing ${\rm Re}[\mathcal{C}(\beta)]$ in (a) and ${\rm Im}[\mathcal{C}(\beta)]$ in (b). Both optimized states have simulated fidelities exceeding $\SI{98}{\percent}$ with their respective target states. Black markers indicate the displacements used to extract the logical Pauli expectation values. \textbf{(c)} Logical expectation values $\ev{\op{I}_L}$, $\ev{\op{X}_L}$, $\ev{\op{Y}_L}$, and $\ev{\op{Z}_L}$ versus storage delay. Exponential fits to $\ev{\op{Z}_L}$ and the transverse magnitude $\sqrt{\ev{\op{X}_L}^2+\ev{\op{Y}_L}^2}$ give $T_Z=\SI{94.2\pm4.0}{\micro\second}$ and $T_{XY}=\SI{123.9\pm5.1}{\micro\second}$, respectively.}
    \label{si-fig: fock lifetimes}
\end{figure}

The Fock qubit is encoded in the two lowest Fock states of the storage resonator, with the code space defined by $\ket{+z_L} = \ket{0}$ and $\ket{-z_L} = \ket{1}$. Using the optimal-control techniques (with $N=4$ blocks and a regularization strength of $\lambda=0.5$) described in Sec.~\ref{sec: gkp state} and Sec.~\ref{sec: offline control compilation}, we prepare the storage resonator in a Fock-qubit cardinal state with the fluxonium in its ground state, and then measure the Fock-qubit logical Pauli expectation values after a variable delay. The values $\ev{\op{X}_L}$ and $\ev{\op{Y}_L}$ are measured after preparing $\ket{-y_L}=(\ket{0}-i\ket{1})/\sqrt{2}$, while $\ev{\op{Z}_L}$ and $\ev{\op{I}_L}$ are measured after preparing $\ket{-z_L}$. The simulated and measured characteristic functions of the optimized $\ket{-z_L}$ and $\ket{-y_L}$ states are shown in Fig.~\ref{si-fig: fock lifetimes}(a,b). The Fock-qubit logical operators can be decomposed into resonator displacements~\cite{Ding2025}:

\begin{align*}
    \op{I}_L &= P D(0) P,\\
    \op{X}_L &= \frac{e}{2i\sqrt{2}} P\left[D(i\sqrt{2}) - D(-i\sqrt{2})\right]P,\\
    \op{Y}_L &= \frac{ie}{\sqrt{2}}P\left[D(\sqrt{2}) - \frac{1}{2}D(i\sqrt{2}) - \frac{1}{2}D(-i\sqrt{2})\right]P,\\
    \op{Z}_L &= \frac{e}{2} P\left[D(i\sqrt{2}) + D(-i\sqrt{2}) \right]P.
\end{align*}
Here, $P=\ketbra{0}{0}+\ketbra{1}{1}$ projects onto the Fock-qubit code space. We assume that the resonator state is in this code space, such that the Fock qubit logical operator expectation values can be measured directly through characteristic-function measurements, $\mathcal{C}(\beta)=\ev{D(\beta)}$. We therefore measure the logical Paulis by sampling the characteristic function at the displacements given above [also marked on the characteristic functions in Fig.~\ref{si-fig: fock lifetimes}(a,b)]. With the fluxonium initialized in $\ket{g}$, the resonator rotates at a rate $-\chi/2$ [see Eq.~\eqref{eq: dispersive system hamiltonian}]. This rotation produces the oscillations in $\ev{\op{X}_L}(t)$ and $\ev{\op{Y}_L}(t)$ shown in Fig.~\ref{si-fig: fock lifetimes}(c).

As expected, the Fock-qubit logical $Z$ lifetime $T_Z$ is approximately equal to the single-photon resonator lifetime $T_1^a$. Unlike in a transmon, where relaxation is typically much faster than thermal excitation, the low transition frequency of a half-flux-biased fluxonium is comparable to the relevant thermal energy scale of the device, resulting in similar upward and downward transition rates. The Fock-qubit logical $X$ and $Y$ lifetimes, $T_X$ and $T_Y$, are therefore limited by the fluxonium bit-flip lifetime $T_1^q$ despite initializing the fluxonium in its ground state. A Lindblad master equation simulation using the central device parameters in Table~\ref{tab: summary of parameters} predicts $T_Z=\SI{103}{\micro\second}$ and $T_{XY}=\SI{145}{\micro\second}$, compared with the measured values of $T_Z=\SI{94.2\pm4.0}{\micro\second}$ and $T_{XY}=\SI{123.9\pm5.1}{\micro\second}$. The tabulated parameters used in this simulation were measured during device characterization and, most likely, differ from the coherences at the time of the Fock-qubit lifetime measurement.

\section{Floquet Modeling of the Fluxonium-Resonator System}
\subsection{DUST Simulations of Multi-Photon Resonances\label{sec: DUST Simulations of the Fluxonium-Resonator System}}

It has been shown experimentally and theoretically that applying strong drives to a circuit QED system can result in drive-induced unwanted state transitions (DUST) arising from the interplay between the drive and the nonlinearities of the system~\cite{Shillito2022, Xiao2023, Dumas2024, floquet2024, Fechant2025, Dai2026, Wang2026, Sank2016, Khezri2023}. The same holds true in our fluxonium-resonator device when populating the storage resonator with a large coherent state $\ket{\alpha_0}$ during an ECD gate. As a result, the intermediate photon number $\bar{n} = |\alpha_0|^2$ used in our ECD gates must be chosen to be well below the onset of DUST. Experimentally, we observe signatures of DUST in out-and-back measurements [Fig.~\ref{si-fig: out and back}(e)] with a feature at $|\alpha_0| \approx 45$ that we attribute to a multi-photon nonlinear resonance between the fluxonium states $\ket{e}$ and $\ket{h}$. We confirm this through numerical simulations of the driven fluxonium via Floquet branch analysis, as described below.

As before, we model the capacitively coupled resonator and fluxonium at half flux using the Hamiltonian
\begin{equation}
    \op{H}_{\rm FR}(t)
    = 4E_C^q \op{n}^2 + \frac{1}{2}E_L^q\op{\varphi}^2 + E_J^q\cos(\op{\varphi})
    + \hbar\omega_a \op{a}^\dagger\op{a}
    + i\hbar g\frac{\op n}{n_{\rm zpf}}\left(\op{a}^\dagger-\op a\right)
    + i\epsilon(t)\left(\op{a}^\dagger-\op a\right)\sin(\omega_d t).
\end{equation}
The circuit operators and parameters follow Eq.~\eqref{eq:fluxonium-resonator-hamiltonian}; $\epsilon(t)$ is the envelope of the storage drive and $\omega_d$ is its carrier frequency. The drive populates the storage to a coherent amplitude $\alpha_0$ and is applied near resonance, $\omega_d\approx\omega_a$. It has been shown in several prior works that the driven resonator dynamics can be accurately modeled in a semiclassical picture by moving to a displaced frame of the resonator, and then tracing out the resonator degree of freedom~\cite{Shillito2022, Dumas2024}. Here, since we go out to many thousands of photons, the use of a semiclassical description considerably reduces the computational resources required to perform simulations. The resulting Hamiltonian is that of a charge-driven fluxonium,
\begin{equation}
    \op{H}_{\rm F}(t;\bar n)
    = \underbrace{4E_C^q \op{n}^2 + \frac{1}{2}E_L^q\op{\varphi}^2 + E_J^q\cos(\op{\varphi})}_{\op{H}_{{\rm F}}^0}
    \,+\, \hbar\Omega_d(\bar n)\frac{\op n}{n_{\rm zpf}}\cos(\omega_d t),
    \label{eq:DUST-floquet-hamiltonian}
\end{equation}
where $\op{H}_{{\rm F}}^0 \coloneqq \op{H}_{\rm F}(t; \bar{n} = 0)$ is the undriven half-flux fluxonium Hamiltonian and the charge-drive amplitude is $\Omega_d(\bar n)=2g\sqrt{\bar{n}}$, where the storage resonator contains $\bar{n} = |\alpha_0|^2$ photons on average when driven to a coherent state of amplitude $\alpha_0$. Since Eq.~\eqref{eq:DUST-floquet-hamiltonian} is time-periodic with period $T=2\pi/\omega_d$, Floquet's theorem gives the one-period propagator $\op{U}(T, 0)$ as
\begin{equation}
    \op U(T,0)=\mathcal T\exp\left[-\frac{i}{\hbar}\int_0^T\op{H}_{\rm F}(t;\bar n)dt\right],
    \quad \text{with}\,\,\,
    \op U(T,0)\ket{\Phi_s}
    =e^{-i\epsilon_s T/\hbar}\ket{\Phi_s} \, \Longleftrightarrow\, \Big[\op{H}_{\rm F}(t;\bar n) - i\hbar\frac{d}{dt}\Big]\ket{\Phi_s} = \epsilon_s \ket{\Phi_s}.
\end{equation}
Here $\ket{\Phi_s}=\ket{\Phi_s(t)}$ is the $T$-periodic Floquet mode indexed by $s$, $\epsilon_s$ is its quasienergy, and $\mathcal T$ denotes time ordering. By diagonalizing the propagator at each value of the drive amplitude $\bar{n}$ and tracking the Floquet modes $\ket{\Phi_s(t=0)}$, we can perform branch analysis following Refs.~\cite{Xiao2023, Dumas2024, floquet2024}. We compute the average excitation $n_f\coloneqq\ev{\op{N}_f}$ versus $|\alpha_0|$ in Fig.~\ref{si-fig: out and back}(c), where $\op{N}_f\coloneqq\sum_\ell \ell\ketbra{\ell}{\ell}$ counts excitations in the ordered eigenbasis $\{\ket\ell\}$ of $\op{H}_{\rm F}^0$. We observe a clear branch swap indicative of a multi-photon resonance between the fluxonium states $\ket{e}$ (with $n_f = 1$ initially) and $\ket{h}$ (with $n_f = 3$ initially) occurring at $|\alpha_0| \approx 45$ and matching the measured out-and-back data from Figs.~\ref{fig: conditional displacement demo}(a) and~\ref{si-fig: out and back}(e). Furthermore, in Fig.~\ref{si-fig: out and back}(d), we compute a ``scar plot'' at varying values of $\omega_d$ showing the hybridization indicator $P_{s\rightarrow}\coloneqq1-|\mathcal O_s|^2$ for the ground- and excited-state Floquet modes ($s\in\{g,e\}$), where $\mathcal O_s$ is the overlap between the tracked Floquet mode and its corresponding ideal displaced state~\cite{Xiao2023}. We observe no large features in $P_{g\rightarrow}$, while the large feature in $P_{e\rightarrow}$ tracks the Stark-shifted $\ket{e}\leftrightarrow \ket{h}$ transition. These results indicate that the onset of DUST is highly dependent on the storage and fluxonium parameters; specifically, the storage frequency must be chosen judiciously in relation to the fluxonium transitions to prevent multi-photon resonances from occurring at low photon numbers. While the Floquet simulations in Figs.~\ref{si-fig: out and back}(c, d) were used here to diagnose DUST in the present device, this type of analysis therefore also serves as a useful benchmark when designing devices. In addition to achieving a target dispersive coupling $\chi$ and suppressing the Kerr nonlinearities $K_{g/e}$, candidate fluxonium and storage parameters can be screened for drive-induced resonances. This allows us to target designs that push the onset of DUST beyond the photon-number range required for fast ECD control.

\subsection{Efficient Numerical Modeling of Out-and-Back Measurements\label{sec: Efficient Numerical Modeling of Out-and-Back Measurements}}

The nonlinear storage rotation measured using out-and-back [Fig.~\ref{si-fig: out and back}(e)] can be predicted using a Floquet extended-Hilbert-space \cite{Shirley1965, Sambe1973, Grifoni1998, Drese1999, Eckardt2015, Rudner2020} calculation that tracks the Floquet modes connected to the fluxonium states $\ket g$, $\ket e$, and $\ket h$ at zero drive, while ignoring the $\ket f$ branch because it does not swap with either the $\ket g$ or $\ket e$ branch. We treat the resonator semiclassically while retaining its nonperturbative backaction through the driven-fluxonium quasienergies. Using the notation of Eq.~\eqref{eq:DUST-floquet-hamiltonian}, the truncated extended-space Floquet Hamiltonian is
\begin{align}
    \op{\mathcal H}_{\rm floquet}(\bar n)={}&\sum_{m=-M}^{M}\left(\op{H}_{{\rm F}}^0+m\hbar\omega_d\right)\otimes\ketbra{m}{m}\nonumber\\
    &+\frac{\hbar\Omega_d(\bar n)}{2}\frac{\op n}{n_{\rm zpf}}\otimes\sum_{m=-M}^{M-1}\left(\ketbra{m+1}{m}+\ketbra{m}{m+1}\right),
    \label{eq:out-and-back-Floquet-hamiltonian}
\end{align}
where the Floquet photon-number index $m$ labels the $m$th Fourier harmonic of the drive and $M$ is the corresponding numerical truncation cutoff; we use $M = 9$ for the displayed pulse-integrated curves. Diagonalizing Eq.~\eqref{eq:out-and-back-Floquet-hamiltonian} at each successive value of $\bar{n}$ gives extended Floquet eigenstates $\floqket{\tilde{\Phi}_{s, m}(\bar{n})}$ with energies $\tilde{E}_{s, m}(\bar{n})$ indexed by the two quantum numbers $s$ (for the fluxonium) and $m$ (for the drive) \cite{shoumik-practical-floquet}. We track the Floquet eigenstates that connect to the zero-drive state $\floqket{\tilde{\Phi}_{s, 0}(\bar{n} = 0)} = \ket{s}\otimes\ket{m=0}$, for $s\in\{g,e,h\}$, and jointly assign labels by maximizing overlaps between the eigenstates at successive values of $\bar{n}$. Following the notation in the preceding section, the associated quasienergies and time-periodic Floquet modes are denoted $\epsilon_s(\bar n) \equiv \tilde{E}_{s, m}(\bar{n})\!\mod \hbar\omega_d$ and $\ket{\Phi_s(t;\bar n)}$, respectively; furthermore, we define $\tilde{E}_s(\bar{n}) \equiv \tilde{E}_{s, 0}(\bar{n})$ as the dressed energy for state $s$ after choosing the $m = 0$ Floquet replica (which we do in all calculations below). Armed with this notation, we can calculate the effective fluxonium-state-dependent resonator rotation rate as
\begin{equation}
    \frac{\Delta_s(\alpha)}{2\pi}=\frac{\Delta_a}{2\pi}+\frac{\partial[\tilde{E}_s(\bar n)/h]}{\partial\bar n},
    \qquad\text{with}\,\, \bar n=|\alpha|^2.
    \label{eq:floquet-storage-rotation}
\end{equation}
Intuitively, this fluxonium-state-dependent resonator rotation reflects the relative phase accumulated between neighboring resonator photon-number components. In the full fluxonium-resonator Hamiltonian [Eq.~\eqref{eq:fluxonium-resonator-hamiltonian}, non-Floquet], this phase evolves at a rate $(E_{n+1, s}-E_{n, s})/\hbar$ for fluxonium-state branch $s$ \cite{Khezri2016}. In the semiclassical, large-$\bar n$ limit, this finite difference becomes the $\bar{n}$ derivative in Eq.~\eqref{eq:floquet-storage-rotation}. Applying this derivative to the Hamiltonian in Eq.~\eqref{eq: dispersive system hamiltonian} with truncated nonlinearity gives expressions for $\Delta_g$ and $\Delta_e$ that are linear in $\bar n$ [Eq.~\eqref{eq: fluxonium-state-dependent resonator rotation}], while applying it to the full tracked Floquet energy retains contributions from all higher-order nonlinearities. The derivative above is evaluated using the Floquet Hellmann--Feynman theorem,
\begin{equation}
    \frac{\partial\tilde{E}_s(\bar n)}{\partial\bar n}
    =\frac{1}{T}\int_0^T\mel{\Phi_s(t;\bar n)}{\frac{\partial\op{H}_{\rm F}(t;\bar n)}{\partial\bar n}}{\Phi_s(t;\bar n)}dt = \floqmel{\tilde{\Phi}_s}{\frac{\partial\op{\mathcal H}_{\rm floquet}(\bar n)}{\partial\bar n}}{\tilde{\Phi}_s} = \frac{\hbar g}{n_{\rm zpf}\sqrt{\bar{n}}} \floqmel{\tilde{\Phi}_s}{[\hat{n}\otimes \cos(\hat{\vartheta})]}{\tilde{\Phi}_s},
    \label{eq:floquet-hellmann-feynman}
\end{equation}
where $\cos(\hat{\vartheta}) \equiv \frac{1}{2}\sum_m\left(\ketbra{m+1}{m}+\ketbra{m}{m+1}\right)$ and the last equality comes from the relation $\Omega_d(\bar{n}) = 2g\sqrt{\bar{n}}$ introduced earlier. We are thus able to compute the derivative as a static matrix element using the extended Floquet eigenstates. The expression is evaluated directly for $\bar n>0$, while its value at $\bar n=0$ is obtained by smooth low-amplitude extrapolation. Unlike Eq.~\eqref{eq: fluxonium-state-dependent resonator rotation}, the Floquet construction of $\Delta_s(\alpha)$ in Eq.~\eqref{eq:floquet-storage-rotation} retains the full nonperturbative dependence on $\alpha$ and does not truncate the response at Kerr or any fixed higher-order nonlinearity.

The measured out-and-back protocol is modeled by propagating the fluxonium-state-dependent coherent storage amplitude $\alpha_s(t)$ in the frame rotating at $\omega_d$ according to
\begin{equation}
    \dot\alpha_s(t)=\left[-\frac{1}{2T_1^a}+i\Delta_s\!\verythinspace\big(|\alpha_s(t)|\big)\right]\alpha_s(t)+\varepsilon(t).
    \label{eq:floquet-out-and-back-evolution}
\end{equation}
Here $\varepsilon(t)$ is the complex storage-drive envelope in the rotating frame and $T_1^a$ is the measured storage energy-relaxation time. The experimental sequence consists of a $\SI{208}{\nano\second}$ cosine-envelope displacement to the target amplitude $\alpha_0$, an out-and-back delay time of $\SI{6.000}{\micro\second}$, and a return displacement scaled to compensate for photon loss during this delay time. We integrate Eq.~\eqref{eq:floquet-out-and-back-evolution} over the full pulse envelope. For each $\alpha_0$ and fluxonium state branch $s$, the predicted return phase minimizes the final residual photon number,
\begin{equation}
    \phi_s^{\rm pred}(\alpha_0)=\underset{\phi}{\operatorname{argmin}}\,
    \left|\alpha_s\!\left(t_f;\alpha_0,\phi\right)\right|^2.
    \label{eq:floquet-return-phase}
\end{equation}
Here $\phi$ is the phase of the return displacement and $t_f$ is the time at the end of the out-and-back sequence.

As in experiment, the optimal return phase $\phi_s^{\rm pred}(\alpha_0)$ is divided by the out-and-back delay time $\SI{6.000}{\micro\second}$ to obtain the fluxonium-state-dependent rotation rate $\Delta_s^{\rm avg}(\alpha_0)$ plotted in Fig.~\ref{si-fig: out and back}(e). Although $\phi_s^{\rm pred}$ includes the phase accumulated during the displacement pulses, $\Delta_s^{\rm avg}$ is calculated using just the out-and-back delay time rather than also including the displacement pulse times to match what is done in the experimental data analysis.

The black crosses in Fig.~\ref{si-fig: out and back}(e) mark peaks corresponding to the experimental optimal return phases $\phi_{s,i}^{\rm peak}$ at six amplitudes satisfying $200<\bar n_i<1200$ for each $s\in\{g,e\}$, with all peaks appearing before the $\ket e$--$\ket h$ DUST feature. Dividing these phases by the $\SI{6.000}{\micro\second}$ out-and-back delay time gives $\Delta_{s,i}^{\rm plot}$. The detuning between the simulation and experimental rotating frames is then calibrated by comparing $\Delta_s(\alpha_{0,i})$ to $\Delta_{s,i}^{\rm plot}$,
\begin{equation}
    \Delta_{\rm frame}
    =\underset{\delta}{\operatorname{argmin}}
    \sum_{s\in\{g,e\}}\sum_{i=1}^{6}
    \left[\Delta_{s,i}^{\rm plot}-\left(\Delta_s(\alpha_{0,i})+\delta\right)\right]^2.
\end{equation}
This gives $\Delta_{\rm frame}/2\pi=\SI{0.291}{\kilo\hertz}$, corresponding to $\SI{0.629}{\degree}$ over the $\SI{6.000}{\micro\second}$ out-and-back delay time. The $\Delta_{\rm frame}$ offset is then applied to the simulation curves plotted in Fig.~\ref{si-fig: out and back}(e). Since $\Delta_{\rm frame}$ shifts the $\ket{g}$, $\ket{e}$, and $\ket{h}$ curves equally and independently of $\alpha_0$, it cannot alter $\Delta_g-\Delta_e$ or produce its amplitude-dependent flattening. Notably, $\Delta_{\rm frame}$ is the only fitted parameter in the comparison between the simulated and measured curves. As such, the excellent match between the theory and experiment shown in Fig.~\ref{si-fig: out and back}(e) supports the completeness and validity of our theory.

In these simulations, we retain 25 fluxonium levels and $2M+1=19$ Floquet harmonics, giving a 475-dimensional Floquet extended-space Hamiltonian matrix that is diagonalized across 1000+ amplitude points over the simulated range $0\leq\alpha_0\leq200$. The overlay curves in Fig.~\ref{si-fig: out and back}(e) are generated with JAXQuantum~\cite{jha2024jaxquantum}. The full amplitude sweep---including diagonalization, branch tracking, and the evaluation of the relevant matrix elements---takes $\SI{104.85}{\second}$ on an Intel i9-14900KF CPU and $\SI{13.52}{\second}$ on an NVIDIA GeForce RTX 4080 SUPER GPU, a $7.75$-fold speedup. The CPU and GPU rotation curves are numerically indistinguishable at the plotted precision.

The Floquet rotation curves also yield the dispersive and nonlinear coefficients, which can be compared to the results of low-photon-number numerical diagonalization of the composite fluxonium-resonator system. Writing $\bar n=|\alpha|^2$ and $K_s\coloneqq K_s^{(2)}$, we fit, for $s\in\{g,e\}$,
\begin{equation}
\begin{aligned}
    \frac{\Delta_s(\alpha)}{2\pi}
    &=\frac{\Delta_s(0)}{2\pi}
    +\sum_{m\geq2}m\frac{K_s^{(m)}}{2\pi}\bar n^{m-1},\\
    \frac{\chi}{2\pi}
    &=\frac{\Delta_e(0)-\Delta_g(0)}{2\pi},\qquad
    \frac{K_s}{2\pi}
    =\frac{1}{2}\left.\frac{\partial[\Delta_s/(2\pi)]}{\partial\bar n}\right|_{\bar n=0}.
\end{aligned}
\end{equation}
The Floquet calculation [in Eq.~\eqref{eq:floquet-storage-rotation}] gives $\chi/2\pi=\SI{23.00159}{\kilo\hertz}$, $K_g/2\pi=\SI{2.41237}{\hertz}$, and $K_e/2\pi=\SI{-7.18617}{\milli\hertz}$. Meanwhile, numerical diagonalization via Eqs.~(\ref{eq:fluxonium-resonator-hamiltonian}--\ref{eq: full numerical eigenenergy finite differences higher kerr}) gives $\SI{23.00000}{\kilo\hertz}$, $\SI{2.41108}{\hertz}$, and $\SI{-7.18626}{\milli\hertz}$, respectively, showing excellent agreement. This approach can also be extended to extract higher-order nonlinearities $K_s^{(m)}$ for $m > 2$, providing an accurate and efficient alternative to modeling nonlinearities via the diagonalization of increasingly large matrices representing the composite fluxonium-resonator system.

The naive approach to computing the high-photon-number nonlinearities up to $\bar n\gtrsim 40,000$ would require tens of thousands of resonator Fock states in the numerical representation of the Hamiltonian in Eq.~\eqref{eq:fluxonium-resonator-hamiltonian}. Even taking a cutoff of 40,001 resonator states, combined with 25 fluxonium levels, would result in a composite fluxonium-resonator Hilbert space dimension of approximately $10^6$: in complex double precision, this dense Hamiltonian and its full eigenvector matrix would require $\approx \SI{16.0}{\tera\byte}$ each. Since dense eigendecomposition scales cubically with dimension, obtaining the full eigensystem would also require on the order of $10^{18}$ floating-point operations and remain prohibitively slow even if sufficient memory were available. In contrast, the 475-dimensional Floquet Hamiltonian occupies $\SI{3.6}{\mega\byte}$, or $\SI{7.2}{\mega\byte}$ together with its full eigenvector matrix, per amplitude point. These memory savings arise because the semiclassical Floquet calculation eliminates the explicit resonator Fock basis while still efficiently capturing the high-photon-number bending, saturation, and branch exchange shown in Fig.~\ref{si-fig: out and back}(e). 

For an alternative way to efficiently model out-and-back measurements using iterative displaced-frame tracking, we encourage the reader to consult Ref.~\cite{Ali2026}.

\section{Offline control compilation \label{sec: offline control compilation}}

\subsection{JAXQuantum State-Preparation Optimization}

Offline control compilation is implemented in JAXQuantum~\cite{jha2024jaxquantum}, an open-source, JAX-native toolkit for quantum hardware design, simulation, and control. Its core provides a QuTiP-like~\cite{johansson2012qutip} interface built on the \texttt{Qarray} objects, which are batched quantum states and operators, representable as dense, sparse, or otherwise custom data formats. Higher-level modules build on this common representation: \texttt{jaxquantum.devices} supports coupled systems of superconducting-circuit devices, including Floquet treatments of periodic drives; \texttt{jaxquantum.codes} constructs cat, binomial, and GKP codewords, logical operators, and phase-space visualizations; and \texttt{jaxquantum.circuits} composes gates into hierarchical registers and layers supporting unitary, Hamiltonian, Lindbladian, and Kraus evolution. Because these modules are all built using the JAX library~\cite{jax2018github}, they can be composed together to create simulations compatible with automatic differentiation, vectorization, just-in-time (JIT) compilation, and hardware acceleration on GPUs and TPUs.

For state preparation, the optimizer targets a particular resonator state with a sequence of native single-qubit rotations and conditional displacements applied in the joint fluxonium-resonator Hilbert space on a given initial state. Following Ref.~\cite{Eickbusch2022}, the rotation-conditional-displacement (RCD) structure used for GKP and Fock-state preparation repeatedly applies $R_x(\theta_{x,n})$, $R_y(\theta_{y,n})$, ${\rm CD}(\beta_n)$, and $R_x(\pi)$ in sequence. In the JAXQuantum convention, the successive ${\rm CD}(\beta_n)$ and $R_x(\pi)$ operations implement the ECD gate defined in the main text. The essential circuit construction and optimization code is written at the end of this section.

The parameter array has shape $(B,4,N)$ for $B$ optimization seeds and $N$ RCD blocks. Its four rows contain $\theta_x/(2\pi)$, $\theta_y/(2\pi)$, $\operatorname{Re}\beta$, and $\operatorname{Im}\beta$. Each trajectory begins from an independently sampled parameter set. At every optimization step, \texttt{jax.vmap} simulates all $B$ circuits and evaluates their target-state fidelities in parallel. The loss averages the logarithmic infidelity and the conditional-displacement penalty in Sec.~\ref{sec: gkp state}, while \texttt{jax.value\_and\_grad} propagates derivatives backward through the entire circuit with respect to every rotation and displacement parameter. An adaptive gradient optimizer updates all candidates simultaneously. The multiple seeds reduce sensitivity to local minima, and the displacement penalty favors solutions with smaller, experimentally accessible ECD amplitudes. After optimization, the minimum-loss solution is reevaluated with a larger oscillator truncation. Changing only the target state allows the same workflow to prepare GKP, Fock, or squeezed states starting from vacuum.

This calculation is particularly well suited to JAX and GPU execution. The circuit topology is identical across seeds and epochs, so \texttt{jax.jit} compiles the complete forward simulation and reverse-mode gradient calculation once and reuses the resulting executable throughout the optimization. Vectorization converts the independent trajectories into large batched linear algebra operations, exposing substantially more parallel work than a single state-preparation trajectory, while automatic differentiation avoids a separate finite-difference simulation for every control parameter. For the $B=500$, $N=7$ GKP compilation in Fig.~\ref{fig: gkp state prep}, the same program runs in minutes on an NVIDIA H200 GPU and yields the $165\times$ speedup over a 10th-generation Intel i7 CPU reported in the main text. Moreover, this workflow can be readily adapted to use other control primitives to target states and unitary gates. With the addition of dissipation resources such as measurement, even non-unitary quantum channels can be realized using these optimization methods. 
\\
\definecolor{jqcodebackground}{HTML}{F7F7F7}
\definecolor{jqcodekeyword}{HTML}{1F4E79}
\definecolor{jqcodecomment}{HTML}{3C763D}
\definecolor{jqcodestring}{HTML}{A31515}
\lstdefinestyle{jaxquantum}{
    language=Python,
    basicstyle=\ttfamily\scriptsize,
    keywordstyle=\color{jqcodekeyword}\bfseries,
    commentstyle=\color{jqcodecomment},
    stringstyle=\color{jqcodestring},
    backgroundcolor=\color{jqcodebackground},
    frame=single,
    rulecolor=\color{black!20},
    framesep=4pt,
    numbers=left,
    numberstyle=\tiny\color{black!45},
    numbersep=6pt,
    showstringspaces=false,
    breaklines=true,
    columns=fullflexible,
    keepspaces=true
}

\begin{lstlisting}[style=jaxquantum]
import jax
import jax.numpy as jnp
import jaxquantum as jqt
import jaxquantum.circuits as jqtc

def run_circuit(params, n_osc):
    theta_x = 2 * jnp.pi * params[0]
    theta_y = 2 * jnp.pi * params[1]
    beta = params[2] + 1j * params[3]

    reg = jqtc.Register([2, n_osc])
    circuit = jqtc.Circuit.create(reg, layers=[])
    for tx, ty, b in zip(theta_x, theta_y, beta):
        circuit.append(jqtc.Rx(tx), 0)
        circuit.append(jqtc.Ry(ty), 0)
        circuit.append(jqtc.CD(n_osc, b), [0, 1])
        circuit.append(jqtc.Rx(jnp.pi), 0)

    psi0 = jqt.basis(2, 0) ^ jqt.basis(n_osc, 0)
    return jqtc.simulate(circuit, psi0)[-1][-1].unit()

def fidelity(params, target):
    n_osc = target.dims[0][1]
    prepared = run_circuit(params, n_osc)
    return jnp.real(jqt.overlap(target, prepared))

batch_fidelity = jax.vmap(fidelity, in_axes=(0, None))

def loss(param_batch, target, lam):
    fids = batch_fidelity(param_batch, target)
    beta = param_batch[:, 2] + 1j * param_batch[:, 3]
    cd_penalty = jnp.mean(jnp.abs(beta)**2, axis=1)
    return jnp.mean(
        jnp.log10(1.0 - fids) + lam * cd_penalty
    )

loss_and_grad = jax.jit(jax.value_and_grad(loss))
\end{lstlisting}

\newpage

\subsection{GKP State-Preparation Compilation Parameters\label{sec: gkp state preparation compilation parameters}}

Table~\ref{tab: gkp state preparation compilations} lists the optimized parameters used to prepare the three GKP codewords shown in Fig.~\ref{fig: gkp state prep}(c). The listed values follow the block structure in Fig.~\ref{fig: gkp state prep}(a): each block applies $R_x(\theta_{x,n})$, $R_y(\theta_{y,n})$, and ${\rm ECD}(\beta_n)$ in sequence, with the fluxonium $\pi$ pulse included in the ECD gate.

\begin{table}[h]
    \centering
    \newcommand{\topspacing}{\rule{0pt}{2.4ex}}
    \newcommand{\bottomspacing}{\rule[-1.0ex]{0pt}{0.6ex}}
    \setlength{\tabcolsep}{12pt}
    \footnotesize
    \begin{tabular}{c c c c c}
        \hline\hline
        \topspacing\bottomspacing\textbf{Target} & $\bm n$ & $\bm{\theta_{x,n}/(2\pi)}$ & $\bm{\theta_{y,n}/(2\pi)}$ & $\bm{\beta_n}$ \\
        \hline
        \multirow{7}{*}{$\ket{+z_L}$}
        & \topspacing 1 & 0.233577 & 0.750000 & $-1.565885+0.828794i$ \\
        & 2 & 0.467114 & 0.324397 & $-0.759427-1.458915i$ \\
        & 3 & 0.529657 & 0.893687 & $-1.218707-0.458406i$ \\
        & 4 & 0.070825 & 0.652047 & $0.274143+0.896559i$ \\
        & 5 & 0.402077 & 0.124053 & $-0.871355+0.512794i$ \\
        & 6 & 0.272429 & 0.623766 & $-0.219406-0.217307i$ \\
        & \bottomspacing 7 & 0.250000 & 0.234211 & $0.000000+0.000000i$ \\
        \hline
        \multirow{7}{*}{$\ket{-z_L}$}
        & \topspacing 1 & 0.942039 & 0.249996 & $0.106873+1.646285i$ \\
        & 2 & 0.667897 & 0.550956 & $0.540238+0.860233i$ \\
        & 3 & 0.886543 & 0.981507 & $-1.341011+0.168722i$ \\
        & 4 & 0.205544 & 0.479635 & $0.179782+0.696264i$ \\
        & 5 & 0.144060 & 0.495411 & $-0.299945+1.140023i$ \\
        & 6 & 0.757505 & 0.003560 & $-0.316746-0.048403i$ \\
        & \bottomspacing 7 & 0.278294 & 0.750167 & $0.000041+0.000101i$ \\
        \hline
        \multirow{7}{*}{$\ket{+y_L}$}
        & \topspacing 1 & 0.750000 & 0.777851 & $-0.783496-0.925643i$ \\
        & 2 & 0.000000 & 0.247440 & $0.762469-0.639110i$ \\
        & 3 & 0.500000 & 0.758695 & $-0.749670-0.618603i$ \\
        & 4 & 0.999998 & 0.270014 & $-0.908635+0.949332i$ \\
        & 5 & 0.000037 & 0.742813 & $0.945703+0.954563i$ \\
        & 6 & 0.484028 & 0.749634 & $0.232630-0.216115i$ \\
        & \bottomspacing 7 & 0.750000 & 0.984028 & $0.000000+0.000000i$ \\
        \hline\hline
    \end{tabular}
    \caption{\textbf{GKP state-preparation compilation parameters.} Optimized rotation angles and conditional displacements for the $\Delta=0.438$ codewords shown in Fig.~\ref{fig: gkp state prep}(c).}
    \label{tab: gkp state preparation compilations}
\end{table}

\section{Improving Logical Lifetimes \label{sec: improving logical lifetimes}}

\subsection{Optimizing error correction \label{sec: optimizing error correction}}

In the presence of control imperfections and experimental noise, it is useful to tune the control parameters of sBs error correction in situ. While the available signal-to-noise ratio in our system appears insufficient for reinforcement learning~\cite{Sivak2023, Ding2023, Almanakly2025}, we were able to improve logical lifetimes by identifying and tuning important control knobs.

We optimize four such knobs of the small-big-small sequence introduced in Sec.~\ref{sec: error correction}. First, the large-displacement knob sets the coherent-state displacement length of the central Big ECD, whose ideal value is $\ell\cosh\Delta^2\approx\ell$ for the square-lattice spacing $\ell=\sqrt{2\pi}$. The next knob, the final resonator rotation, is applied as a small virtual storage-frame update after each sBs stabilizer to compensate residual coherent phase accumulation; this is distinct from the programmed $\pi/2$ change of ECD direction that exchanges the $x$- and $p$-quadrature stabilizers. Finally, the small displacements in sBs are parameterized by two remaining knobs: the finite-energy squeezing parameter $\Delta$ and the ratio of the two small displacements $r\coloneqq\epsilon_2/\epsilon_1$. The squeezing parameter $\Delta$ is related to the first small displacement $\epsilon_1$ by $\epsilon_1=\ell\Delta^2/2$, which is the small-$\Delta$ approximation to $(\ell/2)\sinh\Delta^2$. 

We optimize these controls in three sequential sweeps acquired under the same device conditions. Starting from the original control settings, we varied the large ECD displacement from 2 to 3, then the post-sBs storage-frame rotation from $\SI{-4}{\degree}$ to $\SI{4}{\degree}$, and finally $\Delta$ and $r$ over $0.30\leq\Delta\leq0.60$ and $1.0\leq r\leq2.0$. Within each sweep, all previously selected parameters were held fixed and the sampled maximum was carried into the next stage. At every sampled setting, we measured the decay of $\ev{Z_L}_{-}$ over 15 correction rounds and fit its magnitude to $|\ev{Z_L}_{-}|=A\exp(-t/T_Z)$. Full $\ev{X_L}_{-}$ and $\ev{Z_L}_{-}$ baseline decays before the optimization and checkpoint decays after each stage provided independent evaluations of the sequential optimization process.

\begin{figure}[h]
    \centering
    \includegraphics[width=\linewidth]{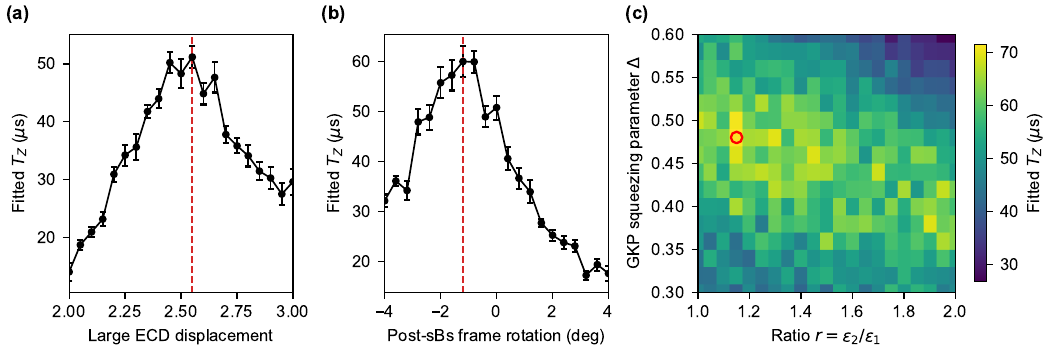}
    \caption{\textbf{Sequential optimization of GKP error-correction controls.} \textbf{(a)} Fitted $T_Z$ versus the central large-ECD displacement length, with the selected sampled value 2.55 indicated by the dashed line. \textbf{(b)} Fitted $T_Z$ versus the virtual storage-frame rotation applied after each sBs stabilizer, with the selected value $\SI{-1.2}{\degree}$ indicated by the dashed line and the large displacement fixed at 2.55. \textbf{(c)} Map of fitted $T_Z$ versus the finite-energy squeezing parameter $\Delta$ of the stabilized GKP qubit and the small-ECD ratio $r=\epsilon_2/\epsilon_1$, where $\Delta$ sets $\epsilon_1$ and $r$ sets $\epsilon_2$, with the selected point $(\Delta,r)=(0.48,1.15)$ marked by the red circle and the preceding controls fixed. Each fit uses the decay of $\ev{Z_L}_{-}$ measured over 15 correction rounds. Error bars are one-standard-deviation fit uncertainties.}
    \label{si-fig: gkp lifetime optimization}
\end{figure}

\par\medskip
Before this optimization, the controls were a large-ECD displacement of 2.37, a post-sBs frame rotation of $\SI{0}{\degree}$, and $(\Delta,r)=(0.42,1.30)$, corresponding to $(\epsilon_1,\epsilon_2)=(0.221,0.287)$, all of which were calibrated in a past optimization. This baseline gave $T_X=\SI{43.3\pm1.1}{\micro\second}$ and $T_Z=\SI{41.5\pm1.0}{\micro\second}$. After optimizing the large displacement, final rotation, and small displacements, the corresponding checkpoints were $(T_X,T_Z)=(\SI{53.8\pm1.6}{\micro\second},\SI{52.4\pm1.4}{\micro\second})$, $(\SI{59.8\pm1.8}{\micro\second},\SI{53.8\pm1.9}{\micro\second})$, and $(\SI{69.0\pm1.6}{\micro\second},\SI{66.4\pm2.3}{\micro\second})$, respectively. These measurements form one internally consistent optimization dataset and are not intended to reproduce the representative or device-averaged lifetime values quoted elsewhere in the manuscript. These sweeps demonstrate that the optimal values of the sBs parameters can deviate significantly from their ideal theoretical counterparts in an actual device.

\subsection{Corrected and Uncorrected Logical Lifetimes \label{sec: corrected and uncorrected logical lifetimes}}

The fitted logical lifetimes and corrected-to-uncorrected ratios corresponding to the bar plot in Fig.~\ref{fig: error correction}(b) of the main text are listed in Table~\ref{tab: gkp logical lifetimes}.

\begin{table}[h]
    \centering
    \newcommand{\topspacing}{\rule{0pt}{2.4ex}}
    \newcommand{\bottomspacing}{\rule[-1.0ex]{0pt}{0.6ex}}
    \setlength{\tabcolsep}{12pt}
    \begin{tabular}{c c c c}
        \hline\hline
        \topspacing\bottomspacing\textbf{Lifetime} & \textbf{Uncorrected} & \textbf{Corrected} & \textbf{Ratio} \\
        \hline
        \topspacing $T_X$ & \SI{49.6\pm1.6}{\micro\second} & \SI{73.6\pm1.2}{\micro\second} & $1.48\pm0.05$ \\
        $T_Y$ & \SI{23.2\pm1.1}{\micro\second} & \SI{39.1\pm1.3}{\micro\second} & $1.69\pm0.10$ \\
        $T_Z$ & \SI{46.2\pm1.3}{\micro\second} & \SI{70.3\pm1.5}{\micro\second} & $1.52\pm0.05$ \\
        \bottomspacing $T_{\rm avg}$ & \SI{35.3\pm0.9}{\micro\second} & \SI{56.2\pm1.0}{\micro\second} & $1.59\pm0.05$ \\
        \hline\hline
    \end{tabular}
    \caption{\textbf{Corrected and uncorrected GKP logical lifetimes.} Following Ref.~\cite{Sivak2023}, the average lifetime is defined from the mean logical decay rate as $T_{\rm avg}=3/(T_X^{-1}+T_Y^{-1}+T_Z^{-1})$. Lifetime uncertainties are one-standard-deviation fit uncertainties; average-lifetime and ratio uncertainties are propagated assuming independent fits.}
    \label{tab: gkp logical lifetimes}
\end{table}

\subsection{Logical Error Budget \label{sec: logical error budget}}

\begin{figure}[h]
    \centering
    \includegraphics[width=\linewidth]{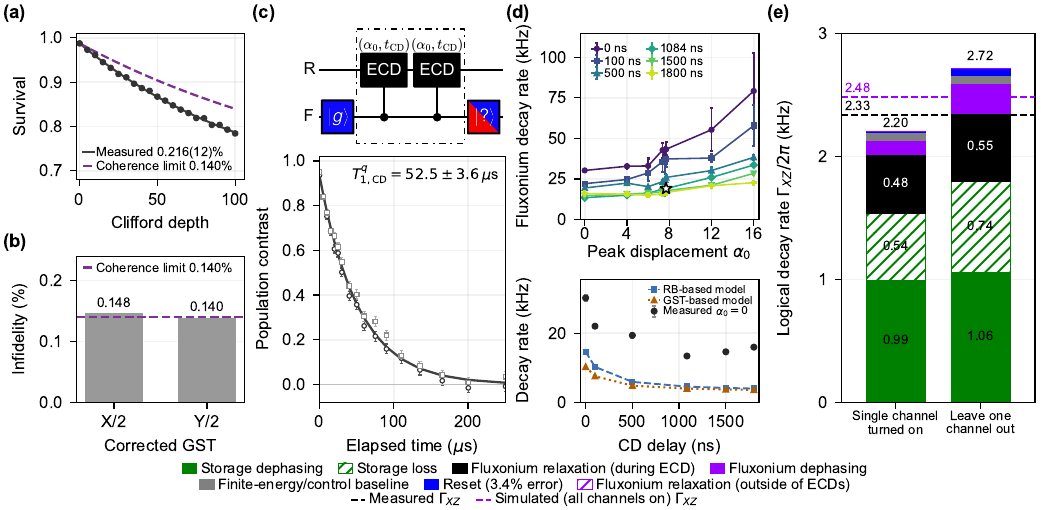}
    \caption{\textbf{Self-consistent characterization and logical error budget.} \textbf{(a)} Fluxonium randomized benchmarking survival probability versus Clifford depth, together with the measured fit and the coherence-limited decay calculated from the contemporaneous fluxonium lifetimes. The quoted infidelities are per native gate. \textbf{(b)} Corrected gate-set tomography (GST) infidelities of the $X/2$ and $Y/2$ gates, with the common coherence limit indicated by the dashed purple line. \textbf{(c)} Fluxonium decay measurement under repeated ${\rm ECD}(\alpha_0,t_{\rm CD})$ pairs at the largest sBs ECD operating point. The two marker sets are independent acquisitions; their combined exponential fit gives $T_{1,{\rm CD}}^q=\SI{52.5\pm3.6}{\micro\second}$. \textbf{(d)} Top: repeated-ECD fluxonium decay rate versus $\alpha_0$ for the indicated values of $t_{\rm CD}$; the star marks the large-ECD operating point, $\alpha_0=7.70$ and $t_{\rm CD}=\SI{1084}{\nano\second}$. Bottom: measured $\alpha_0=0$ rates and independent-error predictions constructed from the contemporaneous RB or GST gate infidelity and $T_1^q$.  \textbf{(e)} Logical-$X/Z$-averaged error-channel attributions, with contributions stacked in descending order. Because the all-channels-on simulated $X$ and $Z$ total rates differ by only $(1.20\times10^{-4})\%$, we plot the logical-$X/Z$-averaged rate $\Gamma_{XZ}\coloneqq(\Gamma_X+\Gamma_Z)/2$. The left bar enables each channel separately for comparison against the finite-energy/control baseline logical decay rate [see details in the text]; the right bar reports the decay-rate drop when each channel is removed from the all-channels-on simulation. These two calculations of physical error channel contributions to the logical lifetime are averaged to obtain the bar plot shown in Fig.~\ref{fig: error correction}(d). The dashed black and purple lines show the measured and all-channels-on simulated $\Gamma_{XZ}$, respectively.}
    \label{si-fig: repeated ecd error budget}
\end{figure}

Figure~\ref{si-fig: repeated ecd error budget} tests the predictive consistency of a pulse-level logical-error model using a single set of contemporaneous characterization measurements. Fluxonium $T_1^q$ and $T_{2E}^q$, randomized benchmarking (RB), gate-set tomography (GST), repeated-ECD decay, reset fidelity, and corrected GKP logical lifetimes were measured under the same device conditions. The storage values $T_1^a=\SI{103\pm2}{\micro\second}$ and $T_\phi^a=\SI{1.08\pm0.16}{\milli\second}$ were retained from the measurements in Figs.~\ref{si-fig: storage spec relaxation} and~\ref{si-fig: storage dephasing}, respectively. The contemporaneous fluxonium values, $T_1^q=\SI{331\pm24}{\micro\second}$ and $T_{2E}^q=\SI{36.1\pm3.6}{\micro\second}$, differ from the longer-term averages reported elsewhere in this work because the fluxonium lifetimes vary over time. The purpose of Fig.~\ref{si-fig: repeated ecd error budget} is therefore not to replace those device-average values, but to compare physical characterization, simulation, and logical decay using one internally consistent parameter set.

The RB and GST measurements in Fig.~\ref{si-fig: repeated ecd error budget}(a,b) provide an independent check of the error model used for the $\SI{144}{\nano\second}$ single-qubit rotations. Using $T_{2E}^q$ as the gate-time coherence estimate and assuming Markovian amplitude damping and dephasing, we can calculate the coherence-limited average gate infidelity as~\cite{OBrien2017}:
\begin{equation}
    r_{\rm coh}(t_g)=\frac{3-2e^{-t_g/T_{2E}^q}-e^{-t_g/T_1^q}}{6}.
\end{equation}
The measured RB infidelity is $\SI{0.216\pm0.012}{\percent}$ per native gate, using an average of $1.875$ native gates per Clifford, compared with the corresponding coherence limit of $\SI{0.140}{\percent}$ [Fig.~\ref{si-fig: repeated ecd error budget}(a)]. The corrected GST compilation, verified against all 448 executed circuits, gives infidelities of $\SI{0.148}{\percent}$ and $\SI{0.140}{\percent}$ for $X/2$ and $Y/2$, respectively, near the same coherence limit [Fig.~\ref{si-fig: repeated ecd error budget}(b)]. These measurements support modeling each single-qubit rotation as an ideal rotation followed by amplitude-damping and pure-dephasing channels applied over the pulse duration and parameterized by the measured $T_1^q$ and $T_\phi^q=[1/T_{2E}^q-1/(2T_1^q)]^{-1}$.

Fluxonium decay during an ECD cannot be inferred reliably from the undriven $T_1^q$ alone because the fluxonium is repeatedly inverted while the storage follows a large phase-space trajectory. It is characterized directly by initializing the fluxonium in the ground state and the resonator in vacuum and repeating the pair ${\rm ECD}(\alpha_0,t_{\rm CD})\,{\rm ECD}(-\alpha_0,t_{\rm CD})$, which applies and then undoes the conditional displacement. The fluxonium population is then measured versus the number of ECD pairs and fit to an exponential decay by accounting for the total time taken by each ECD gate. The population data from two independent 1000-shot repetitions at the large-ECD operating point were averaged before fitting, giving $T_{1,{\rm CD}}^q=\SI{52.5\pm3.6}{\micro\second}$ [Fig.~\ref{si-fig: repeated ecd error budget}(c)]. Figure~\ref{si-fig: repeated ecd error budget}(d) shows that this effective fluxonium decay rate increases with $\alpha_0$, particularly at short $t_{\rm CD}$, and decreases as $t_{\rm CD}$ increases.

For the $\alpha_0=0$ comparison in the bottom subplot, each two-ECD pair has duration $t_{\rm pair}=2(2t_{\rm CD}+4t_D+2t_{\rm pad}+t_\pi)$, where $t_D=\SI{40}{\nano\second}$ is the displacement pulse length in this dataset, $t_{\rm pad}=\SI{16}{\nano\second}$ is the padding time on either side of the ECD $\pi$ pulse for pulse alignment, and $t_\pi=\SI{144}{\nano\second}$ is the fluxonium $\pi$-pulse length. Here, $r_g=1-F_g$ denotes the measured per-gate infidelity used for each model curve: the native-gate RB infidelity for the RB curve, or the mean corrected $X/2$ and $Y/2$ GST infidelity used as a native-$X$ proxy for the GST curve. Under a depolarizing-error approximation, this gate infidelity $r_g$ is incorporated by multiplying the difference between the ground- and excited-state populations by $\lambda_g=1-2r_g$, giving
\begin{equation}
    \Gamma_{\alpha_0=0}(t_{\rm CD})=-\frac{1}{t_{\rm pair}}\ln\!\left[\lambda_g^2\exp\!\left(-\frac{t_{\rm pair}-2t_\pi}{T_1^q}\right)\right].
\end{equation}
In the above expression, we first multiply the $T_1^q$ population-contrast decay accumulated outside the two $\pi$-pulse intervals by the infidelity factor $\lambda_g^2$ corresponding to the two imperfect $\pi$ pulses. Then, we convert the total contrast reduction per ECD pair into an effective decay rate using $-\ln(\cdot)/t_{\rm pair}$. Despite this comprehensive modeling, the measured $\alpha_0=0$ decay rate exceeds both the RB- and GST-based predictions at every $t_{\rm CD}$. Thus, the measured gate fidelities and $T_1^q$ do not account for the full repeated-ECD decay, and the origin of this gap remains unresolved. The effective $T_{1,{\rm CD}}^q$ empirically includes any relaxation, excitation, leakage, or other errors that change the population contrast and should not be interpreted as the bare fluxonium relaxation time.

In the logical simulation, the operating-point value of $T_{1,{\rm CD}}^q=\SI{52.5\pm3.6}{\micro\second}$ replaces the bare $T_1^q$ during each of the three ECD gates. This replacement explains the small bare-$T_1^q$ contribution in Fig.~\ref{si-fig: repeated ecd error budget}(e), since fluxonium transitions during the ECD gates are instead attributed to the ``Fluxonium relaxation (during ECD)'' channel. For each modeled half-ECD of duration $\tau_{{\rm CD}/2}=t_{\rm CD}+2t_D+t_{\rm pad}+t_\pi/2$, the simulation uses four possible transition times $t_k=(k+1/2)\tau_{{\rm CD}/2}/4$, with $k\in\{0,1,2,3\}$, and equal excitation and relaxation probabilities
$p_\uparrow=p_\downarrow=[1-\exp(-\tau_{{\rm CD}/2}/T_{1,{\rm CD}}^q)]/2$. These probabilities apply to the full half-ECD and are divided equally among the four potential fluxonium bit-flip times. The half-ECD map contains the ideal net conditional displacement and a displacement modification produced by a fluxonium bit-flip error at one of the four possible times. The Kraus operators for the first half of ${\rm ECD}(\beta)$ are then
\begin{align}
    \op{K}_0
    &=
    \sqrt{1-p_\uparrow}\ketbra{g}{g}\otimes D(\beta/4)
    +\sqrt{1-p_\downarrow}\ketbra{e}{e}\otimes D(-\beta/4),\\
    \op{K}_{\downarrow,k}
    &=
    \sqrt{\frac{p_\downarrow}{4}}\,
    \op{\sigma}_-\otimes
    D\!\left[-\frac{\beta}{4}\left(\frac{2t_k}{\tau_{{\rm CD}/2}}-1\right)\right],\\
    \op{K}_{\uparrow,k}
    &=
    \sqrt{\frac{p_\uparrow}{4}}\,
    \op{\sigma}_+\otimes
    D\!\left[\frac{\beta}{4}\left(\frac{2t_k}{\tau_{{\rm CD}/2}}-1\right)\right],
\end{align}
where $\op{\sigma}_-=\ketbra{g}{e}$ and $\op{\sigma}_+=\ketbra{e}{g}$. Fluxonium pure dephasing is applied after each half-ECD using
\begin{equation}
    \op{L}_0=\sqrt{\frac{1+e^{-\tau_{{\rm CD}/2}/T_\phi^q}}{2}}\,\op{I},
    \qquad
    \op{L}_1=\sqrt{\frac{1-e^{-\tau_{{\rm CD}/2}/T_\phi^q}}{2}}\,\op{\sigma}_z,
\end{equation}
where identities on the storage are implicit. Storage relaxation is applied over the same interval through the zero-temperature bosonic-loss channel
\begin{equation}
    \mathcal{A}_a(\rho)
    =\sum_{r=0}^{\infty}\op{A}_r\rho\op{A}_r^\dagger,
    \qquad
    \op{A}_r=
    \sqrt{\frac{(1-\eta_a)^r}{r!}}\,
    \eta_a^{\op{a}^\dagger\op{a}/2}\op{a}^r,
    \qquad
    \eta_a=e^{-\tau_{{\rm CD}/2}/T_1^a},
\end{equation}
and storage dephasing is applied as in Ref.~\cite{KyungjooPhDThesis}:
\begin{equation}
    \bra{m}\mathcal{D}_a(\rho)\ket{n}
    =
    e^{-\tau_{{\rm CD}/2}(m-n)^2/T_\phi^a}
    \bra{m}\rho\ket{n}.
\end{equation}
The complete map for one half-ECD is
\begin{equation}
    \mathcal{E}_{\beta}^{(1/2)}(\rho)
    =\sum_{j=0}^{1}\op{L}_j \left\{
    \left(
        \mathcal{D}_a
        \circ
        \mathcal{A}_a
    \right)
    \left[
        \op{K}_0\rho\op{K}_0^\dagger+
        \sum_{k=0}^{3}\left(
            \op{K}_{\downarrow,k}\rho\op{K}_{\downarrow,k}^\dagger+
            \op{K}_{\uparrow,k}\rho\op{K}_{\uparrow,k}^\dagger
        \right)
    \right]\right\}\op{L}_j^\dagger.
\end{equation}
The ideal echoing $\pi$ pulse $\op{R}_x(\pi)$ is applied separately between the two halves, giving the following expression for the full-ECD channel:
\begin{equation}
    \mathcal{E}_{\rm ECD}
    =\mathcal{E}_{-\beta}^{(1/2)}
    \circ\mathcal{U}_{R_x(\pi)}
    \circ\mathcal{E}_{\beta}^{(1/2)}.
\end{equation}

The logical simulation evolves the joint fluxonium-storage density matrix through the measured sBs sequence, applying coherent operations as unitary transformations and error channels as Kraus maps or mathematically equivalent direct density-matrix updates. The model includes the experimental pulse durations and alternating phase-space directions used for GKP stabilization. Implemented in JAXQuantum, the simulations benefit from just-in-time compilation, vectorization over noise variants, and GPU execution.

The physical error channels we consider in the simulation are storage photon loss, storage pure dephasing, fluxonium relaxation outside the ECD gates, fluxonium pure dephasing, the effective ``Fluxonium relaxation (during ECD)'' channel described above, and imperfect fluxonium reset. Storage loss and dephasing use the independently measured values $T_1^a$ and $T_\phi^a$ described above. Fluxonium relaxation and dephasing use the contemporaneous $T_1^q$ and $T_{2E}^q$ measured by the tune-up protocol in Sec.~\ref{sec: Fluxonium Daily Tune-Up}. The reset protocol and fidelity estimate are described in Sec.~\ref{sec: Pulsed f0g1 Reset of Fluxonium}; this measured reset error is applied after each stabilizer. Storage and fluxonium amplitude damping and dephasing are applied over their respective pulse segments, while the effective fluxonium decay rate taken from repeated-ECD measurements is implemented over the conditional displacement trajectory in simulation.

To construct the error budget in Fig.~\ref{si-fig: repeated ecd error budget}(e), the logical lifetime is first simulated with all measured physical noise channels disabled. We report $\Gamma_{XZ}$, the average of the simulated logical-$X$ and logical-$Z$ decay rates. The resulting decay, which surprisingly has a small but nonzero baseline rate $\Gamma_{XZ,0}/(2\pi)=\SI{0.061}{\kilo\hertz}$, is not due to any physical decoherence channel, but instead arises from an interplay between numerical effects, the finite-energy nature of the GKP states, and the trotterization of the sBs error correction protocol (seen also in Refs. \cite{Royer2020, Chowdhury2024Thesis}). This finite-energy/control decay rate nevertheless provides a baseline to compare against in the following simulations.

The physical error channels in our device do not independently affect the decay of logical information. So, we attempt to isolate the impact of each channel in two ways. First, in the \textit{single-channel-turned-on} simulation, each physical channel $i$ is enabled separately and contributes $\Delta\Gamma_{XZ,i}^{\rm on}=\Gamma_{XZ,i}-\Gamma_{XZ,0}$. The resulting averaged decay-rate contributions $\Delta\Gamma_{XZ,i}^{\rm on}/(2\pi)$ are $\SI{0.993}{\kilo\hertz}$ from storage dephasing, $\SI{0.543}{\kilo\hertz}$ from storage loss, $\SI{0.476}{\kilo\hertz}$ from the ``Fluxonium relaxation (during ECD)'' channel, $\SI{0.121}{\kilo\hertz}$ from fluxonium dephasing, $\SI{0.0085}{\kilo\hertz}$ from reset, and $\SI{0.0016}{\kilo\hertz}$ from fluxonium relaxation outside of ECDs.

Next, the complementary \textit{leave-one-channel-out} method compares $\Gamma_{XZ,{\rm all}\setminus i}$ during a simulation with only one channel turned off against the all-channels-on rate $\Gamma_{XZ}^{\rm all}$. In particular, we calculate $\Delta\Gamma_{XZ,i}^{\rm out}=\Gamma_{XZ}^{\rm all}-\Gamma_{XZ,{\rm all}\setminus i}$. The averaged decay-rate contributions $\Delta\Gamma_{XZ,i}^{\rm out}/(2\pi)$ are $\SI{1.06}{\kilo\hertz}$ from storage dephasing, $\SI{0.741}{\kilo\hertz}$ from storage loss, $\SI{0.546}{\kilo\hertz}$ from the ``Fluxonium relaxation (during ECD)'' channel, $\SI{0.251}{\kilo\hertz}$ from fluxonium dephasing, $\SI{0.0576}{\kilo\hertz}$ from reset, and $\SI{0.0033}{\kilo\hertz}$ from fluxonium relaxation outside of ECDs.

Including the common baseline, the \textit{single-channel-turned-on} and \textit{leave-one-channel-out} stacks sum to $\Gamma_{XZ}^{\rm total,on}/(2\pi) = \SI{2.20}{\kilo\hertz}$ and $\Gamma_{XZ}^{\rm total,out}/(2\pi) = \SI{2.72}{\kilo\hertz}$, respectively. The all-channels-on simulation gives $\Gamma_{XZ}^{\rm all}/(2\pi) = \SI{2.48}{\kilo\hertz}$, while the measured logical-$X/Z$ average is $\Gamma_{XZ}/(2\pi) = \SI{2.33\pm0.03}{\kilo\hertz}$, corresponding to $T_{XZ,\mathrm{avg}} = \SI{68.18\pm0.94}{\micro\second}$. Because the fitted logical decay rate depends nonlinearly on the strengths of simultaneously enabled noise channels, the two attribution sums need not equal the all-channels-on rate. In this calculation, the two sums bracket the all-channels-on rate, and their difference quantifies how channel interactions are assigned by the two constructions. No fitted residual is added. The black dashed line in Fig.~\ref{si-fig: repeated ecd error budget}(e) directly reports the measured logical rate, while the purple dashed line shows the independent all-channels-on simulation. Agreement at this level, despite the reduced and time-varying fluxonium coherence, demonstrates that the independently measured physical loss rates describe the logical decay without tuning an additional error channel.

\section{Other Advantages of the Fluxonium Control Qubit \label{sec: other advantages of the fluxonium control qubit}}

\begin{figure}[h]
    \centering
    \includegraphics[width=\linewidth]{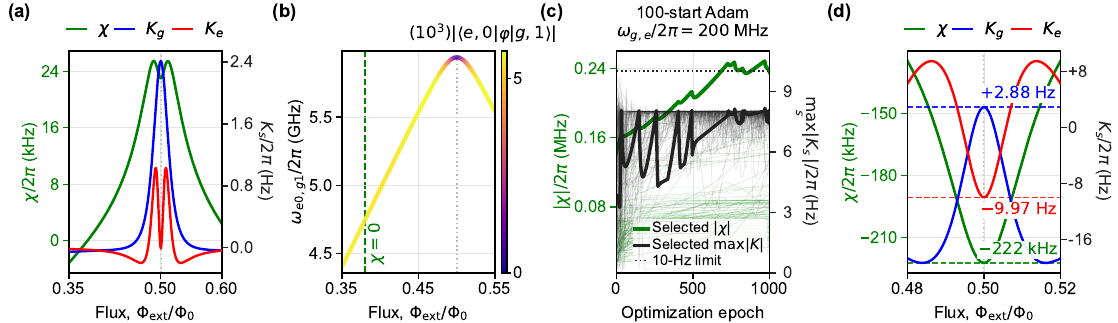}
    \caption{\textbf{Other advantages of the fluxonium control qubit.} \textbf{(a) Flux-tunable storage interaction for the primary measured device.} Storage-fluxonium dispersive coupling $\chi$ (left axis) and storage self-Kerr coefficients $K_g, K_e$ (right axis) from full numerical diagonalization. The simulated storage coupling is calibrated to the measured half-flux $\chi/2\pi=\SI{23}{\kilo\hertz}$; all other device parameters are taken from Table~\ref{tab: summary of parameters}. \textbf{(b) Direct sideband cooling reset.} Frequency of the $\ket{e,0}\leftrightarrow\ket{g,1}$ sideband transition, with line color indicating the dimensionless phase matrix element $|\mel{e,0}{\op{\varphi}}{g,1}|$. The sideband is parity forbidden at half flux and enabled away from it. The green dashed line marks the storage-fluxonium zero-$\chi$ bias point. \textbf{(c) Differentiable device optimization.} Metrics $|\chi|$ (left axis) and $\max(|K_g|,|K_e|)$ (right axis) versus epoch for 100 independent starts over 1000 epochs at half flux, with a constraint enforcing the bare fluxonium frequency to be $\omega_{g,e}/2\pi=\SI{200}{\mega\hertz}$. Thin curves show the 96 trajectories that remain below the $\SI{10}{\hertz}$ rejection threshold throughout; the bold curves show the trajectory containing the optimal, selected parameter set. The Kerr trajectories plateau near the $\SI{8}{\hertz}$ soft-penalty onset, while the dotted black line marks the $\SI{10}{\hertz}$ rejection threshold. \textbf{(d) Optimized-device sweet-spot sweep.} A dense full-numerical sweep over $0.48\leq\Phi_{\rm ext}/\Phi_0\leq0.52$ resolves the dispersive coupling and state-dependent Kerr coefficients near half flux. The dashed horizontal lines label the half-flux value of each solid curve.}
    \label{si-fig: fluxonium advantages}
\end{figure}

Flux tunability of the fluxonium provides an in situ interaction switch that is unavailable in fixed-frequency control qubits without an additional tunable coupler mode. Figure~\ref{si-fig: fluxonium advantages}(a) uses the measured-device parameters reported in Table~\ref{tab: summary of parameters} and varies the external flux bias $\Phi_{\rm ext}$. The storage-fluxonium coupling $g$ in Eq.~\eqref{eq:fluxonium-resonator-hamiltonian} is inferred by matching the measured half-flux $\chi/2\pi=\SI{23}{\kilo\hertz}$ under numerical diagonalization, yielding $g/2\pi=\SI{1.626}{\mega\hertz}$. Notably, we observe that the dispersive coupling strength $\chi$ crosses through zero near $\Phi_{\rm ext}=0.38\,\Phi_0$, while the Kerr nonlinearity coefficients remain small in magnitude. This zero-$\chi$ bias point is accessible via fast-flux control and could be used to decouple the storage and fluxonium in real time to suppress fluxonium backaction during reset \cite{Chowdhury2024Thesis, Atanasova2025}.

The readout mode provides another opportunity to exploit the flux tunability of the fluxonium control qubit. Again, in the following simulations, the readout-fluxonium coupling is inferred by matching the measured $|\chi_{qr}|/2\pi=\SI{7.0}{\mega\hertz}$ under numerical diagonalization, yielding $g_{qr}/2\pi=\SI{39.9}{\mega\hertz}$. The direct $\ket{e,0}\leftrightarrow\ket{g,1}$ sideband transition is parity forbidden at half flux, as it is for a fixed-frequency transmon. Unlike a fixed-frequency transmon, the fluxonium can be tuned away from its symmetry point with a fast-flux pulse to enable this sideband [Fig.~\ref{si-fig: fluxonium advantages}(b)]. In particular, it is allowed at the storage-fluxonium zero-$\chi$ bias, permitting sideband cooling reset while suppressing storage backaction~\cite{NajeraSantos2024}. This could provide a simpler and potentially faster reset than the $\ket{f,0}\leftrightarrow\ket{g,1}$ protocol used in this work.

The same model can also be used for gradient-based device co-design. JAXQuantum~\cite{jha2024jaxquantum} exposes the dressed eigenenergy metrics as JIT-compilable and automatically differentiable functions, allowing the Hamiltonian construction, constraints, and gradients to be expressed within the same numerical model. As one example, we optimize $E_C^q$, $E_J^q$, $E_L^q$, the storage frequency, and the storage-fluxonium coupling $g$ while enforcing the following constraints: $\omega_{g,e}/2\pi=\SI{200}{\mega\hertz}$ at half flux, a storage frequency above $\SI{1}{\giga\hertz}$, and $|K_g|,|K_e|\leq\SI[parse-numbers=false]{2 \pi \times 10}{\hertz}$. The objective applies a soft Kerr penalty above $\SI{8}{\hertz}$, whereas $\SI{10}{\hertz}$ is used only to reject parameter sets. The optimized trajectories therefore tend to plateau near $\SI{8}{\hertz}$ rather than at the rejection threshold. A deterministic 2048-point prescreen supplies 100 feasible initial parameter sets to a 1000-epoch batched Adam optimization [Fig.~\ref{si-fig: fluxonium advantages}(c)]. All 100 parameter vectors are advanced in parallel by applying one JIT-compiled value-and-gradient function over the batch of system parameters. Each initial seed retains its own Adam moments and objective value, so batching accelerates evaluation without coupling the independent trajectories. The highest-$|\chi|$ checkpoint satisfying the Kerr constraints is selected. This checkpoint occurs at epoch 609 of the highlighted trajectory and gives
\begin{equation}
    \frac{(E_C^q,E_J^q,E_L^q)}{h}=(\SI{0.966}{\giga\hertz},\SI{4.281}{\giga\hertz},\SI{0.535}{\giga\hertz}),\quad
    \frac{\omega_a}{2\pi}=\SI{2.468}{\giga\hertz},\quad
    \frac{g}{2\pi}=\SI{12.512}{\mega\hertz}.\label{eq: optimized fluxonium parameters}
\end{equation}
This set of device parameters (all of which are experimentally accessible within our fabrication process) yields $\chi/2\pi=\SI{-222.091}{\kilo\hertz}$, $K_g/2\pi=\SI{2.883}{\hertz}$, and $K_e/2\pi=\SI{-9.971}{\hertz}$. This optimized example reaches $9.7$ times the half-flux $|\chi|$ of the primary measured device while satisfying both half-flux Kerr constraints, illustrating that a high-coupling regime is accessible to a fluxonium-resonator system without necessarily imposing a large resonator nonlinearity. The sweep in Fig.~\ref{si-fig: fluxonium advantages}(d) shows the stability of $\chi$, $K_g$, and $K_e$ near half flux.

The ECD rate scales as $\chi|\alpha|$ for an intermediate displacement $\alpha$, whereas the fluxonium Stark shift (which quantifies how effectively storage photons ``drive'' the qubit) scales as $\chi|\alpha|^2$. The onset of DUST likewise occurs at some given Stark shift threshold $|\alpha|^2 = C/\chi$, where the constant $C$ is approximately fixed for a given set of fluxonium and storage parameters across coupling strengths. In the strong-dispersive limit of Eq.~\eqref{eq: photon shot noise dephasing}, described for the storage resonator in Sec.~\ref{sec: Resonator Spectroscopy and Relaxation}, the photon-shot-noise dephasing rate is $\Gamma_\phi\approx|\alpha|^2/T_1^a$. For a fixed target ECD rate $\Omega_{\rm ECD}\propto\chi|\alpha|$, this gives $\Gamma_\phi\propto\Omega_{\rm ECD}^2/(\chi^2T_1^a)$. Increasing $\chi$ therefore achieves the same ECD rate with fewer photons, which, in turn, decreases the shot-noise dephasing rate on the fluxonium during operation. The optimization above serves only to illustrate the accessible design space and demonstrate a JAXQuantum-based workflow; it does not establish that this parameter set is globally optimal. Candidate fluxonium-resonator designs still require analyses of DUST, coherence limits, fabrication-sensitivity, and reset, along with consideration of other practical constraints.

Finally, the low fluxonium transition frequency enables direct control with baseband flux pulses rather than resonant microwave driving~\cite{Campbell2020, Zhang2021}. SFQ and AQFP architectures may enable local cryogenic pulse generation, further reducing control-line and room-temperature hardware overhead~\cite{Li2019SFQ,Kannan2024AQFP,Takeuchi2024AQFP}.

\putbib[citations]
\end{bibunit}

\fi

\end{document}